\documentclass[12pt]{article}
\usepackage{subfigure}
\usepackage{amssymb}
\usepackage{color}
\usepackage{calc}
\usepackage[latin1]{inputenc}
\usepackage{amsfonts}
\usepackage{amsmath}
\usepackage[english]{babel}
\usepackage[T1]{fontenc}
\usepackage{verbatim}
\usepackage{subfig}
\usepackage{enumerate}
\usepackage{graphicx}
\usepackage{natbib}
\usepackage{url}

\def\Bmp#1{ \begin{minipage}{#1} }
\def\Bmpc#1{ \begin{minipage}[c]{#1} }
\def\Bmpt#1{ \begin{minipage}[t]{#1} }
\def\Bmpb#1{ \begin{minipage}[b]{#1} }
\def\Emp{ \end{minipage} }

\def\E{{\mathcal{E}}}
\def\G{{\mathcal{G}}}

\def\O{\mbox{\textit{O}}}
\def\P{{\mathcal{P}}}
\def\R{{\mathcal{R}}}

\def\K{{\mathcal{K}}}
\def\S{{\mathcal{S}}}
\def\Q{{\mathcal{Q}}}

\def\tpEP{\tilde{\psi}_{\E_0,\P_0}}
\def\tpKP{\tilde{\psi}_{\K_0,\P_0}}
\def\tpP{\tilde{\psi}_{\P_0}}
\def\tp{\tilde{\psi}}
\def\tf0{\tilde{\varphi}_{0}}

\def\PP{{\mathbb{P}}}
\def\RR{{\mathbb{R}}}

\def\x{{\bf x}}

\def\v{{\bf v}}
\def\u{{\bf u}}
\def\0{{\bf 0}}

\def\bnabla{\boldsymbol{\nabla}}

\def\bxi{\boldsymbol{\xi}}

\def\Dpartial#1#2{ {\frac{\partial #1}{\partial #2} }}

\newcommand{\uvec}{\mathbf{u}}

\newcommand{\argmax}{\operatorname{argmax}}

\newcommand{\Id}{\operatorname{Id}}
\newcommand{\Ker}{\operatorname{Ker}}

\begin{document}
\title{Maximum Palinstrophy Growth in 2D Incompressible Flows}

\author{Diego Ayala and Bartosz Protas\thanks{Email address for correspondence: bprotas@mcmaster.ca}
\\
Department of Mathematics and Statistics, McMaster University \\
Hamilton, Ontario, L8S 4K1, Canada
}
\date{\today}
\maketitle

\begin{abstract}
  In this study we investigate vortex structures which lead to the
  maximum possible growth of palinstrophy in two-dimensional
  incompressible flows on a periodic domain.  The issue of
  palinstrophy growth is related to a broader research program
  focusing on extreme amplification of vorticity-related quantities
  which may signal singularity formation in different flow models.
  Such extreme vortex flows are found systematically via numerical
  solution of suitable variational optimization problems.  We identify
  several families of maximizing solutions parameterized by their
  palinstrophy, palinstrophy and energy and palinstrophy and
  enstrophy. Evidence is shown that some of these families saturate
  estimates for the instantaneous rate of growth of palinstrophy
  obtained using rigorous methods of mathematical analysis, thereby
  demonstrating that this analysis is in fact sharp. In the limit of
  small palinstrophies the optimal vortex structures are found
  analytically, whereas for large palinstrophies they exhibit a
  self-similar multipolar structure.  It is also shown that the time
  evolution obtained using the instantaneously optimal states with
  fixed energy and palinstrophy as the initial data saturates the
  upper bound for the maximum growth of palinstrophy in finite time.
  Possible implications of this finding for the questions concerning
  extreme behavior of flows are discussed.

\end{abstract}

\begin{flushleft}
Keywords: 2D Navier-Stokes equation, maximum growth, palinstrophy, variational optimization, vortex dynamics
\end{flushleft}

\section{Introduction}
\label{sec:intro}

This work makes a contribution to a broader research effort concerning
systematic characterization of extreme events in hydrodynamic systems.
In addition to their independent physical interest, such questions are
intrinsically related to the problem of finite-time singularity
formation in various flow models, which is one of the issues at the
center of mathematical fluid mechanics. In the context of the
three-dimensional (3D) Navier-Stokes system in an unbounded or
periodic domain, the key problem concerns the existence for
arbitrarily large times of smooth solutions corresponding to arbitrary
initial data \citep{d09}. To date, global in time existence has been
established for weak solutions only, which need not be smooth. On the
other hand, for initial data of arbitrary size, smooth (classical)
solutions are guaranteed to exist up to certain finite times only, and
loss of regularity, referred to as ``blow-up'', cannot be ruled out.
The importance of this issue has been recognized by the Clay
Mathematical Institute which identified it as one of the ``millennium
challenges'' for the mathematics community with a suitable monetary
prize \citep{f00}. Similar questions concerning existence of smooth
solutions also pertain to the Euler equations in 3D. While the problem
is essentially one of mathematical analysis, a number of computational
investigations have been undertaken
\citep[e.g.,][]{bmonmu03,b91,k93,h09,p01,oc08,o08,ghdg08,gbk08,opc12},
to understand whether or not blow-up may occur in finite time.
Although some of these studies indicated the possibility of a blow-up,
the results obtained to date are not conclusive and their
interpretation remains the subject of a debate. Another related
research direction involves the study of complex-valued extensions of
the Euler and Navier-Stokes equations. The idea is that solutions to
the equations which at some fixed time are real-analytic functions of
the space variables possess singularities in the complex plane, and
the distance from the real axis to the nearest singularity, referred
to as the width of the analyticity strip \citep{ssf83}, further
characterizes the smoothness of the solution.  Therefore, migration of
such complex-plane singularities towards the real line might be a
signature of an approaching blow-up. In the context of this approach
we only mention recent studies by \cite{mbf08,sc09,bb12}, and
refer the reader to the references quoted therein for further details.
A common limitation of these earlier attempts is that the candidates
for blow-up (given in terms of the initial data for the Navier-Stokes
or Euler equations) were chosen in a rather ad-hoc manner based on
some heuristic arguments. A long-term goal of the present research
program is to conduct the search for potential finite-time
singularities in hydrodynamic systems more systematically, leveraging
modern methods of numerical optimization.

In mathematical analysis there are many different lines of attack on
the Navier-Stokes regularity problem. One important approach relies on
estimates for the growth of the {\em enstrophy} $\E(t) :=
\frac{1}{2}\int_{\Omega} | \bnabla \times \u(t,\x)|^2 \, d\Omega$, where
$\u(t,\cdot) \; : \; \Omega \rightarrow \RR^3$ is the velocity field and
$\Omega$ is a 3D domain (periodic or unbounded). It is well known
\citep{ft89} that the loss of regularity will manifest itself by the
enstrophy becoming unbounded $\E(t) \rightarrow \infty$ as
$t\rightarrow t_0$, where $t_0$ is the blow-up time. Therefore, it is
essential to provide tight bounds on how rapidly the enstrophy can
grow, and the sharpest estimate available to date for the 3D
Navier-Stokes system has the form \citep{dg95,d09}
\begin{equation}
\frac{d\E(t)}{dt} \le \frac{27 C^3}{32 \nu^3} \E(t)^3,
\label{eq:dEdt3D}
\end{equation}
where $\nu$ is the kinematic viscosity and $C>0$ is a constant
(hereafter $C$ will denote a generic positive constant which may
assume different numerical values in different instances).  Results
similar to \eqref{eq:dEdt3D} were developed earlier by \citet{s63} and
\citet{l69b}.  Since upon integration with respect to time this upper
bound blows up at $t_0 = \frac{16 \nu^3} {27 C^3 \E_0^2}$, where
$\E_0$ is the initial value of the enstrophy, the regularity problem
can be rephrased as the question whether or not estimate
\eqref{eq:dEdt3D} can be saturated uniformly during the system
evolution over a {\em finite} window of time $[0,T]$, where $T < t_0$.
In other words, the question is whether there exists initial data
$\u_0$ with some prescribed enstrophy $\E_0$ such that the
corresponding system evolution will realize estimate \eqref{eq:dEdt3D}
over a finite time window $[0,T]$. Such initial data can be sought via
solution of a suitably formulated variational optimization problem for
partial differential equation (PDEs) in which the objective is to
maximize the growth of the enstrophy.

In order to investigate the possibility of a finite-time
blow-up, two questions need to be addressed, namely:

\begin{itemize}
\item[(P1)] Sharpness of {\em instantaneous} estimate \eqref{eq:dEdt3D},
  and

\item[(P2)] the maximum growth of enstrophy over {\em finite
    time} window $[0,T]$, which is mathematically defined as
\begin{equation}
\mathop{\max}_{\u_0 \in H^1(\Omega),\; \E(0) = \E_0} \E(T).
\label{eq:maxET}
\end{equation}

\end{itemize}

By solving optimization problem \eqref{eq:maxET} over a set of time
windows with increasing length $T$ one could assess whether or not the
worst-case growth of enstrophy indeed exhibits a tendency towards
blow-up in finite time. Moreover, this will also shed light on the
structure of the most singular initial data which can lead to new
conjectures in the mathematical analysis of the problem.  In the
context of the 3D Navier-Stokes system problem P1 was already
addressed in the seminal study by \citet{ld08} \citep[see
also][]{l06}, where it was demonstrated using computations that
estimate \eqref{eq:dEdt3D} is in fact sharp (up to a prefactor).  From
the computational point of view, solution of problems P1 and P2 is
based on a form of the discrete gradient flow. Needless to say, this
approach is much more complicated in the case of open problem P2,
since in order to compute the gradient directions, the time-dependent
Navier-Stokes system and its suitably defined adjoint have to be
solved. While this is a formidable computational task, it does appear
within reach of the computational techniques and resources available
to date. At this point it should be made clear that, although solving
problem P2 is the long-term objective of the present research program,
accomplishing this task will not resolve the Clay Millennium Problem
where a rigorous mathematical proof is required \citep{f00}.

In analogy with problems P1 and P2, questions concerning the maximum
possible growth of various quantities can also be formulated in regard
to the two-dimensional (2D) Navier-Stokes and one-dimensional (1D)
Burgers equations. While for both of these systems it is well known
that smooth solutions exist {\em globally} in time for arbitrary
smooth initial data \citep{kl04}, one can also obtain estimates for
both the instantaneous and finite-time growth of the relevant
quadratic quantities and it is important to know whether these
estimates are sharp and can be attained during the nonlinear evolution
of the system. Our interest is justified by the fact that these
estimates are obtained using similar techniques as employed in the
analysis of the 3D Navier-Stokes problem. The quantities of interest
are the ``enstrophy'' $\E(t) := \frac{1}{2} \int_0^1 (\partial_x
v(t,x))^2\, dx$ in 1D, where $v \; : \; \RR^+ \times [0,1] \rightarrow
\RR$ is the solution of the Burgers equation, and the palinstrophy
$\P(t) := \frac{1}{2}\int_{\Omega} | \bnabla \omega(t,\x)|^2 \,
d\Omega$ in 2D, where $\Omega := [0,1]\times[0,1]$ and $\omega \; : \;
\RR^+ \times \Omega \rightarrow \RR$ is the scalar vorticity (as will
be discussed further below, enstrophy is not interesting in 2D, since
in the absence of any right-hand side forcing, it may not increase in
flows on periodic and unbounded domains). 

In this investigation we are interested in assessing the sharpness of
certain analytic estimates for the instantaneous and finite-time
growth of palinstrophy with respect to variations of the palinstrophy
in the presence of suitable side constraints. In other words, we will
seek upper bounds which are slowest-growing functions of the
palinstrophy with some other quantity fixed. An estimate is thus
declared ``sharp'' if there exists a family of fields unconstrained by
flow evolution which exhibits the same growth of $d\P / dt$ or
$\max_{t\ge 0} \P(t)$ with increasing palinstrophy as predicted by the
estimate (up to a constant prefactor). Since such notion of estimate
sharpness does not explicitly involve flow evolution (which is
considered only a posteriori), an estimate may be sharp even if it
does not have an optimal structure with respect to the time evolution
(this point is further discussed in section \ref{sec:NS} below).  We
also remark that, in principle, depending on the structure of the side
constraints, estimates exhibiting different power-law dependence on
the palinstrophy may simultaneously be sharp.  In addition to being
physically relevant, such formulation leads to precise and
computationally verifiable criteria for sharpness.

Our focus will be on upper bounds expressed in terms of quadratic
quantities, namely, energy $\K$, enstrophy $\E$ and palinstrophy $\P$.
Selected estimates for problems analogous to problems P1 and P2
formulated for the 1D Burgers and 2D Navier-Stokes systems, together
with the aforementioned results for the 3D Navier-Stokes system, are
summarized in table \ref{tab:estimates}.  Determining whether or not
the estimates listed in table \ref{tab:estimates} are sharp in the
sense made precise above and, if so, identifying the solutions which
saturate these estimates constitutes the long-term goal of this
research program. In fact, significant progress has already been made
addressing some of these questions. The instantaneous bound on
$d\E/dt$ for the 1D Burgers problem was shown to be sharp by
\citet{ld08} \citep[see also][]{l06}, and a remarkable feature of this
result is that it was obtained analytically.  Finite-time estimates
for the 1D Burgers problem were probed computationally by
\citet*{ap11a}, where it was shown that they are not in fact sharp.
This result is important, as it suggests that the standard way for
performing analysis based on integrating (sharp) instantaneous bounds
over time might not be optimal and might lead to significant
overestimates. The results obtained numerically by \citet{ap11a} were
then justified rigorously by \citet{p12b,p12}.  We emphasize that,
based on the results in the 1D and 2D cases, it is not possible to
speculate about the maximum growth of various quantities in 3D flows.
We add here that variational optimization methods have recently been
employed to study other fundamental problems in hydrodynamics such as
the realizability of the Kraichnan-Leith-Batchelor theory of the 2D
turbulence \citep{fkp11} and also involving the growth of quadratic
quantities, e.g., optimal perturbations in the laminar-turbulent
transition \citep{rck12}.

\begin{table}
\begin{center}
\hspace*{-1.1cm}
\begin{tabular}{l|c|c}      
  &  \Bmp{3.0cm} \small \begin{center} {\sc Estimate} \\ \smallskip \end{center} \Emp   
  & \Bmp{3.5cm} \small \begin{center} {\sc Sharpness}  \end{center} \Emp \\  
  \hline
  \Bmp{2.5cm}  \small {\begin{center} \smallskip 1D Burgers  \\ instantaneous \smallskip \end{center}} \Emp &  
  \small {$\frac{d\E}{dt} \leq \frac{3}{2}\left(\frac{1}{\pi^2\nu}\right)^{1/3}\E^{5/3}$}  & 
  \Bmp{3.5cm} \footnotesize {\begin{center} \smallskip {\sc Yes} \\ \citep{ld08}  \smallskip  \end{center}} \Emp \\ 
  \hline 
  \Bmp{3.0cm} \small {\begin{center} \smallskip 1D Burgers  \\ finite-time \smallskip \end{center}} \Emp &  
  \small {$\max_{t \in [0,T]} \E(t) \leq \left[\E_0^{1/3} + \frac{1}{16}\left(\frac{1}{\pi^2 \nu}\right)^{4/3}\E_0\right]^{3}$} &  
  \Bmp{3.5cm} \small {\begin{center} \smallskip {\sc No} \\ \citep{ap11a} \smallskip  \end{center}} \Emp \\ 
  \hline 
  \Bmp{3.0cm} \small {\begin{center} \smallskip 2D Navier-Stokes  \\ instantaneous \smallskip\end{center}} \Emp &  
  \Bmp{7.0cm} \smallskip \centering \small $\frac{d\P(t)}{dt}  \le -\nu\frac{\P^2}{\E} + \frac{C_1}{\nu} \E\,\P$ \\ \smallskip $\frac{d\P(t)}{dt} \le \frac{C_2}{\nu} \K^{1/2}\P^{3/2}$ \smallskip \Emp& 
  \Bmp{3.5cm} \small {\begin{center} \smallskip present work \smallskip  \end{center}} \Emp  \\ 
  \hline 
  \Bmp{3.0cm} \small \begin{center} \smallskip 2D Navier-Stokes \\ finite-time \smallskip\end{center} \Emp &  
  \Bmp{7.0cm} \smallskip \centering \small $\max_{t>0} \P(t) \le \P_0 + \frac{C_1}{2\nu^2}\E_0^2$ \\ \smallskip $\max_{t>0} \P(t) \le \left(\P_0^{1/2} + \frac{C_2}{4\nu^2}\K_0^{1/2}\E_0\right)^2$ \smallskip \Emp  &  present work  \\ 
  \hline 
  \Bmp{3.0cm} \small {\begin{center} \smallskip 3D Navier-Stokes  \\ instantaneous  \smallskip \end{center}} \Emp &  
  \small {$\frac{d\E(t)}{dt} \le \frac{27 C^2}{32 \nu^3} \E(t)^3$} & \Bmp{3.5cm} \small {\begin{center} \smallskip {\sc Yes} \\ \citep{ld08} \smallskip  \end{center}} \Emp  \\ 
  \hline 
  \Bmp{3.0cm} \small \begin{center} \smallskip\smallskip 3D Navier-Stokes  \\ finite-time \smallskip \end{center} \Emp &  
  \small $\E(t) \le \frac{\E(0)}{\sqrt{1 - 4 \frac{C \E(0)^2}{\nu^3} t}}$ & \Bmp{3.5cm} \centering {???} \smallskip\smallskip \Emp
\end{tabular}
\end{center}
\caption{Summary of selected estimates for the instantaneous 
  rate of growth and the growth over finite time of enstrophy and palinstrophy 
  in 1D Burgers, 2D and 3D Navier-Stokes systems.}
\label{tab:estimates}
\end{table}

In this study we report new results concerning the realizability of
analytic bounds for $d\P/dt$ and $\max_{t>0}\P(t)$ in the 2D
Navier-Stokes problem. It should be noted that, given the structure of
the corresponding extreme vortex states, these results are also quite
interesting from the physical point of view, outside the context of
the sharpness of mathematical analysis.  The structure of the paper is
as follows: in the next section we discuss a number of rigorous
estimates of the palinstrophy growth in the 2D Navier-Stokes system.
In section \ref{sec:probe} we demonstrate how questions about the
sharpness of these estimates can be framed in terms of suitable
variational optimization problems. A gradient-based approach to
solution of such problems is discussed in section \ref{sec:grad},
whereas some analytical insights concerning the solutions of the
maximization problems in the limit of small palinstrophies are
presented in section \ref{sec:smallP}.  Computational results are
presented in section \ref{sec:results} and discussed in section
\ref{sec:discuss}.  Conclusions and a discussion of some future
research directions are deferred to section \ref{sec:final}. Some
technical material is collected in an appendix.
 
\section{Two-Dimensional Navier-Stokes System}
\label{sec:NS}

We consider a viscous incompressible fluid on a 2D periodic
domain $\Omega = [0,1] \times [0,1]$. Its motion is governed by the
Navier-Stokes equation, written here in the form
\begin{subequations}
\label{eq:NS2D}
\begin{alignat}{2}
& \Dpartial{\omega}{t} + J(\omega,\psi) = \nu \Delta \omega &  \qquad &
\textrm{in} \ (0,\infty) \times \Omega, \label{eq:NS2Da} \\
& - \Delta \psi = \omega & \quad &
\textrm{in} \ (0,\infty) \times \Omega, \label{eq:NS2Db} \\
& \omega(0) = \omega_0 & \quad &
\textrm{in} \ \Omega, \label{eq:NS2Dc}
\end{alignat}
\end{subequations}
where $\psi$ and $\omega$ are, respectively, the streamfunction and
(scalar) vorticity, whereas $\omega_0$ is the initial condition. In
system \eqref{eq:NS2D} $\nu$ denotes the kinematic viscosity (assumed
fixed), $\Delta$ is the Laplacian operator and $J(f,g) := \partial_x f
\, \partial_y g - \partial_y f \, \partial_x g$, defined for $f,g \; :
\; \Omega \rightarrow \RR$, is the Jacobian determinant. Discussion
concerning various aspects of formulation \eqref{eq:NS2D} can be
found, for example, in \citet{mb02}.

We are interested in studying the growth of the following quadratic
quantities characterizing the evolution of system \eqref{eq:NS2D}
\begin{alignat}{2}
&\text{kinetic energy} & \qquad\quad
\K(\psi(t)) & = \frac{1}{2}\int_{\Omega} |\bnabla \psi(t,\x)|^2 \, d\Omega,
\label{eq:K} \\
&\text{enstrophy} &
\E(\psi(t)) & = \frac{1}{2}\int_{\Omega} (\Delta \psi(t,\x))^2 \, d\Omega,
\label{eq:E} \\
&\text{palinstrophy} &
\P(\psi(t)) & = \frac{1}{2}\int_{\Omega} |\bnabla \Delta \psi(t,\x)|^2 \, d\Omega
\label{eq:P}
\end{alignat}
which, to simplify our analysis, are rewritten here in terms of
streamfunction as the state variable.  Without loss of generality, we
will assume that the streamfunction fields have zero mean. As regards
enstrophy \eqref{eq:E}, we note that multiplying \eqref{eq:NS2Da} by
$\omega$, integrating the resulting expression over $\Omega$,
performing integration by parts and making necessary simplifications,
we arrive at
\begin{equation}
\frac{d\E}{dt} = - 2 \nu \, \P \; \le \; 0
\label{eq:dEdt}
\end{equation}
which implies that, unlike in the dimension one or three, in 2D flows
on periodic domains the enstrophy cannot increase (it will, in fact,
decrease unless the vorticity is constant, or the fluid is inviscid,
i.e., $\nu=0$). This result (which also holds on unbounded domains) is
a consequence of the absence of the ``vortex stretching'' term in the
2D vorticity equation \eqref{eq:NS2Da}. On the other hand, the
phenomenon of stretching is observed (in the form of the last term on
the right-hand side) in the evolution equation for the vorticity
gradient $\bnabla\omega$ which is obtained by applying the gradient
operator $\bnabla$ to equation \eqref{eq:NS2Da}
\begin{equation}
  \Dpartial{\bnabla\omega}{t} + (\u\cdot\bnabla)\bnabla\omega
  = \nu \Delta \bnabla\omega - \left[\bnabla\u\right]^T \cdot \bnabla\omega.
\label{eq:gradw}
\end{equation}
For clarity, this equation is written using the velocity field $\u =
\left[\Dpartial{\psi}{y}, - \Dpartial{\psi}{x}\right]^T$.
Palinstrophy \eqref{eq:P} is the quadratic quantity associated
with equation \eqref{eq:gradw}, and a relation characterizing its
evolution in time is obtained by dotting equation
\eqref{eq:gradw} with $\bnabla\omega$, integrating over $\Omega$,
then integrating by parts and simplifying
\begin{equation}
\frac{d\P(t)}{dt} = \int_{\Omega} J(\Delta\psi,\psi) \Delta^2 \psi\, d\Omega
- \nu \, \int_{\Omega} (\Delta^2 \psi)^2 \, d\Omega =: \R_{\P}(\psi),
\label{eq:R}
\end{equation}
where the subscript $\P$ indicates the value of the palinstrophy for
which the expression is evaluated.  We note that now, unlike in
equation \eqref{eq:dEdt}, the right-hand side (RHS) features a cubic
term representing stretching in addition to the negative-definite
dissipative term. We add that an equivalent expression for
$d\P/dt$ was also obtained by \citet{td06}.

Since palinstrophy may exhibit nontrivial behavior, we now go
on to discuss various rigorous bounds available for the palinstrophy
rate of growth \eqref{eq:R}. The following estimate was recently
obtained by \citet{dl11}
\begin{equation}
\frac{d\P}{dt}  \le - \nu\frac{\P^2}{\E} + \frac{C_1}{\nu} \E\, \P.
\label{eq:dPdt_EP}
\end{equation}
A different estimate is derived in appendix \ref{sec:dPdt} and has the
form
\begin{equation}
\frac{d\P}{dt}  \le \frac{C_2}{\nu} \,\K^{\frac{1}{2}}\, \P^{\frac{3}{2}}.
\label{eq:dPdt_KP}
\end{equation}
We observe that, in comparison to the corresponding estimates
available in 1D and in 3D (see table \ref{tab:estimates}), bounds
\eqref{eq:dPdt_EP} and \eqref{eq:dPdt_KP} have a different structure,
since the RHS expressions depend on {\em two} quadratic quantities,
respectively, $\E$ and $\P$ in \eqref{eq:dPdt_EP}, and $\K$ and $\P$
in \eqref{eq:dPdt_KP}, rather than just one. In principle, the
second quantity ($\E$ or $\K$) can be eliminated using Poincar\'e's
inequality [$\,\K \le (2\pi)^{-2} \, \E \le (2\pi)^{-4} \, \P\,$]
yielding
\begin{equation}
\frac{d\P}{dt}  \le \frac{C}{\nu} \, \P^2
\label{eq:dPdt_P}
\end{equation}
(transforming \eqref{eq:dPdt_EP} into \eqref{eq:dPdt_P} also requires
dropping the negative-definite quadratic term).  We note however that,
since the only functions saturating Poincar\'e's inequality are the
eigenfunctions of the Laplacian, such transformation of
\eqref{eq:dPdt_EP} and \eqref{eq:dPdt_KP} into \eqref{eq:dPdt_P} may
not be optimal, resulting in the possible loss of sharpness.
Establishing whether or not upper bounds \eqref{eq:dPdt_EP},
\eqref{eq:dPdt_KP} and \eqref{eq:dPdt_P} are sharp with respect to
variations of palinstrophy $\P$, and determining the structure of the
corresponding maximizing fields is the main goal of the present study.

As regards the maximum growth of palinstrophy over finite time, we
notice that, although straightforward integration of
\eqref{eq:dPdt_EP} and \eqref{eq:dPdt_KP} leads to unbounded increase
of $\P$, when additional relations are used in the process, namely
$d\K/dt = - 2 \nu \E$ and $d\E/dt = - 2 \nu \P$ (cf.~\eqref{eq:dEdt}),
then bounded growth is in fact obtained in finite time.  Starting from
estimate \eqref{eq:dPdt_EP}, \citet{dl11} found that
\begin{equation}
\max_{t > 0} \P(t) \, \le \, \P(0) + \frac{C_1}{2\nu^2} \E(0)^2.
\label{eq:maxPt_Doering}
\end{equation}
Similarly, it follows from estimate \eqref{eq:dPdt_KP} that, cf.~appendix \ref{sec:dPdt},
\begin{equation}
\max_{t > 0} \P(t) \, \le \, \left(\P^{1/2}(0) + \frac{C_2}{4\nu^2} \K^{1/2}(0) \E(0)\right)^2.
\label{eq:maxPt_Ayala}
\end{equation}
It is worth noticing that, although finite-time estimates
\eqref{eq:maxPt_Doering} and \eqref{eq:maxPt_Ayala} are obtained from
two different instantaneous estimates, they both give the same
power-law behavior $\max_{t > 0}\P(t) \sim \P(0)$ in the limit
$\P(0)\to 0$. Similarly, the two estimates reduce to $\max_{t>0}\P(t)
\sim \E(0)^2$ in the limit $\E(0)\to\infty$ (assuming that $\K(0)$ is
fixed in \eqref{eq:maxPt_Ayala}).

We add that some other estimates for the growth of palinstrophy were
also derived in the literature. For example, the following bounds were
established by \citet{td06}
\begin{align} 
\frac{d\P}{dt} & \le  \frac{\|\triangle\omega\|_{L_2(\Omega)}}{\sqrt{2 \E}} 
\left( \|\omega\|_{L_{\infty}(\Omega)} \, \E - \nu\P \right), \label{eq:dPdt_TD}\\
\P(t) & \leq  \frac{\|\omega_0\|_{L_{\infty}(\Omega)}\E(0)}{\nu},\quad\textrm{for} \ t>0, \label{eq:Pmax_TD}
\end{align}
whereas using a logarithmic bound on
$\|\bnabla\u\|_{L_{\infty}(\Omega)}$, cf.~\citet{dg95}, it is possible
to deduce
\begin{equation} 
\frac{d\P}{dt}  \le  C\, \| \omega\|_{L_{\infty}(\Omega)}\, \ln(1+\|\Delta\omega\|_{L_2(\Omega)})\, \P
 \label{eq:dPdt_log}
\end{equation}
in which RHS exhibits dependence on $\P$ involving a smaller exponent
than in bounds \eqref{eq:dPdt_EP}--\eqref{eq:dPdt_KP} when $\|
\omega\|_{L_{\infty}(\Omega)}$ is assumed fixed.  We remark that,
unlike estimates \eqref{eq:dPdt_EP}--\eqref{eq:dPdt_P}, upper bounds
\eqref{eq:dPdt_TD} and \eqref{eq:dPdt_log} rely on the control of
higher (second) derivatives of vorticity through
$\|\triangle\omega\|_{L_2(\Omega)}$ and of the vorticity maximum
through $\|\omega\|_{L_{\infty}(\Omega)}$. While both these quantities
are known to be bounded with respect to time in 2D, they are not
necessarily bounded as $\P$ increases, which is the sense of sharpness
we are concerned with in this study. In addition, in the absence of a
priori bounds on $\|\omega\|_{L_{\infty}(\Omega)}$ in 3D, estimate
\eqref{eq:Pmax_TD} has a rather different structure from the available
finite-time estimate in 3D (table \ref{tab:estimates}).  We also
emphasize that the presence of the norm
$\|\omega\|_{L_{\infty}(\Omega)}$, which is not a smooth function of
the vorticity $\omega$, in estimates
\eqref{eq:dPdt_TD}--\eqref{eq:Pmax_TD} would complicate the
formulation of the variational optimization problems designed to probe
their sharpness (cf.~section \ref{sec:probe}), as these problems would
be nonsmooth. On the other hand, we note that finite-time estimate
\eqref{eq:Pmax_TD} features a milder dependence on the viscosity
$\nu$, and hence is likely more optimal with respect to viscosity than
bounds \eqref{eq:maxPt_Doering} and \eqref{eq:maxPt_Ayala}.  Sharpness
of estimates with respect to variations of $\nu$ is however outside
the scope of the present study.  Some early results concerning the
maximum growth of palinstrophy were also obtained by \citet{plab75},
whereas estimates for the rate of growth in the presence of body
forcing were studied by \citet{dfj10}.

Aside from the questions concerning the sharpness of estimates
\eqref{eq:dPdt_EP}--\eqref{eq:dPdt_P}, there is also independent
interest in the structure of the vorticity fields leading to the
maximum possible palinstrophy production because of their relevance
for the enstrophy cascade in 2D turbulence. In fact, various processes
related to the stretching of vorticity gradients described by equation
\eqref{eq:gradw} have already received some attention in the
literature \citep[e.g.,][]{pkb99}.

\section{Probing Sharpness of Estimates Using Variational Optimization}
\label{sec:probe}

We now go on to discuss how the question of the sharpness of estimates
\eqref{eq:dPdt_EP}, \eqref{eq:dPdt_KP} and \eqref{eq:dPdt_P} can be
framed in terms of solutions of suitably-defined optimization
problems. Analogous questions pertaining to problems in 1D and 3D, cf.
table \ref{tab:estimates}, have already been addressed by
\citet{ap11a} and \citet{ld08}, respectively.  As regards estimate
\eqref{eq:dPdt_EP}, the approach consists in finding, for fixed values
of $\E = \E_0$ and $\P = \P_0$, the streamfunction field
$\tilde{\psi}_{\E_0,\P_0}$ which achieves the greatest rate of
palinstrophy production $\R_{\P_0}(\tpEP)$, and then studying how it
depends on the parameters $\E_0$ and $\P_0$ to see whether or not this
dependence follows the predictions of estimate \eqref{eq:dPdt_EP} (the
use of streamfunction $\psi$, rather than the vorticity or velocity
field, as the control variable leads to a simpler formulation of the
optimization problem).  As regards estimates \eqref{eq:dPdt_KP} and
\eqref{eq:dPdt_P}, the approach is the same, except that,
respectively, $\K=\K_0$ and $\P$ or just $\P$ are fixed. Thus, we
arrive at the following two optimization problems corresponding to
estimates \eqref{eq:dPdt_EP} and \eqref{eq:dPdt_KP}
\begin{equation}
\begin{aligned}
\tilde{\psi}_{\E_0,\P_0} & = \mathop{\arg\max}_{\psi\in\S_{\E_0,\P_0}} \,\R_{\P_0}(\psi) \\ 
\S_{\E_0,\P_0} & = \left\{\psi \in H^4(\Omega) : \frac{1}{2}\int_\Omega(\Delta\psi)^2\,d\Omega = \E_0, \ \frac{1}{2}\int_\Omega|\bnabla\Delta\psi|^2\,d\Omega = \P_0 \right\}, 
\end{aligned}
\label{eq:optR_EP}
\end{equation}
\begin{equation}
\begin{aligned}
\tilde{\psi}_{\K_0,\P_0} & = \mathop{\arg\max}_{\psi\in\S_{\K_0,\P_0}} \,\R_{\P_0}(\psi)  \\ 
\S_{\K_0,\P_0} & = \left\{\psi \in H^4(\Omega) : \frac{1}{2}\int_\Omega|\bnabla\psi|^2\,d\Omega = \K_0, \ \frac{1}{2}\int_\Omega|\bnabla\Delta\psi|^2\,d\Omega = \P_0 \right\},  
\end{aligned}
\label{eq:optR_KP}
\end{equation}
where ``$\arg\max$'' denotes ``the argument of the maximum''
and maximization is performed over the Sobolev space $H^4(\Omega)$
of doubly-periodic functions with square-integrable fourth-order
derivatives \citep{af05}. This regularity requirement plays a key
role in the solution of optimization problems
\eqref{eq:optR_EP}--\eqref{eq:optR_KP} as it ensures that the
expression for the rate of growth of palinstrophy $\R_{\P_0}(\psi)$,
cf. \eqref{eq:R}, is well-defined. We remark that the pairs of
constraints in problems \eqref{eq:optR_EP} and \eqref{eq:optR_KP} are
not in fact independent and must satisfy Poincar\'e's inequalities,
i.e., $\E_0 \le (2\pi)^{-2} \P_0$ and $\K_0 \le (2\pi)^{-4} \P_0$. The
maximization problem corresponding to estimate \eqref{eq:dPdt_P} then
takes the form
\begin{equation}
\begin{aligned}
\tilde{\psi}_{\P_0} & = \mathop{\arg\max}_{\psi\in\S_{\P_0}} \,\R_{\P_0}(\psi)\\ 
\S_{\P_0} & = \left\{\psi \in H^4(\Omega) : \frac{1}{2}\int_\Omega|\bnabla\Delta\psi|^2\,d\Omega = \P_0 \right\}   
\end{aligned}
\label{eq:optR_P}
\end{equation}
in which only one constraint is present (in view of the earlier
remark, we note that fixing palinstrophy $\P(\psi) = \P_0$ also
provides upper bounds, via the aforementioned Poincar\'e's
inequalities, on both enstrophy $\E(\psi)$ and energy $\K(\psi)$). As
will be shown in the next section, single-constraint problem
\eqref{eq:optR_P} is in fact fairly straightforward to solve
numerically given the isotropic nature of the constraint. On the other
hand, two-constraint problems \eqref{eq:optR_EP} and
\eqref{eq:optR_KP} are much harder to solve, since the maximizers are
to be sought at the intersection of two nonlinear constraint manifolds
which may have a fairly complicated structure, both locally and
globally.

We note that, due to the presence of the cubic term in the expression
for $\R_{\P_0}(\psi)$, cf. \eqref{eq:R}, optimization problems
\eqref{eq:optR_EP}--\eqref{eq:optR_P} may be nonconvex, and hence the
presence of multiple local maxima may be expected. We remark here that
rescaling the domain $\Omega$ by rational factors will lead to a
trivial multiplicity of the optimizing solutions. To demonstrate this,
we consider system \eqref{eq:NS2D} on a rescaled domain $\Omega_L :=
[0,L] \times [0,L]$, where $L>0$. The new independent variables become
$\bxi := L \x \in \Omega_L$ and $\tau := L^{\beta} t$ for some $\beta
\in \RR$, whereas the corresponding solution can be expressed as
$\Psi(\tau,\bxi) := L^{\alpha} \, \psi(t(\tau),\x(\bxi))$ for some
$\alpha \in \RR$. Transforming system \eqref{eq:NS2D} to the new
variables, we observe that its form remains unchanged, provided that
$\alpha = 0$ and $\beta = 2$. This shows that
$\tilde{\Psi}_{\E_0,\P_0} := \tpEP(\x(\bxi))$ is a solution of
optimization problem \eqref{eq:optR_EP} rescaled to the new domain
$\Omega_L$. In the particular case when $L=1/2,1/3,\dots$ the domain
$\Omega_L$ is periodically embedded in $\Omega$. In such situation the
maximizers $\tilde{\Psi}_{\E_0,\P_0}$ are nothing, but
higher-wavenumber copies of the ``main'' maximizer $\tpEP$ and this
trivial multiplicity of maximizing solutions will not be separately
considered here.

\section{Gradient-Based Solution of Maximization Problems}
\label{sec:grad}

In this section we describe key elements of the computational
algorithm for the solution of maximization problems stated in section
\ref{sec:probe}. We do this in the spirit of the
``optimize-then-discretize'' approach \citep{g03}. Solutions of
maximization problems \eqref{eq:optR_EP}, \eqref{eq:optR_KP} and
\eqref{eq:optR_P} are characterized by the first-order optimality
condition
\begin{equation}\label{eq:dR}
\R'_{\P_0}(\tilde{\psi};\psi') + \sum_{i=1}^{m}\lambda_i\Q'_i(\tilde{\psi};\psi') = 0 \qquad \forall \; \psi'\in H^4(\Omega)
\end{equation}
where 
\begin{eqnarray*}
\R'_{\P_0}(\psi;\psi') & := & \lim_{\epsilon\rightarrow 0} \left[
  \R_{\P_0}(\psi + \epsilon\psi') - \R_{\P_0}(\psi)\right]/\epsilon \qquad\mbox{and}\\
\Q'_i(\psi;\psi') & := & \lim_{\epsilon\rightarrow 0}  \left[ \Q_i(\psi +
  \epsilon\psi') - \Q_i(\psi)\right]/\epsilon,\quad i=1,\dots,m
\end{eqnarray*}
are the G\^ateaux (directional) differentials \citep{l69} of,
respectively, the objective function and the individual constraints
$\Q_i(\psi)$ defining the manifolds $\S_{\E_0,\P_0}$, $\S_{\K_0,\P_0}$
and $\S_{\P_0}$, cf. \eqref{eq:optR_EP}, \eqref{eq:optR_KP} and
\eqref{eq:optR_P}. The field $\psi'$ represents an arbitrary direction
of differentiation in the space $H^4(\Omega)$, and $\lambda_i \in \RR$
are the Lagrange multipliers associated with the constraints $\Q_i$,
$i=1,\dots,m$ ($m=2$ for problems \eqref{eq:optR_EP} and
\eqref{eq:optR_KP}, and $m=1$ for problem \eqref{eq:optR_P}). We add
here that an optimality condition such as \eqref{eq:dR} could not be
easily stated for a variational problem designed to probe the
sharpness of estimate \eqref{eq:dPdt_TD} with a constraint on
$\|\Delta\psi\|_{L_{\infty}(\Omega)}$ because of the
nondifferentiability of this constraint.

The maximizer $\tilde{\psi}$ can be found using the following
iterative gradient-ascent algorithm which can be interpreted as a
discretization of a continuous gradient flow
\begin{equation}
\begin{aligned}
\tilde{\psi} & =  \lim_{n\to\infty} \psi^{(n)} \\
\psi^{(n+1)} & =  \mathbb{P}\S\left(\;\psi^{(n)} + \tau_n \nabla\R_{\P}(\psi^{(n)})\;\right) \\ 
\psi^{(1)} & =  \psi_0,
\end{aligned}
\label{eq:desc}
\end{equation}
where $\psi^{(n)}$ is the approximation of the maximizer $\tp$
obtained at the $n$-th iteration, $\psi_0$ is the initial guess,
$\tau_n$ the length of the step and $\mathbb{P}\S:H^4\to\S$ is the
projection operator onto the constraint manifold $\S$ (without
subscripts, symbol $\S$ denotes a generic manifold). We emphasize that
the use of the projection $\mathbb{P}\S$ ensures that, at every step
$n$ in optimization iteration \eqref{eq:desc}, the constraint
$\psi^{(n)}\in\S$ is satisfied up to machine precision.  From the
computational point of view, such formulation is in fact preferred to
the more standard approach based on Lagrange multipliers and
projections onto the tangent space $T\S$ which involve linearization
of the constraints and hence result in accumulation of errors.

A key ingredient of algorithm \eqref{eq:desc} is evaluation of the
gradient $\nabla\R_{\P_0}$ of objective function $\R_{\P_0}(\psi)$,
cf. \eqref{eq:R}, representing its (infinite-dimensional) sensitivity
to perturbations of the streamfunction $\psi$. It is essential that
the gradient be characterized by the required regularity,
namely, $\nabla\R_{\P_0}(\psi) \in H^4(\Omega)$. This is, in fact,
guaranteed by Riesz representation theorem \citep{l69}, applicable
because the G\^ateaux differential $\R'_{\P_0}(\psi;\cdot) \; : \;
H^4(\Omega) \rightarrow \RR$ is a bounded linear functional on
$H^4(\Omega)$.  Thus, we have
\begin{equation}
\R'_{\P_0}(\psi;\psi') 
= \Big\langle \nabla\R_{\P_0}(\psi), \psi' \Big\rangle_{H^4(\Omega)}
= \Big\langle \nabla^{L_2}\R_{\P_0}(\psi), \psi' \Big\rangle_{L_2(\Omega)}
\label{eq:riesz}
\end{equation}
in which the Riesz representers $\nabla\R_{\P_0}(\psi)$ and
$\nabla^{L_2}\R_{\P_0}(\psi)$ are the gradients computed with respect
to the $H^4$ and $L_2$ topology, respectively. The corresponding inner
products are defined as follows
\begin{alignat}{2}
& \forall\,z_1, z_2 \in L_2(\Omega)& \qquad 
\Big\langle z_1, z_2 \Big\rangle_{L_2(\Omega)}
&= \int_{\Omega} z_1 z_2 \, d\Omega,  \label{eq:ipL2} \\
& \forall\,z_1, z_2 \in H^4(\Omega)& \qquad 
\Big\langle z_1, z_2 \Big\rangle_{H^4(\Omega)}
&= \int_{\Omega} z_1 z_2 
+ \ell_1^2 \,\bnabla z_1 \cdot \bnabla z_2
+ \ell_2^4 \,\Delta z_1 \Delta z_2 \nonumber \\
&&& \hspace*{1.75cm}
+ \ell_3^6 \,\bnabla\Delta z_1 \cdot \bnabla\Delta z_2
+ \ell_4^8 \,\Delta^2 z_1 \Delta^2 z_2  \, d\Omega,  \label{eq:ipH4} 
\end{alignat}
where $\ell_1,\ell_2, \ell_3 \ge 0$, $\ell_4 >0$ are adjustable
parameters with the meaning of length-scales \citep{pbh04}.  While the
$H^4$ inner products \eqref{eq:ipH4} corresponding to different values
of $\ell_1,\ell_2, \ell_3, \ell_4$ are mathematically equivalent (in
the sense of norm equivalence, \citet{l69}), these parameters will
play a role in tuning the performance of the discrete optimization
algorithm discussed in section \ref{sec:results}. We remark that while
the $H^4$ gradient is used exclusively in the actual computations, cf.
\eqref{eq:desc}, the $L_2$ gradient is computed first as an
intermediate step. Calculating the G\^ateaux differential of
$\R_{\P_0}(\psi)$ and identifying it with the $L_2$ inner product
\eqref{eq:ipL2} we thus obtain
\begin{equation}
\begin{aligned}
\R'_{\P_0}(\psi;\psi') & = \int_{\Omega}\left[\Delta^2 J(\Delta \psi, \psi) +
 \Delta J(\psi, \Delta^2\psi) + J(\Delta^2 \psi, \Delta\psi) -
2 \nu \Delta^4\psi\right] \, \psi' \, d\Omega \\
& = \Big\langle \nabla^{L_2}\R_{\P_0}(\psi), \psi' \Big\rangle_{L_2(\Omega)}
\end{aligned}
\label{eq:dR2}
\end{equation}
from which it follows that
\begin{equation}
\nabla^{L_2}\R_{\P_0}(\psi) = \Delta^2 J(\Delta \psi, \psi) +
 \Delta J(\psi, \Delta^2\psi) + J(\Delta^2 \psi, \Delta\psi) -
2 \nu \Delta^4\psi.
\label{eq:gradRL2}
\end{equation}
Identifying the left-hand side (LHS) of \eqref{eq:dR2} with the $H^4$
inner product \eqref{eq:ipH4}, integrating by parts and using
\eqref{eq:gradRL2}, we obtain the required $H^4$ gradient as the
solution of the following elliptic boundary-value problem
\begin{equation}
\begin{aligned}
&\left[ \Id \, - \,\ell_1^2 \,\Delta + \,\ell_2^4 \,\Delta^2 - \,\ell_3^6 \,\Delta^3 + \,\ell_4^8 \,\Delta^4 \right] \nabla\R_{\P_0}
= \nabla^{L_2} \R_{\P_0}  \qquad \text{in} \ \Omega, \\
& \text{Periodic Boundary Conditions}.
\end{aligned}
\label{eq:gradRH4}
\end{equation}
As shown by \citet{pbh04}, extraction of gradients in spaces of
smoother functions such as $H^4(\Omega)$ can be interpreted as
low-pass filtering of the $L_2$ gradients with parameters
$\ell_1,\ell_2, \ell_3,\ell_4$ acting as cut-off length-scales.

The step size $\tau_n$ in algorithm \eqref{eq:desc} is calculated as
\begin{equation}\label{eq:tau_n}
\tau_n = \mathop{\arg\max}_{\tau>0} \left\{ \R_{\P}\left[\mathbb{P}\S\left( \;\psi^{(n)} + \tau\nabla\R_{\P}(\psi^{(n)}) \;\right)\right] \right\}
\end{equation}
which is done using a derivative-free line search algorithm
\citep{r06}. Equation \eqref{eq:tau_n} can be interpreted as a
modification of a standard line search method where the optimization
is performed following an arc (a geodesic) lying on the constraint
manifold $\S$, rather than a straight line. This approach was already
successfully employed to solve a similar problem in \citet{ap11a}. The
projection $\phi \mapsto \mathbb{P}\S(\phi)$ for some $\phi\in
H^4(\Omega)$ is calculated by solving an optimization subproblem with
form depending on the type of the constraint as follows.

\begin{itemize}
\item
\textbf{Single Constraint:} problem \eqref{eq:optR_P} involves the constraint manifold
\begin{equation}
\S_{\P_0} = \left\{\phi \in H^4(\Omega) : \frac{1}{2}\int_\Omega|\bnabla\Delta\phi|^2\,d\Omega = \P_0 \right\}.   
\label{eq:SP}
\end{equation}
Then, the projection operator $\PP\S_{\P_0}$ is defined as the renormalization
\begin{equation}
\mathbb{P}\S_{\P_0}(\phi) = \sqrt{\frac{\P_0}{\P(\phi)}}\,\phi.
\label{eq:PSP}
\end{equation}

\item \textbf{$(\E_0,\P_0)$-Constraint:} problem \eqref{eq:optR_EP}
  involves the constraint manifold
\begin{equation}
\S_{\E_0,\P_0} = \left\{\phi \in H^4(\Omega) : \frac{1}{2}\int_\Omega(\Delta\phi)^2\,d\Omega = \E_0, \ \frac{1}{2}\int_\Omega|\bnabla\Delta\phi|^2\,d\Omega = \P_0 \right\}. 
\label{eq:SEP}
\end{equation}
Then, the projection $\phi \mapsto \mathbb{P}\S_{\E_0,\P_0}(\phi)$ is
computed as $\mathbb{P}\S_{\E_0,\P_0}(\phi) = \lim_{k
    \rightarrow \infty} \varphi^{(k)}$, where 
\begin{subequations}
\label{eq:FixE0P0} 
\begin{align}
\varphi^{(k+1)} &= \varphi^{(k)} - \tau_k \nabla\Q(\varphi^{(k)}), \ k=1,2,\dots, \label{eq:FixE0P0a} \\
\varphi^{(1)} &= \phi  \label{eq:FixE0P0b}
\end{align}
\end{subequations}
in which $\Q(\varphi) = \frac{1}{2}\left[\P(\varphi) - \P_0\right]^2$
and $\nabla\Q(\varphi)$ is the corresponding gradient. That is, the
projection onto $\S_{\E_0,\P_0}$ is obtained by solving a
single-constraint optimization problem with cost functional
$\Q(\varphi)$ penalizing the deviation from the second constraint and
the first constraint enforced using \eqref{eq:PSP}.

\item \textbf{$(\K_0,\P_0)$-Constraint:} problem \eqref{eq:optR_KP}
  involves the constraint manifold
\begin{equation}
\S_{\K_0,\P_0} = \left\{\phi \in H^4(\Omega) : \frac{1}{2}\int_\Omega|\bnabla\phi|^2\,d\Omega = \K_0, \ \frac{1}{2}\int_\Omega|\bnabla\Delta\phi|^2\,d\Omega = \P_0 \right\}. 
\label{eq:SKP}
\end{equation}
In analogy to the $(\E_0,\P_0)$-constraint, the projection onto
$\S_{\K_0,\P_0}$ is obtained by solving a single-constraint
  optimization problem of the type \eqref{eq:FixE0P0} with cost
  functional $\Q(\varphi) = \frac{1}{2}\left[\K(\varphi) -
    \K_0\right]^2$.
\end{itemize}
We add that, since none of the manifolds defined in \eqref{eq:SP},
\eqref{eq:SEP} and \eqref{eq:SKP} has the structure of a linear space,
the projections defined above are not orthogonal.

Families of maximizers parameterized by their palinstrophy $\P_0$ are
found by solving problems \eqref{eq:optR_EP}, \eqref{eq:optR_KP} and
\eqref{eq:optR_P} for values of $\P_0$ progressively
incremented or decremented by $\pm \Delta \P_0$ and using the
previously obtained maximizer $\tilde{\psi}_{\P_0}$ as the initial
guess $\psi_0$ for $\tilde{\psi}_{\P_0\pm \Delta \P_0}$ in
\eqref{eq:desc}. In order to carry out an exhaustive search for all
possible maximizing fields, this process was initiated in a variety of
ways, including different random fields and closed-form solutions to
the limiting problems described in the next section.

\section{Solution of the Maximization Problems in the Limit of Small Palinstrophies}
\label{sec:smallP}

In this section we investigate the structure of the maximizing
solutions in the limiting cases when $\P_0 \rightarrow 0$ in
\eqref{eq:optR_P}, $\P_0 \rightarrow (2\pi)^{2} \E_0$ in
\eqref{eq:optR_EP} and $\P_0 \rightarrow (2\pi)^{4} \K_0$ in
\eqref{eq:optR_KP}. Motivated by the properties of the vortex states
found numerically for large values of $\P_0$ (see section
\ref{sec:results}), we will consider optimal vorticity distributions
in the form of periodic vortex lattices with 2-fold rotational
symmetry, i.e., possessing the property $\psi(x,y) = - \psi(y,-x)$.

\begin{figure}
\begin{center}
\mbox{
\subfigure[]{\includegraphics[width=0.45\textwidth]{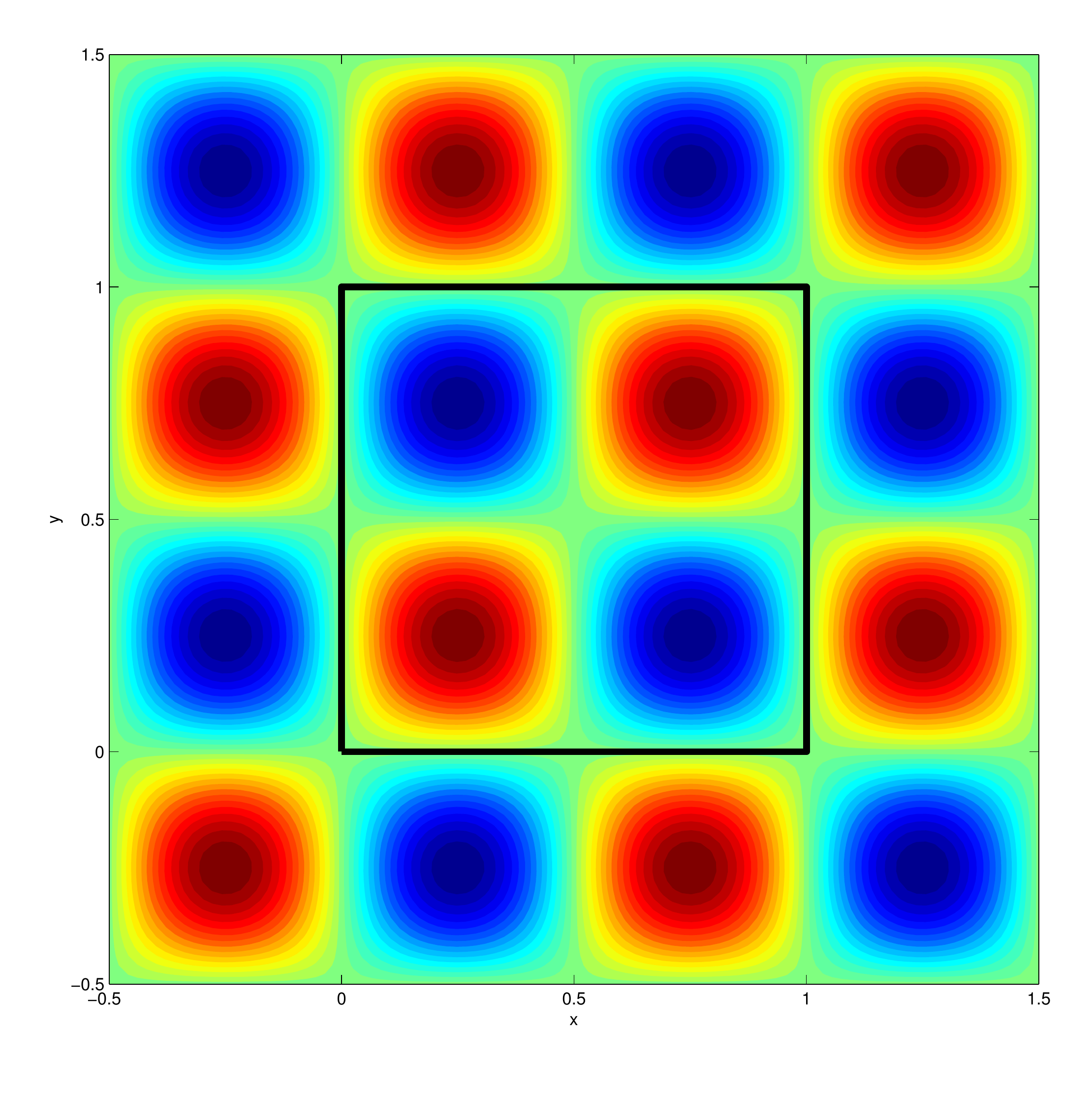}}\qquad
\subfigure[]{\includegraphics[width=0.45\textwidth]{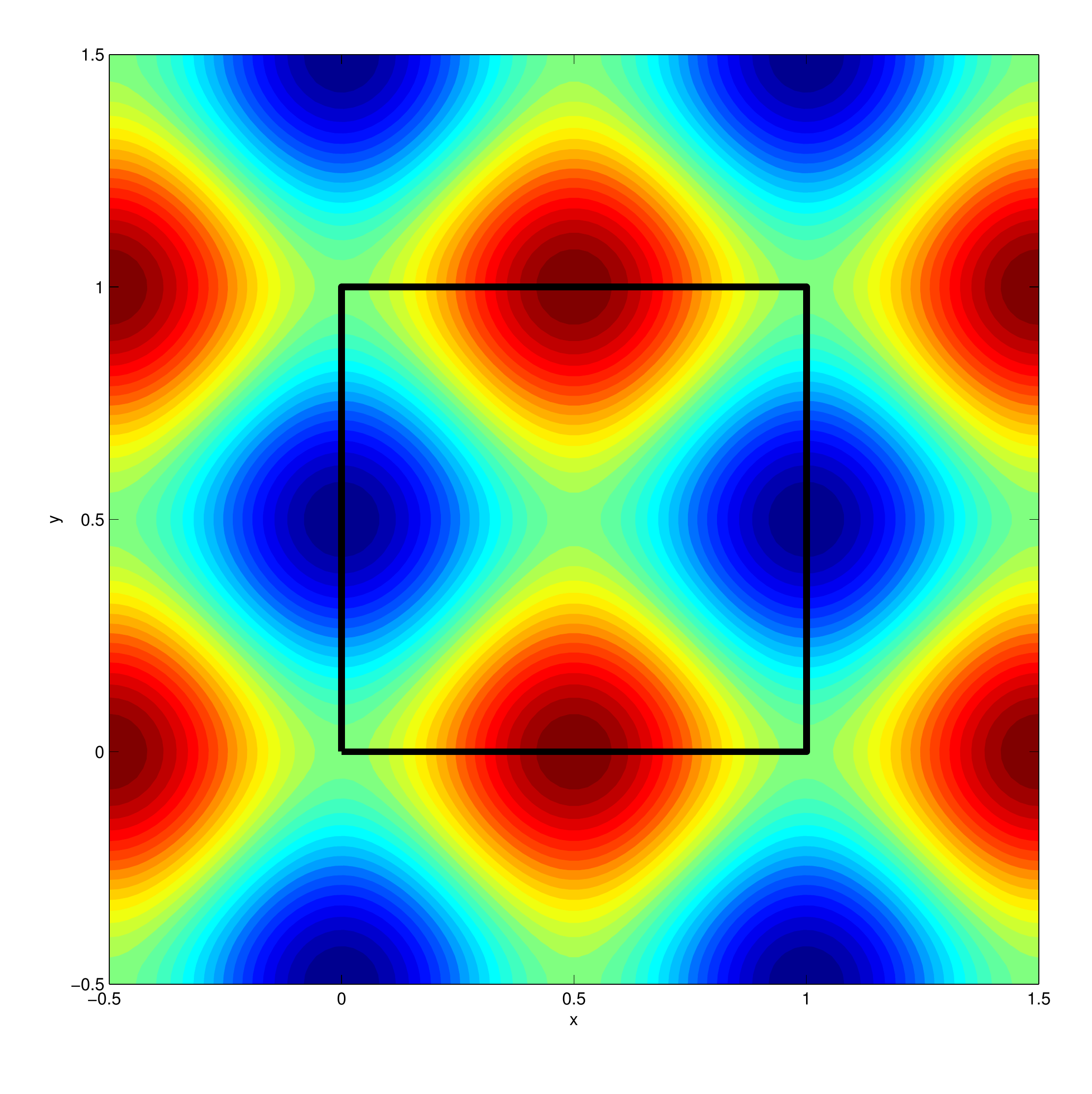}}
}
\caption{Vorticity fields characterized by (a) aligned and (b)
    staggered arrangement of vortex cells obtained as the solutions of
    maximization problem \eqref{eq:optR_P} in the limit $\P_0
    \rightarrow 0$. They are given by, respectively, expressions
  \eqref{eq:tfa} with $p=1$ and \eqref{eq:tfs} with $p=1/2$. The
  boundary of the domain $\Omega$ is marked with the black solid
  line.}
\label{fig:tf0}
\end{center}
\end{figure}

We begin the discussion by analyzing the single-constraint
optimization problem \eqref{eq:optR_P} in the limit $\P_0 \rightarrow
0$. The Euler-Lagrange equation characterizing the solutions of
this problem is, cf.~\eqref{eq:R},
\begin{subequations}
\label{eq:EL} 
\begin{align}
\G(\psi) + 2 \nu \Delta^4 \psi + \lambda \Delta^3 \psi &= 0, \label{eq:ELa} \\
\frac{1}{2} \int_{\Omega} \left( \bnabla\Delta \psi \right)^2\, d\Omega &= \P_0, \label{eq:ELb}
\end{align}
\end{subequations}
where $\lambda \in \RR$ is the Lagrange multiplier associated with
constraint \eqref{eq:ELb}, equation \eqref{eq:ELa} is subject to the
doubly-periodic boundary conditions and we denoted $\G(\psi) :=
\Delta^2 J(\Delta \psi, \psi) + \Delta J(\psi, \Delta^2\psi) +
J(\Delta^2 \psi, \Delta\psi)$. In order to obtain insights about the
behavior of solutions $(\psi,\lambda)$ of \eqref{eq:EL} in the limit
$\P_0 \rightarrow 0$, we use the following series expansion
\begin{subequations}
\label{eq:series} 
\begin{align}
\psi &= \psi_0 + \P_0^{1/2} \, \psi_1 + \P_0^{1} \, \psi_2 + \O(\P_0^{3/2}), \label{eq:psi_s} \\
\lambda &= \lambda_0 + \P_0^{1/2} \, \lambda_1 + \P_0^{1} \, \lambda_2 + \O(\P_0^{3/2}). \label{eq:lam_s} 
\end{align}
\end{subequations}
Introducing ansatz \eqref{eq:series} in \eqref{eq:EL} and collecting
terms proportional to different powers of $\P_0^{1/2}$, we obtain at
the leading order
\begin{subequations}
\label{eq:EL0} 
\begin{alignat}{2}
& \P_0^0: & \qquad 2 \nu \Delta^4 \psi_0 + \lambda_0 \Delta^3 \psi_0 &= - \G(\psi_0), \label{eq:EL0a} \\
&& \frac{1}{2} \int_{\Omega} \left( \bnabla\Delta \psi_0 \right)^2\, d\Omega &= 0, \label{eq:EL0b}
\end{alignat}
\end{subequations}
from which it follows immediately that $\psi_0 \equiv 0$. Using this
result, at the next order we have
\begin{subequations}
\label{eq:EL1} 
\begin{alignat}{2}
& \P_0^{1/2}: & \qquad 2 \nu \Delta^4 \psi_1 + \lambda_0 \Delta^3 \psi_1 &=  0, \label{eq:EL1a} \\
&& \frac{1}{2} \int_{\Omega} \left( \bnabla\Delta \psi_1 \right)^2\, d\Omega &= 1, \label{eq:EL1b}
\end{alignat}
\end{subequations}
where we note that the vanishing of the contribution from $\G(\psi)$
in \eqref{eq:EL1a} is due to $\psi_0$ being identically zero. While
continuing this process might lead to some interesting insights, for
our present purposes it is in fact sufficient to truncate expansions
\eqref{eq:series} at the order $\O(\P_0)$. The corresponding
approximation of our objective function \eqref{eq:R} thus becomes
\begin{equation}
\R_{\P_0} = - \P_0 \, \nu \int_{\Omega} \left( \Delta \psi_1 \right)^2 \, d\Omega + \O(\P_0^{3/2}).
\label{eq:R0}
\end{equation}
As regards problem \eqref{eq:EL1} defining $(\psi_1,\lambda_0)$, we
note that, since for zero-mean functions defined on doubly-periodic
domains, $\Ker(\Delta^3) = \{ 0 \}$, equation \eqref{eq:EL1a} becomes
an eigenvalue problem $\Delta\psi_1 = \lambda_0' \psi_1$, where
$\lambda_0' := - \lambda_0 / (2 \nu) < 0$. It can be shown via direct
calculation that $\R_{\P_0} \approx 2 \nu \lambda_0' \, \P_0$ as $\P_0
\rightarrow 0$, cf.~\eqref{eq:R0}, and we are therefore interested in
the eigenfunctions associated with the largest, i.e., the least
negative, eigenvalues. There are two distinct possibilities
corresponding to different arrangements of vortices with 2-fold
rotational symmetry in the domain $\Omega$, namely (cf. figure
\ref{fig:tf0}),
\begin{itemize}
\item
aligned arrangement where 
\begin{equation}
\psi_{1,a}(x,y) = \frac{1}{4}\sin(2\pi p \,x) \, \sin(2\pi q \,y), 
\quad p=1,2,\dots
\label{eq:tfa}
\end{equation}
with the eigenvalue $\lambda_0' = - 8\pi^2 p^2$ which is maximized
for $p=1$ resulting in
\begin{equation}
\R_{\P_0} \approx  -16\pi^2\nu \, \P_0,
\label{eq:maxRa}
\end{equation}

\item
staggered arrangement where 
\begin{equation}
\psi_{1,s}(x,y) = \frac{1}{4}\sin\left[2\pi p \,(x+y)\right] \, 
\sin\left[2\pi p \,(x-y)\right], 
\quad p=1/2,1,3/2,\dots
\label{eq:tfs}
\end{equation}
with the eigenvalue $\lambda_0' = - 16\pi^2 p^2$ which is maximized
for $p=1/2$ resulting in
\begin{equation}
\R_{\P_0} \approx  -8\pi^2\nu \, \P_0.
\label{eq:maxRs}
\end{equation}
\end{itemize}

We therefore conclude that, while $\R_{\P_0}(\psi_1)$ is
negative-definite for both arrangements, it assumes larger (i.e., less
negative) values for the staggered configuration. To prove that
$\psi_1$ is indeed a local maximizer of $\R_{\P_0}(\psi)$ in the
limit $\P_0 \rightarrow 0$, rather than just a saddle point,
it would be necessary to demonstrate the negative definiteness of the
Hessian of $\R_{\P_0}(\psi)$ at $\psi_1$.  This, however,
becomes technically complicated and we will not study it here.
Computational results presented in section \ref{sec:results} provide
evidence that indeed the maximizers of the single-constraint problem
\eqref{eq:optR_P} approach the field $\P_0^{1/2}\,\psi_1$ in the limit
$\P_0 \rightarrow 0$. As regards the negativity of $\R_{\P_0}(\tpP)$
in this limit, we remark that $J(\varphi,\Delta\varphi) \equiv 0$ when
$\varphi$ is an eigenfunction of the Laplacian and then the cubic part
of $\R_{\P_0}(\tpP)$ vanishes leaving just the dissipative term.

As regards the two-constraint problems \eqref{eq:optR_EP} and
\eqref{eq:optR_KP} in the respective limits $\P_0 \rightarrow
(2\pi)^{2} \E_0$ and $\P_0 \rightarrow (2\pi)^{4} \K_0$, we note that,
as the leading eigenfunction of the Laplacian, the maximizer
$\psi_{1,s}$ defined in \eqref{eq:tfs} saturates Poincar\'e's
inequality, i.e., $\P(\psi_{1,s}) = (2\pi)^{2} \E(\psi_{1,s})$. This
means that second constraints $\E(\psi) = \E_0$ and $\K(\psi) = \K_0$
are satisfied automatically by $\psi_{1,s}$ and therefore need
not be enforced through the introduction of another Lagrange
multiplier. This allows us to conclude that $\psi_{1,s}$ is
also the solution of the limiting forms of two-constraint
optimization problems \eqref{eq:optR_EP} and \eqref{eq:optR_KP}.

\section{Computational Results}
\label{sec:results}

In the two subsections below we describe the results obtained from the
numerical solution of optimization problems \eqref{eq:optR_P},
\eqref{eq:optR_EP} and \eqref{eq:optR_KP} using the computational
approaches described in section \ref{sec:grad} for a broad range of
constraint parameters $\P_0$, $(\E_0,\P_0)$ and $(\K_0,\P_0)$.
In subsection \ref{sec:results_finite_time} we discuss how the palinstrophy
growth in finite time corresponding to the instantaneous maximizers
found in sections \ref{sec:results1} and \ref{sec:results2} used as
the initial data compares with the available finite-time estimates.

In the calculations described in sections \ref{sec:results1} and
\ref{sec:results2} the key element is the evaluation of the gradient
$\nabla\R_{\P_0}(\psi)$, first in the $L_2$ and then in the $H^4$
topology, cf.~\eqref{eq:gradRL2}--\eqref{eq:gradRH4}, which is done
using a pseudospectral Fourier-Galerkin technique with standard
dealiasing. The resolution varied from $128^2$ to $1024^2$ grid points
in the low-palinstrophy and high-palinstrophy cases, respectively.
Convergence of all calculations with respect to the grid refinement
was carefully verified. The rate of convergence of discrete gradient
flow \eqref{eq:desc} depends, among other factors, on the values of
the length-scale parameters characterizing the Sobolev inner product
\eqref{eq:ipH4}. Based on our extensive numerical test, we used
$\ell_1 = \ell_2 = 0$, $\ell_3 \in [10^{-2},10^{-1}]$ and $\ell_4 \in
[10^{-4},10^{-2}]$ (with smaller values of $\ell_3$ and $\ell_4$
corresponding to higher numerical resolutions). We remark that,
  given the equivalence of the inner products \eqref{eq:ipH4}
  corresponding to different values of $\ell_1, \ell_2, \ell_3,
  \ell_4$, these choices do not affect the maximizers found, but only
  how rapidly they are approached by algorithm \eqref{eq:desc}. In
all calculations the value of the viscosity coefficient was $\nu =
10^{-3}$.

As regards the results for the time-dependent problem presented in
section \ref{sec:results_finite_time}, we used a numerical approach
combining a standard pseudospectral discretization in space with the
Krylov subspace method described by \citet{edwards:KrylovMethod} for
the time discretization to numerically solve system \eqref{eq:NS2D}
for the vorticity evolution. As regards the initial data $\omega_0$,
it was chosen as the vorticity corresponding to the solutions of
optimization problems \eqref{eq:optR_EP}, \eqref{eq:optR_KP} and
\eqref{eq:optR_P}, i.e., $-\Delta\tpEP$, $-\Delta\tpKP$ and
$-\Delta\tpP$. In these simulations, the resolution varied from
$512^2$ to $4096^2$ grid points depending on the characteristic length
scales of the initial data which ensured that all calculations were
well-resolved. The length of the time window over which the problem
was solved was chosen for all initial data long enough to capture the
initial amplification of the palinstrophy followed by its viscous
decay. Power-law exponents mentioned below were computed by fitting a
linear polynomial to the $\log_{10} (d\P/dt)$ versus $\log_{10}
(\P_0)$ relationship.

\subsection{Optimization Problems with a Single Constraint on $\P_0$ }
\label{sec:results1}

\begin{figure}
\setcounter{subfigure}{0}
\begin{center}
\subfigure[]{\includegraphics[width=0.45\textwidth]{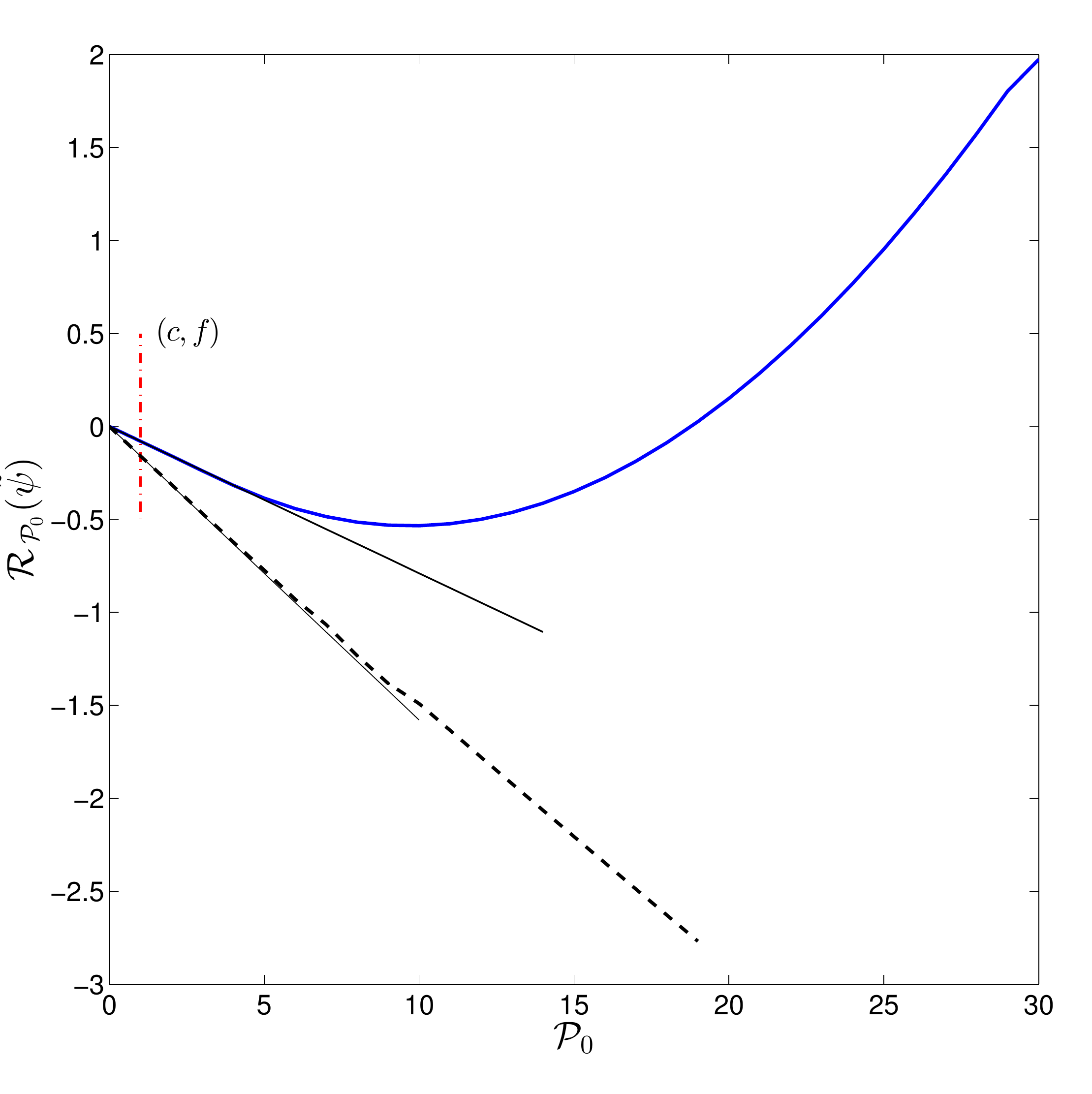}}
\subfigure[]{\includegraphics[width=0.45\textwidth]{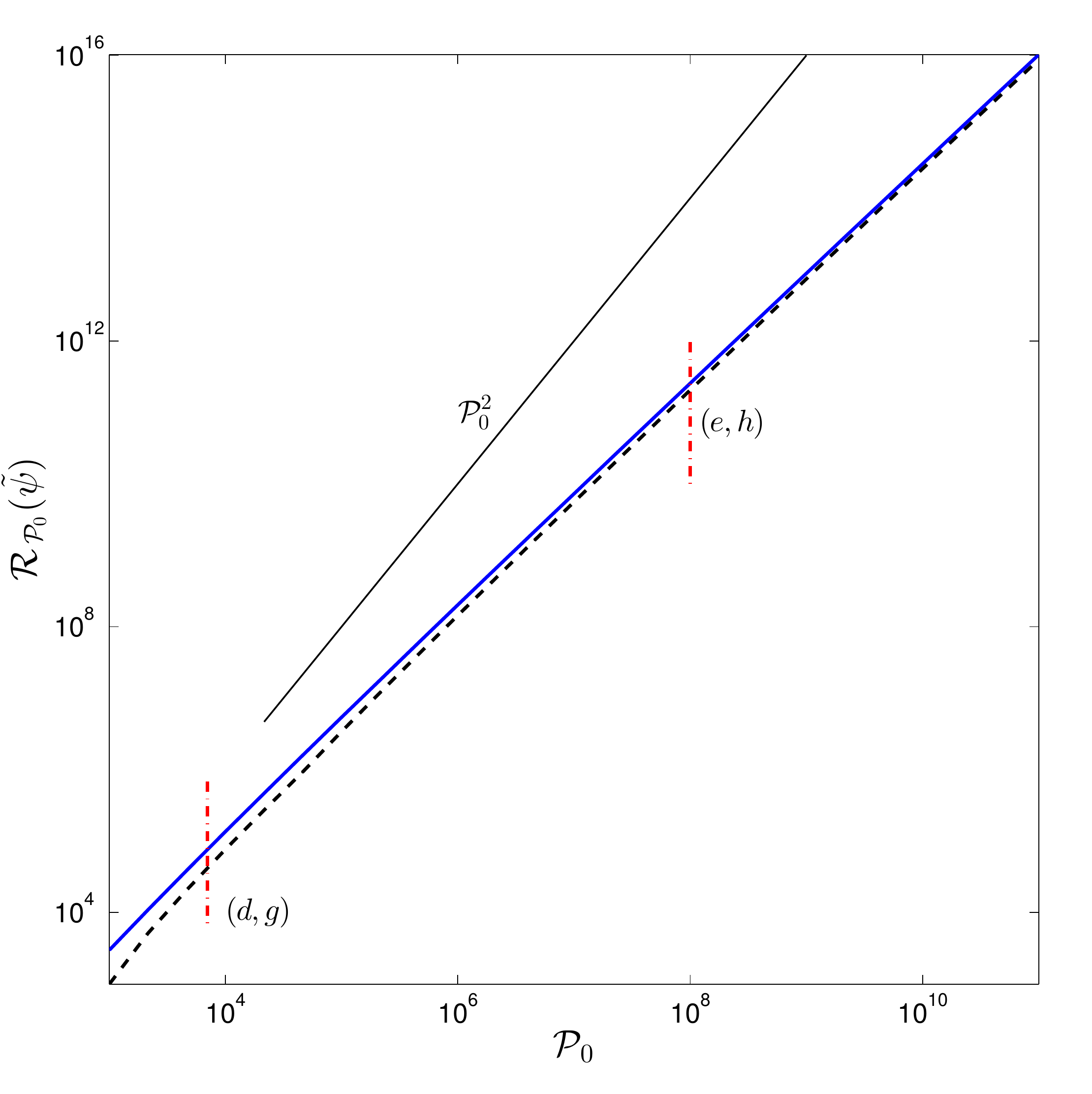}}\\
\subfigure[]{\includegraphics[width=0.3\textwidth]{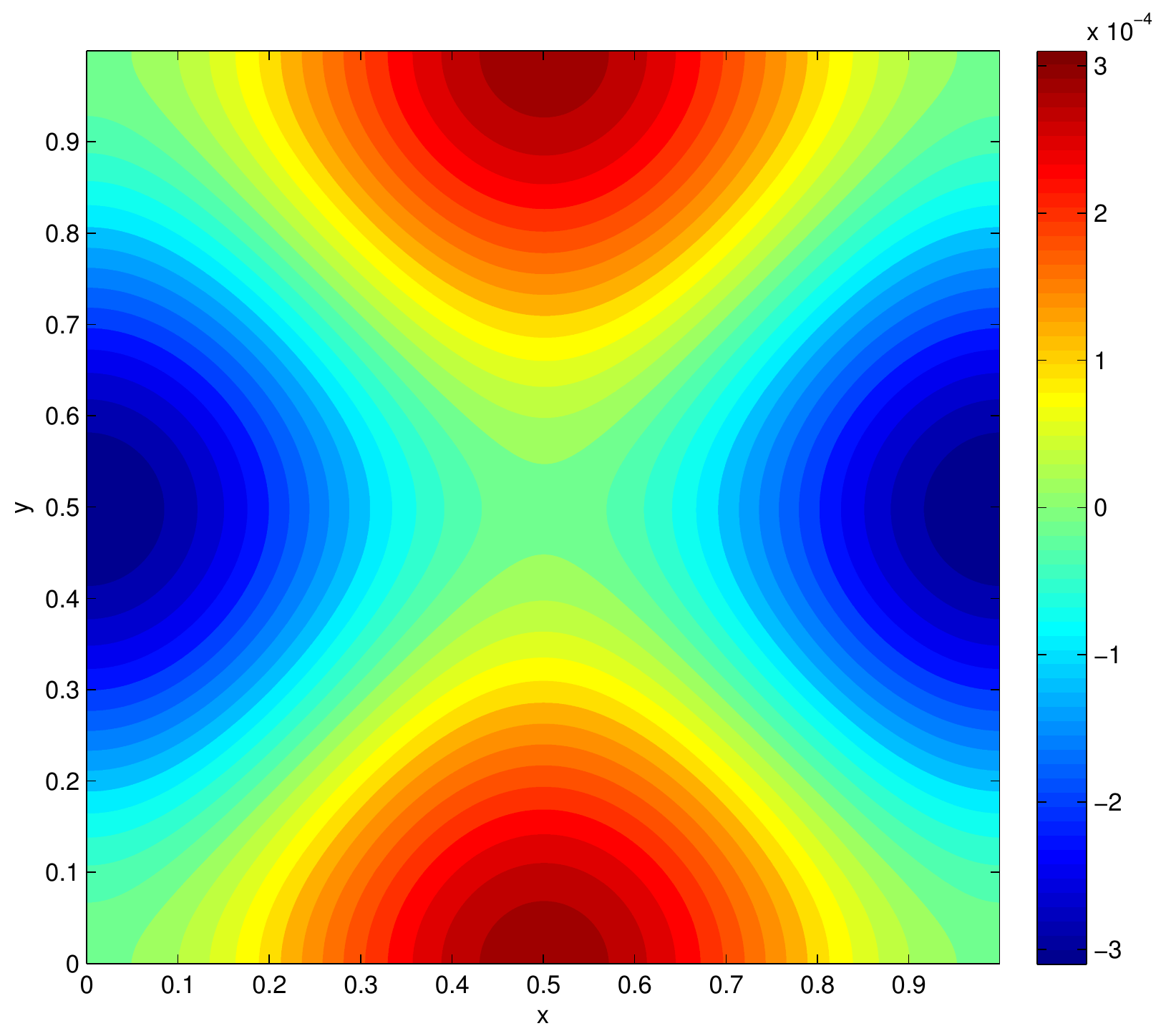}}
\subfigure[]{\includegraphics[width=0.3\textwidth]{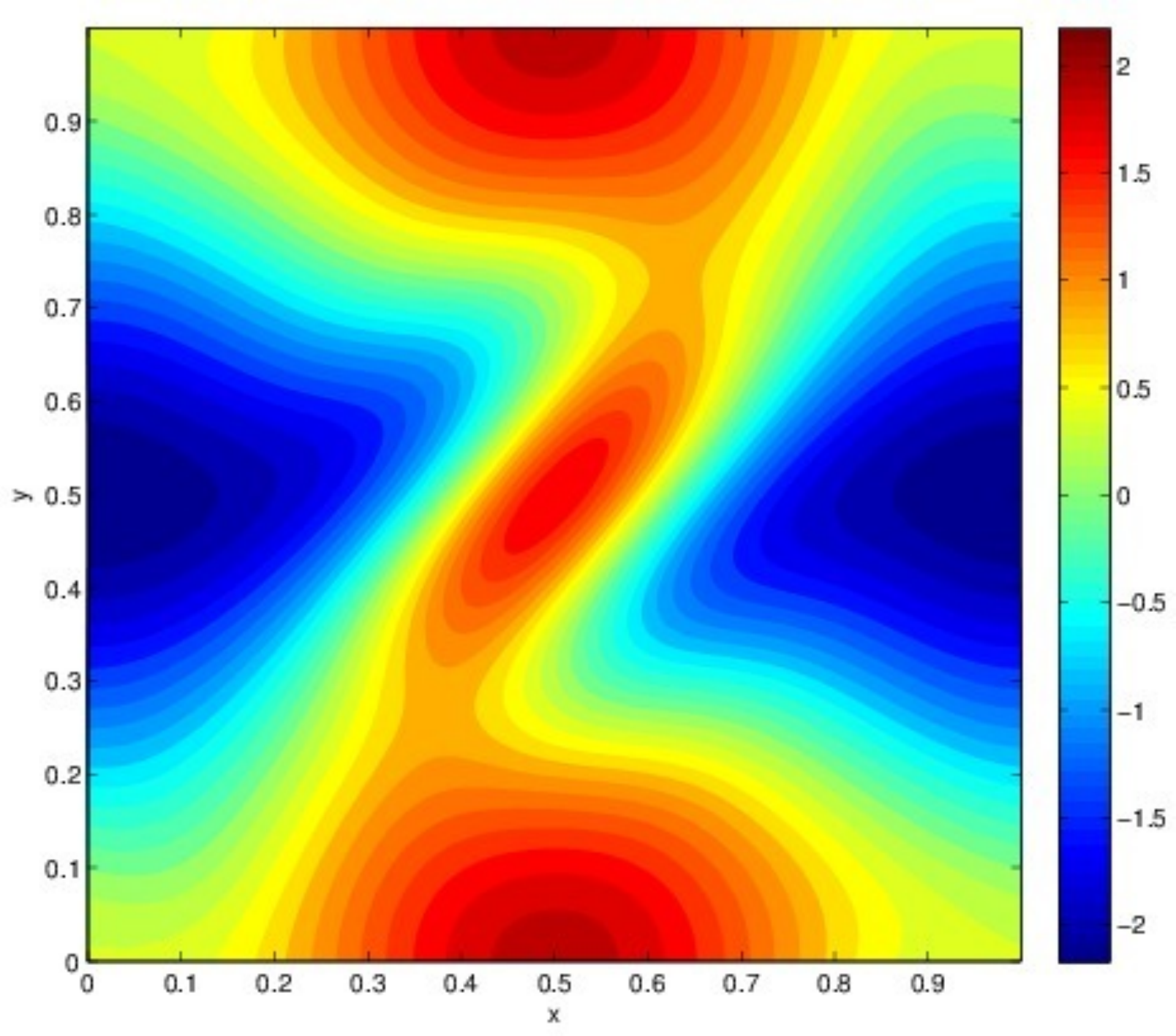}}
\subfigure[]{\includegraphics[width=0.3\textwidth]{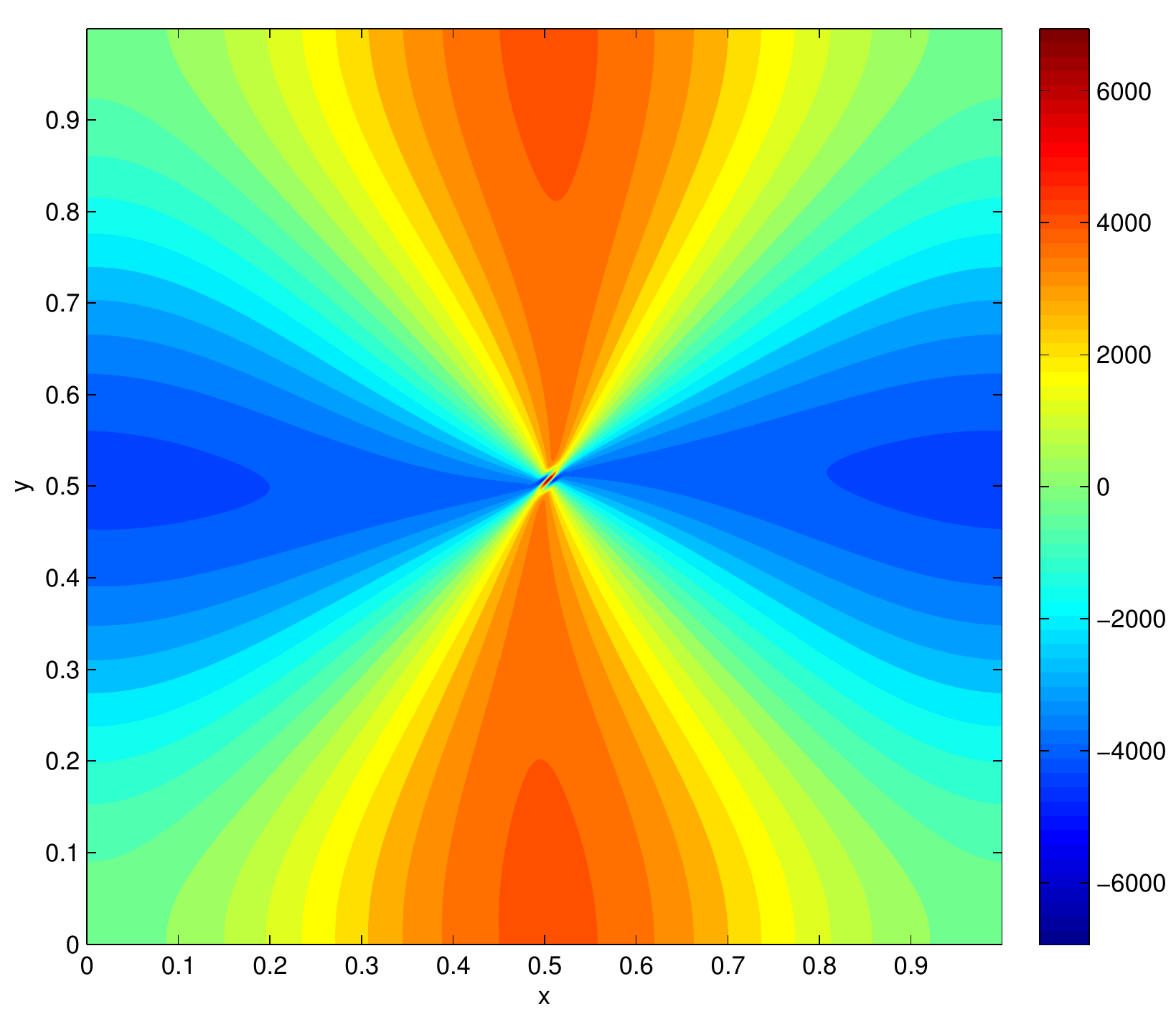}}\\
\subfigure[]{\includegraphics[width=0.3\textwidth]{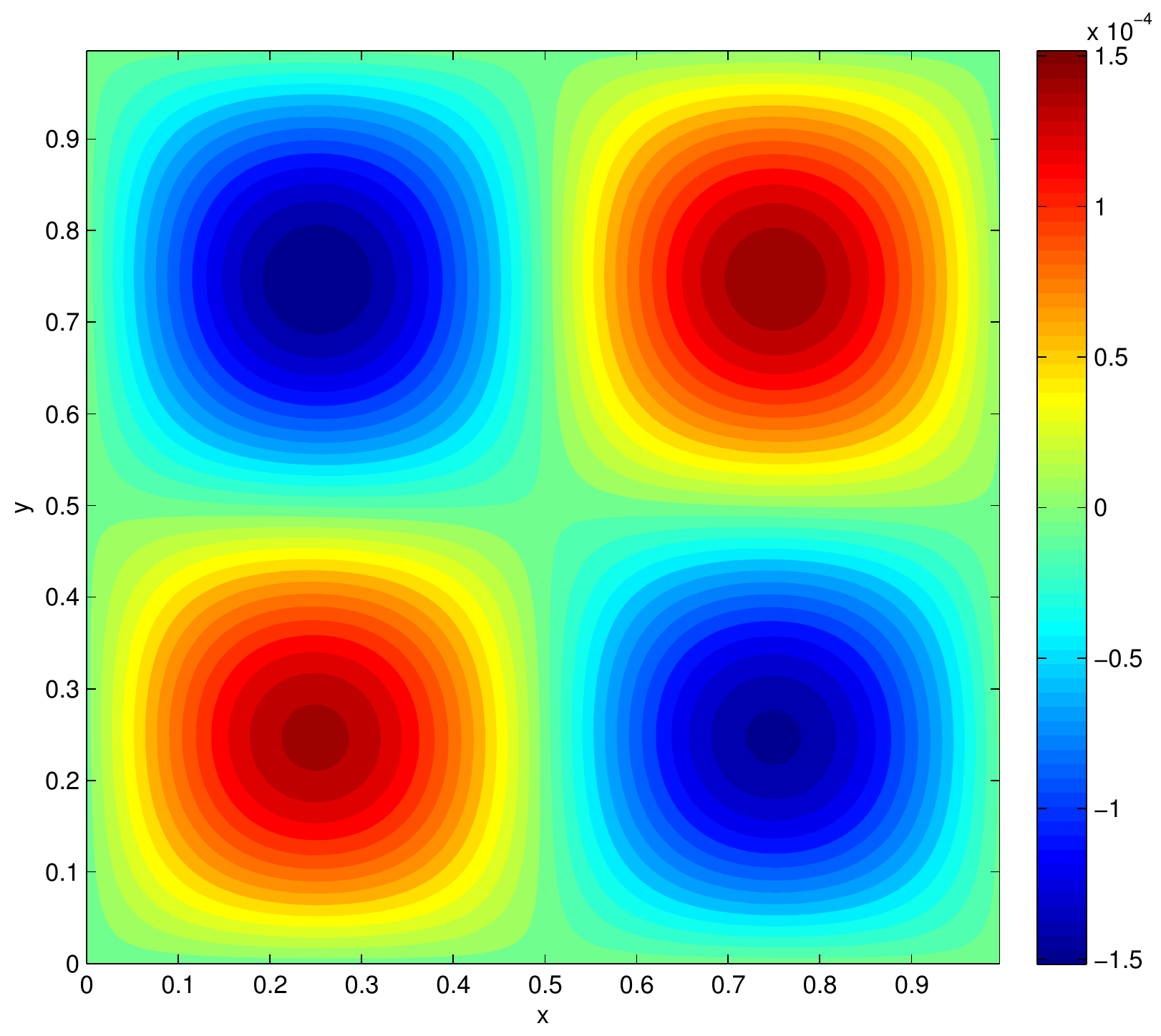}}
\subfigure[]{\includegraphics[width=0.3\textwidth]{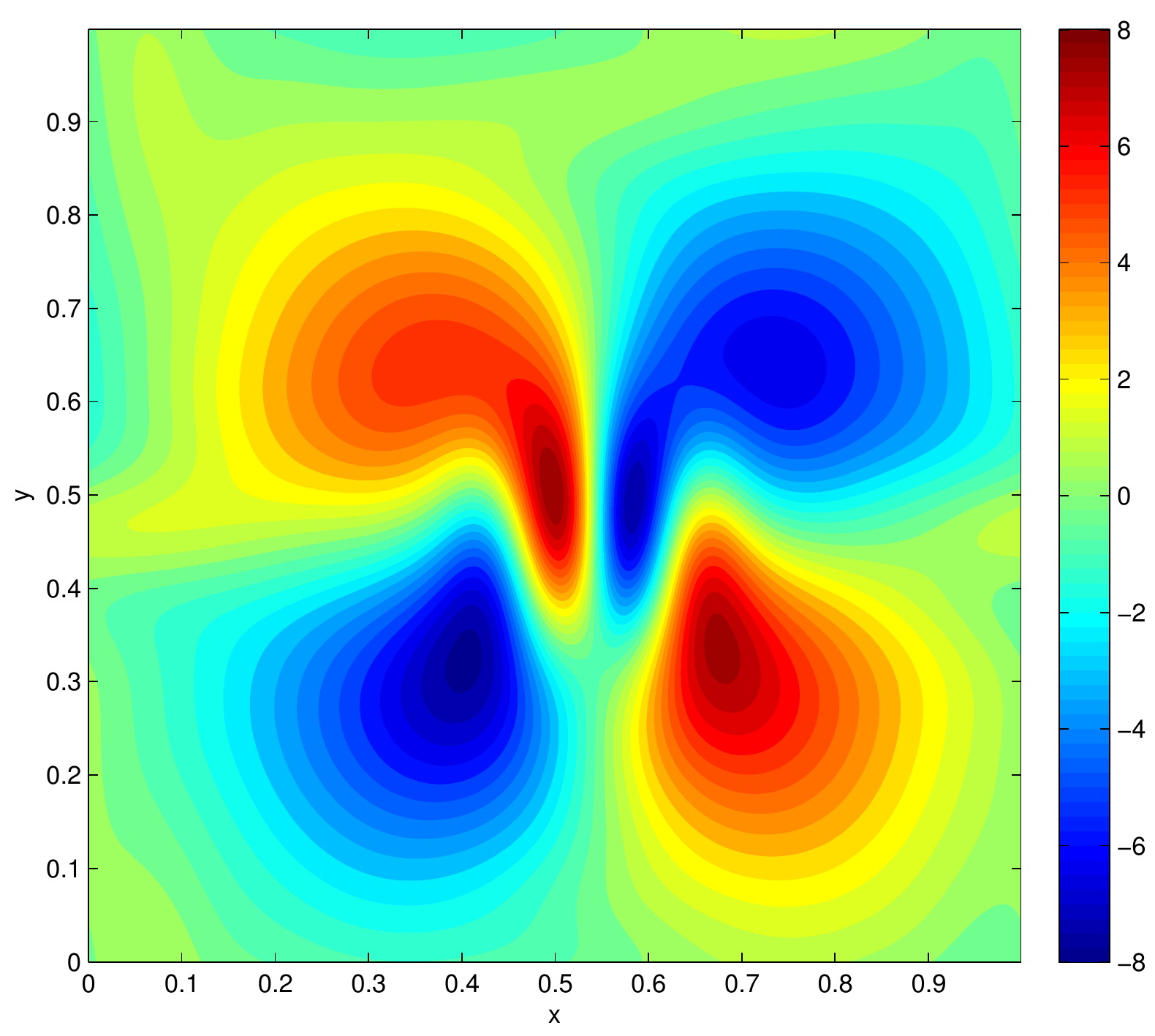}}
\subfigure[]{\includegraphics[width=0.3\textwidth]{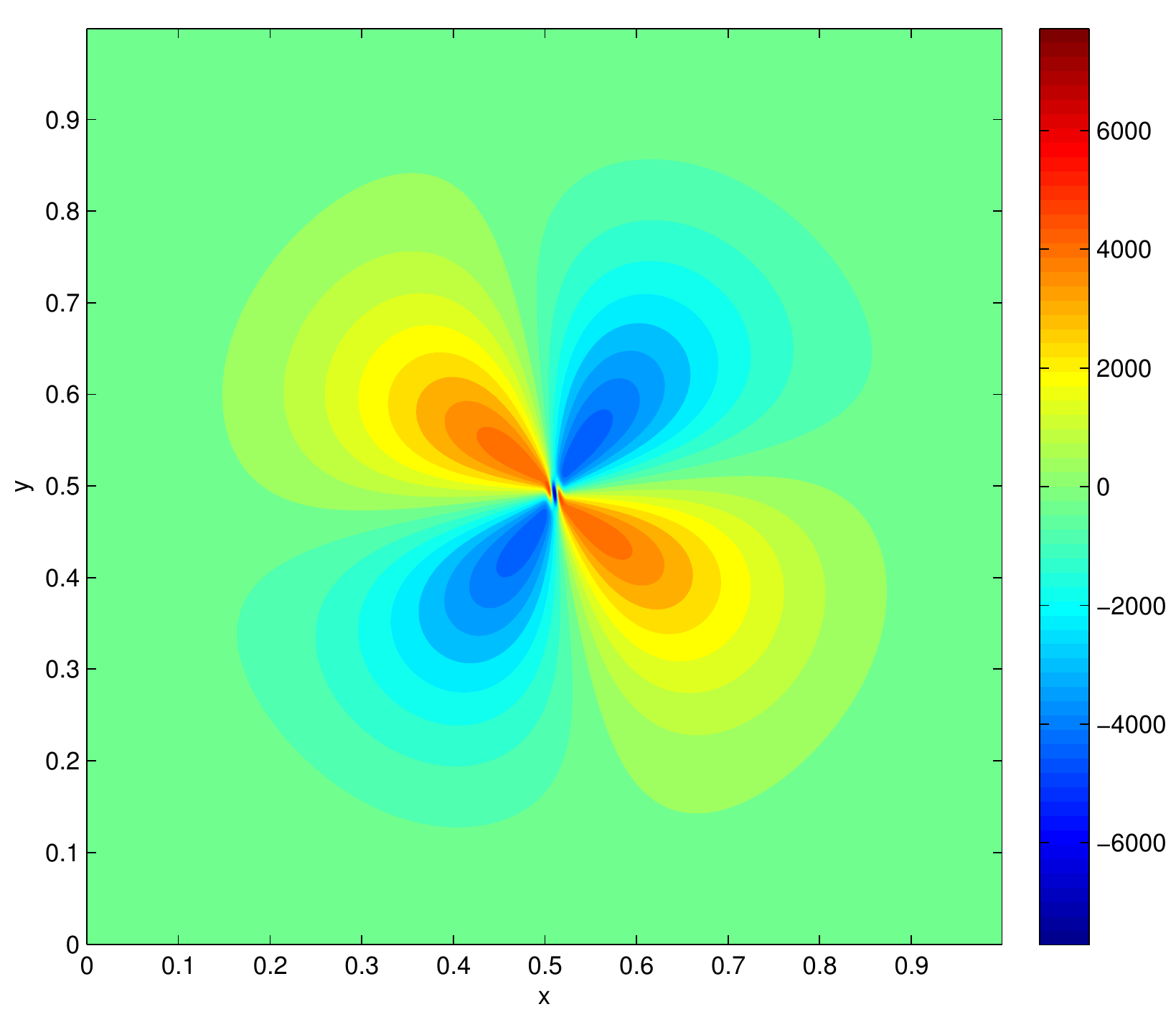}}
\caption{Dependence of the maximum palinstrophy rate of growth
  $\R_{\P_0}(\tpP)$ on $\P_0$ for (a) small $\P_0$ and (b) large
  $\P_0$; (c--h) optimal vortex states corresponding to the branch
  with (c--e) staggered and (f--h) aligned arrangement of the vortex
  cells for the values of $\P_0$ marked with short vertical lines in
  figures (a) and (b); since the maximizing vortex states
  corresponding to the lower branch proved very difficult to compute
  accurately for intermediate values of $\P_0$, this branch is not
  complete in figure (a). }
\label{fig:RvsP0_1constr}
\end{center}
\end{figure}

In this section we discuss solutions of the single-constraint
optimization problem \eqref{eq:optR_P} with the goal of assessing the
sharpness of estimate \eqref{eq:dPdt_P}. In figures
\ref{fig:RvsP0_1constr}(a) and \ref{fig:RvsP0_1constr}(b) we show the
maximum palinstrophy rate of growth $\R_{\P_0}(\tpP)$ obtained,
respectively, for small and large values of $\P_0$. We note the
presence of two distinct branches of maximizing solutions, and the
corresponding vorticity fields $-\Delta\tpP$ are shown in figures
\ref{fig:RvsP0_1constr}(c,d,e) and \ref{fig:RvsP0_1constr}(f,g,h).
The two branches arise, via continuation with respect to parameter
$\P_0$, from the two limiting maximizers discussed in section
\ref{sec:smallP} and characterized by the staggered and aligned
arrangement of the vortex cells (cf.~figure \ref{fig:tf0}), with the
latter case always giving a larger value of $\R_{\P_0}(\tpP)$ for a
given $\P_0$.  The localized vortex structure present in the field
shown in figure \ref{fig:RvsP0_1constr}(e) is magnified in figure
\ref{fig:vortex}(a).  In agreement with the discussion in section
\ref{sec:smallP}, for small $\P_0$ the values of the maximum
palinstrophy rate of growth $\R_{\P_0}(\tpP)$ on both branches are
negative and in the limit $\P_0 \rightarrow 0$, $\R_{\P_0}(\tpP) \sim
-8\pi^2\nu\P_0$ and $\R_{\P_0}(\tpP) \sim -16\pi^2\nu\P_0$ in the
two cases (figure \ref{fig:RvsP0_1constr}(a)).  On the other hand,
in figure \ref{fig:RvsP0_1constr}(b) we see that for large values of
$\P_0$ the maximum rate of growth $\R_{\P_0}(\tpP)$ exhibits a clear
power-law behavior with respect to $\P_0$ for both branches. In both
cases the exponent is $1.57 \pm 0.05$ which is less than the
exponent $2$ appearing in estimate \eqref{eq:dPdt_P}, cf. table
\ref{tab:exp}.

\subsection{Optimization Problems with Two Constraints on $(\E_0,\P_0)$ and $(\K_0,\P_0)$ }
\label{sec:results2}

\begin{figure}
\setcounter{subfigure}{0}
\begin{center}
\subfigure[]{\includegraphics[width=0.45\textwidth]{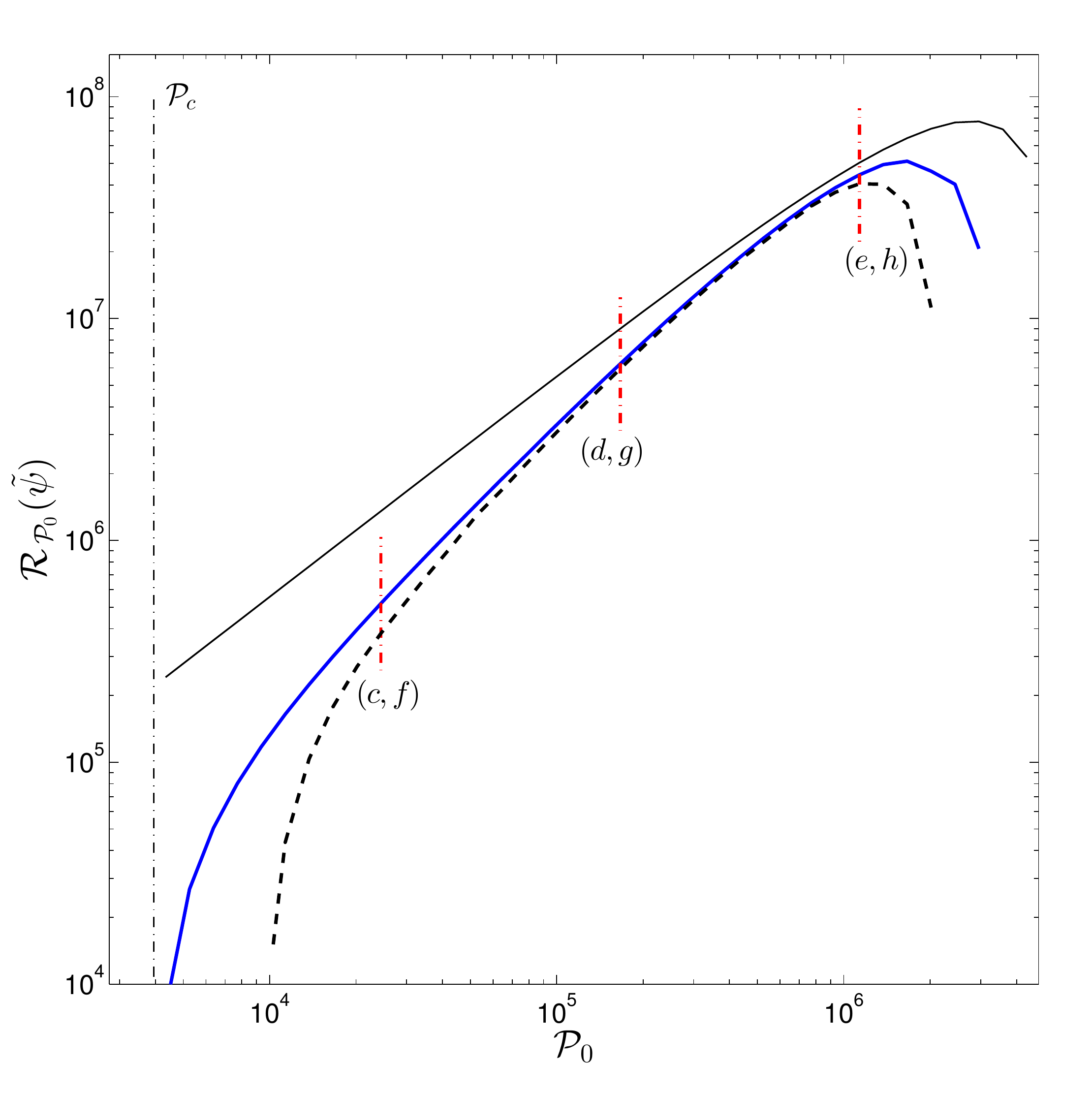}}
\subfigure[]{\includegraphics[width=0.45\textwidth]{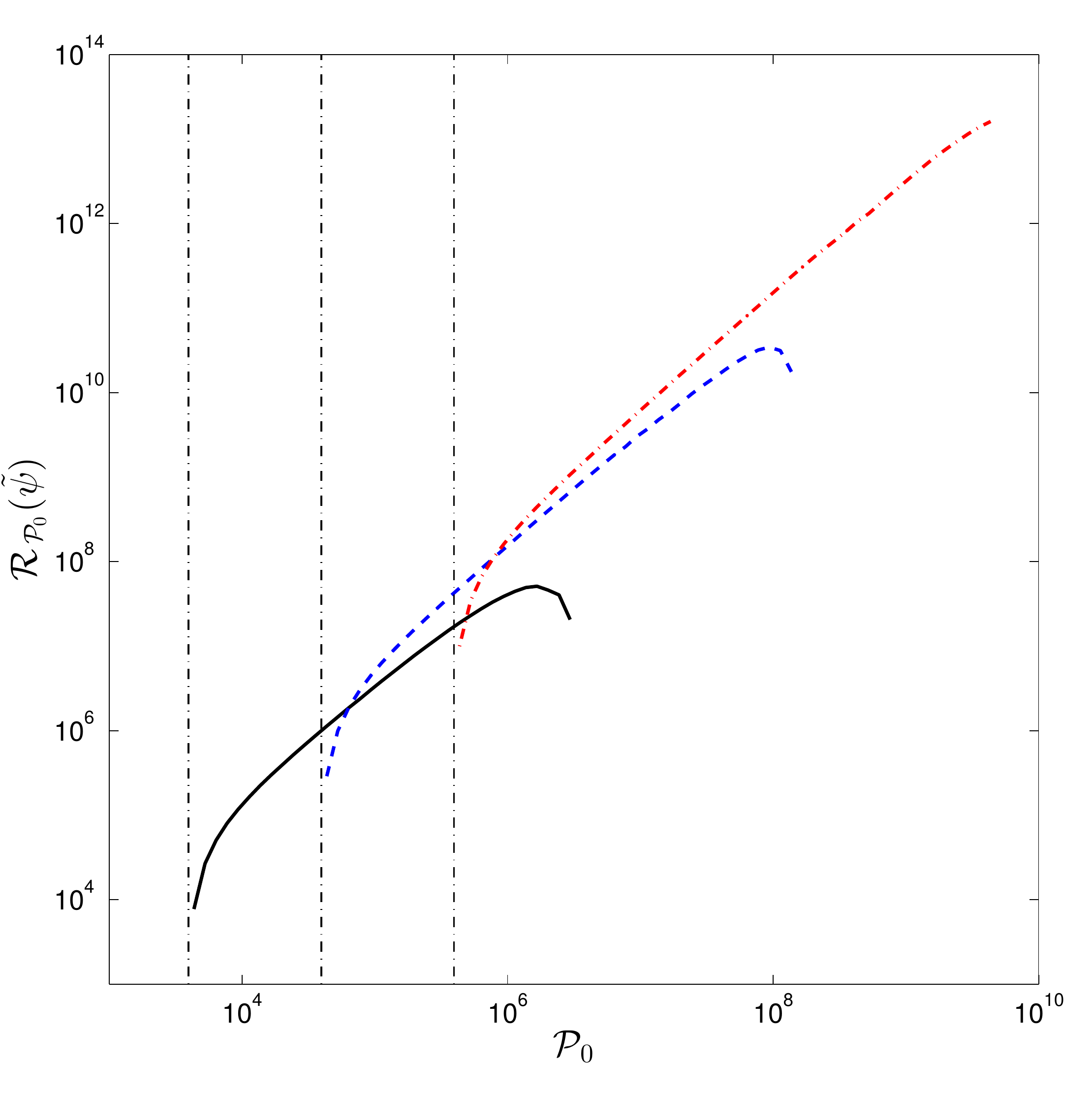}}\\
\subfigure[]{\includegraphics[width=0.3\textwidth]{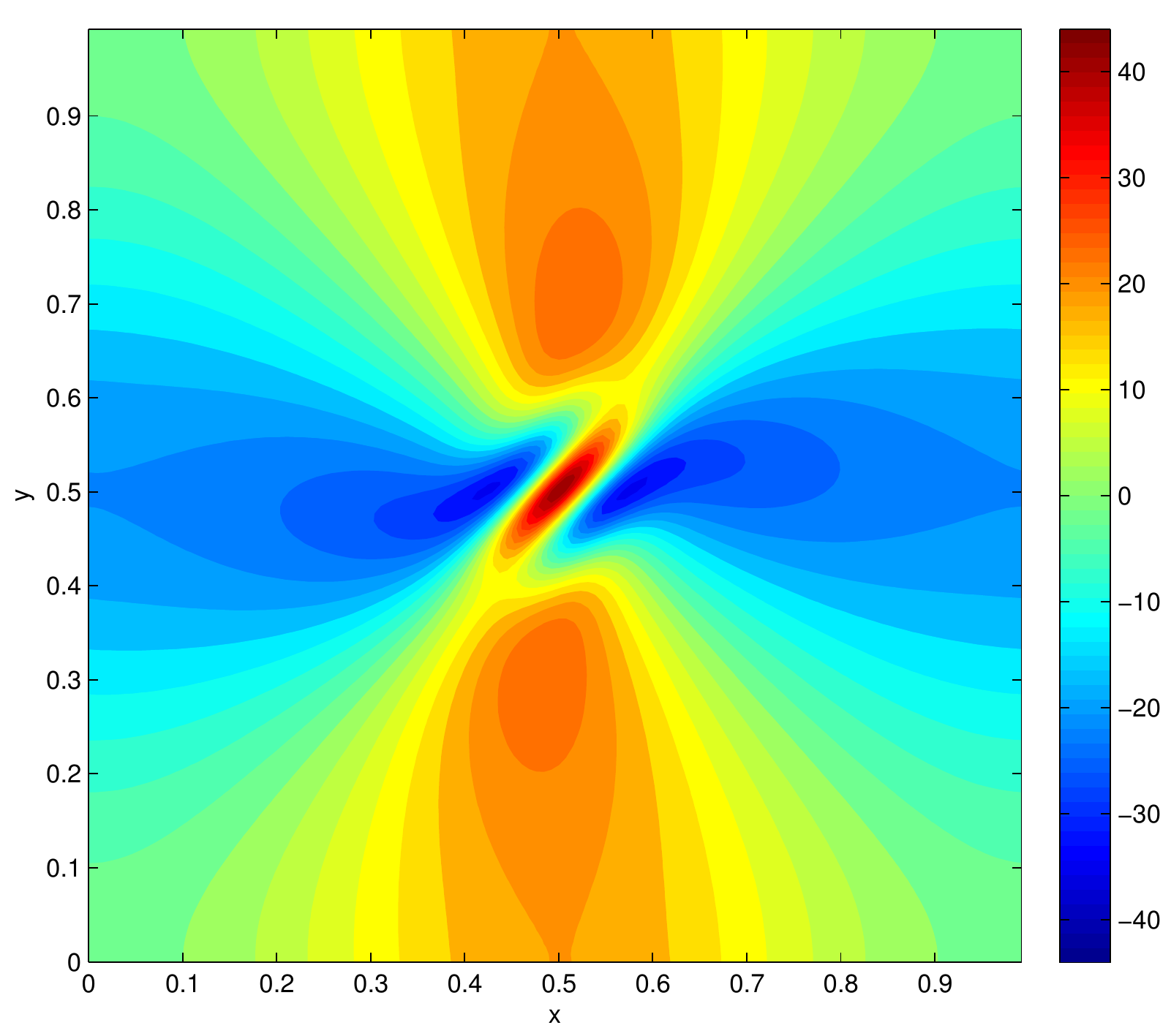}}
\subfigure[]{\includegraphics[width=0.3\textwidth]{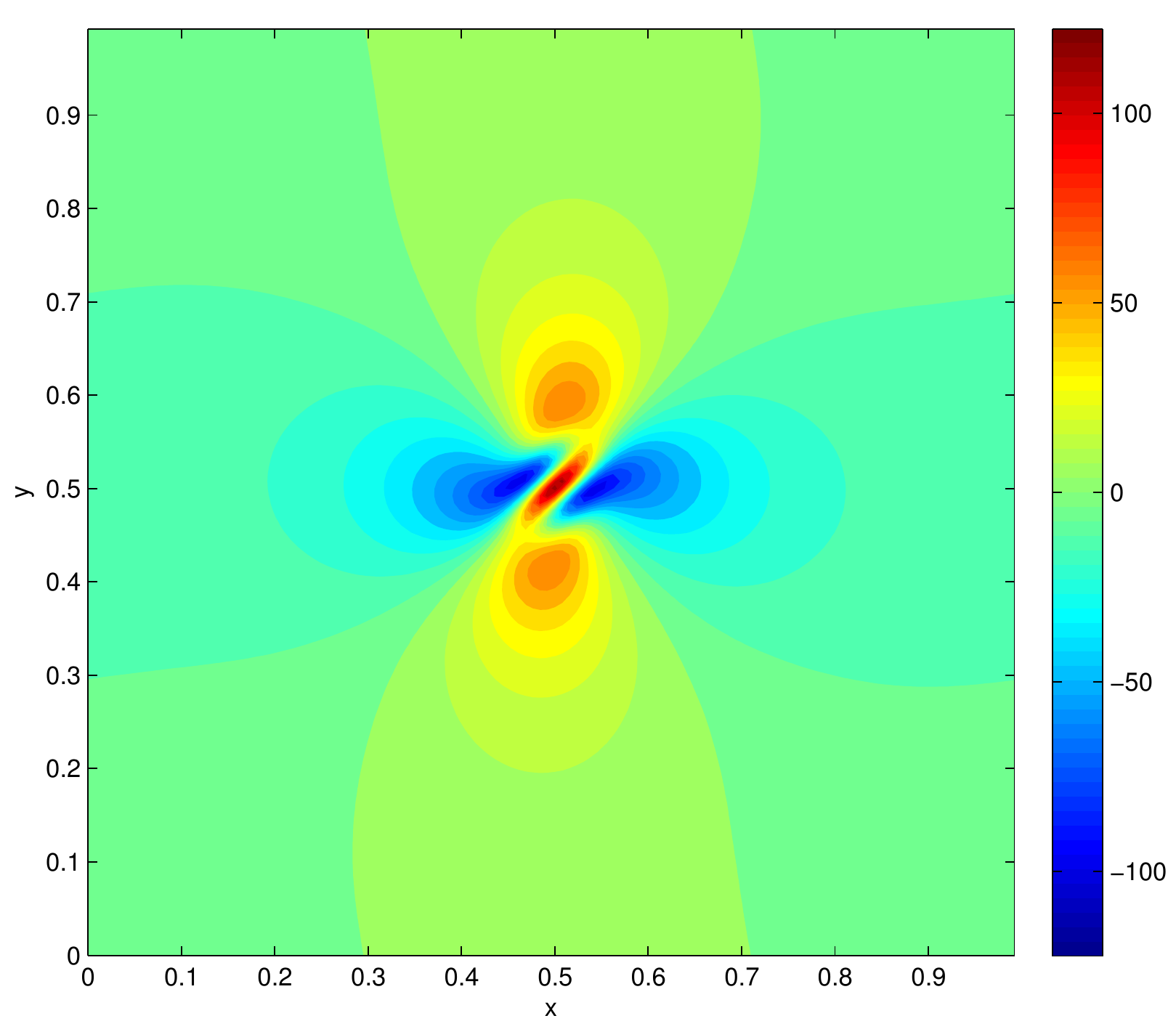}}
\subfigure[]{\includegraphics[width=0.3\textwidth]{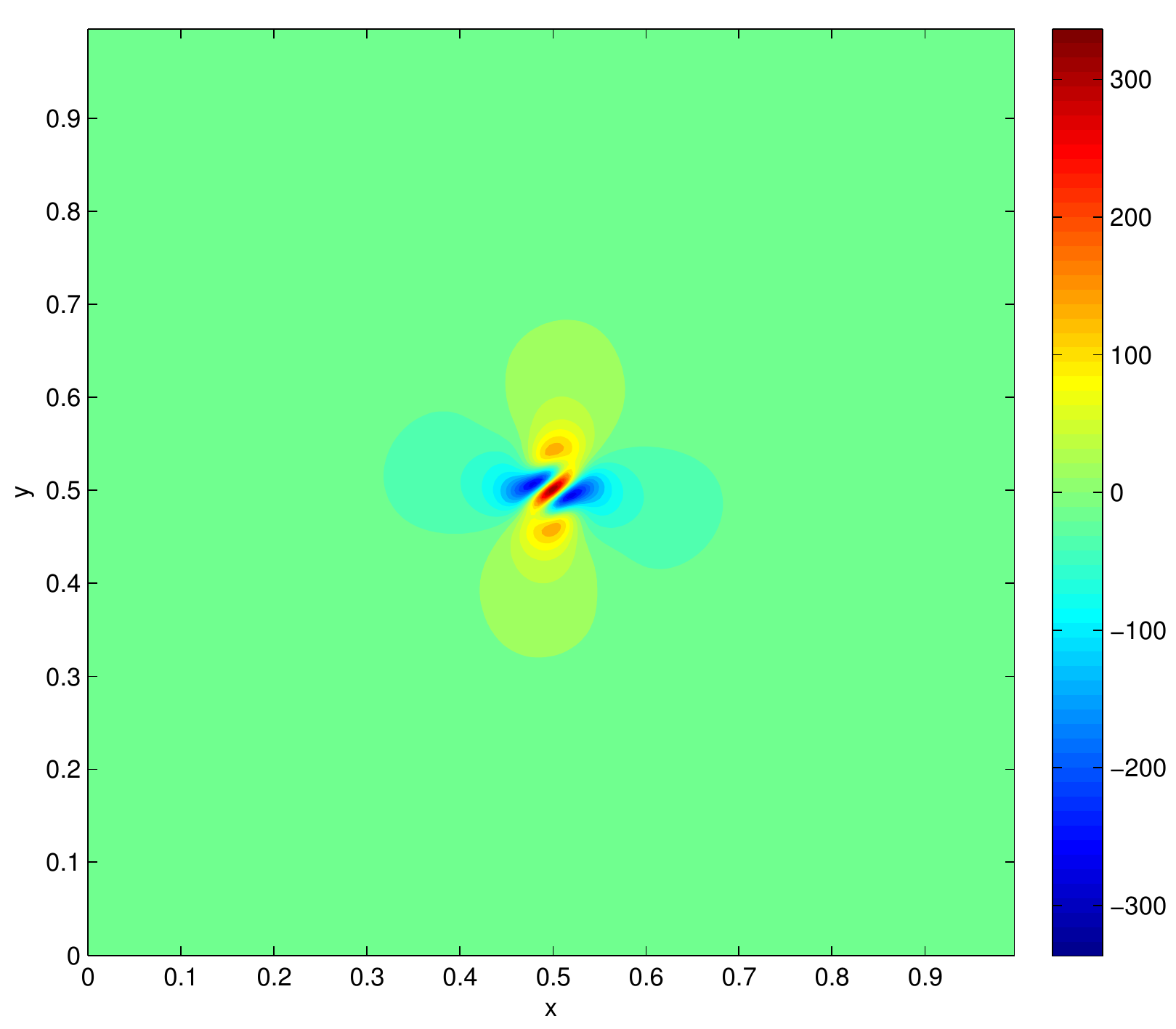}}\\
\subfigure[]{\includegraphics[width=0.3\textwidth]{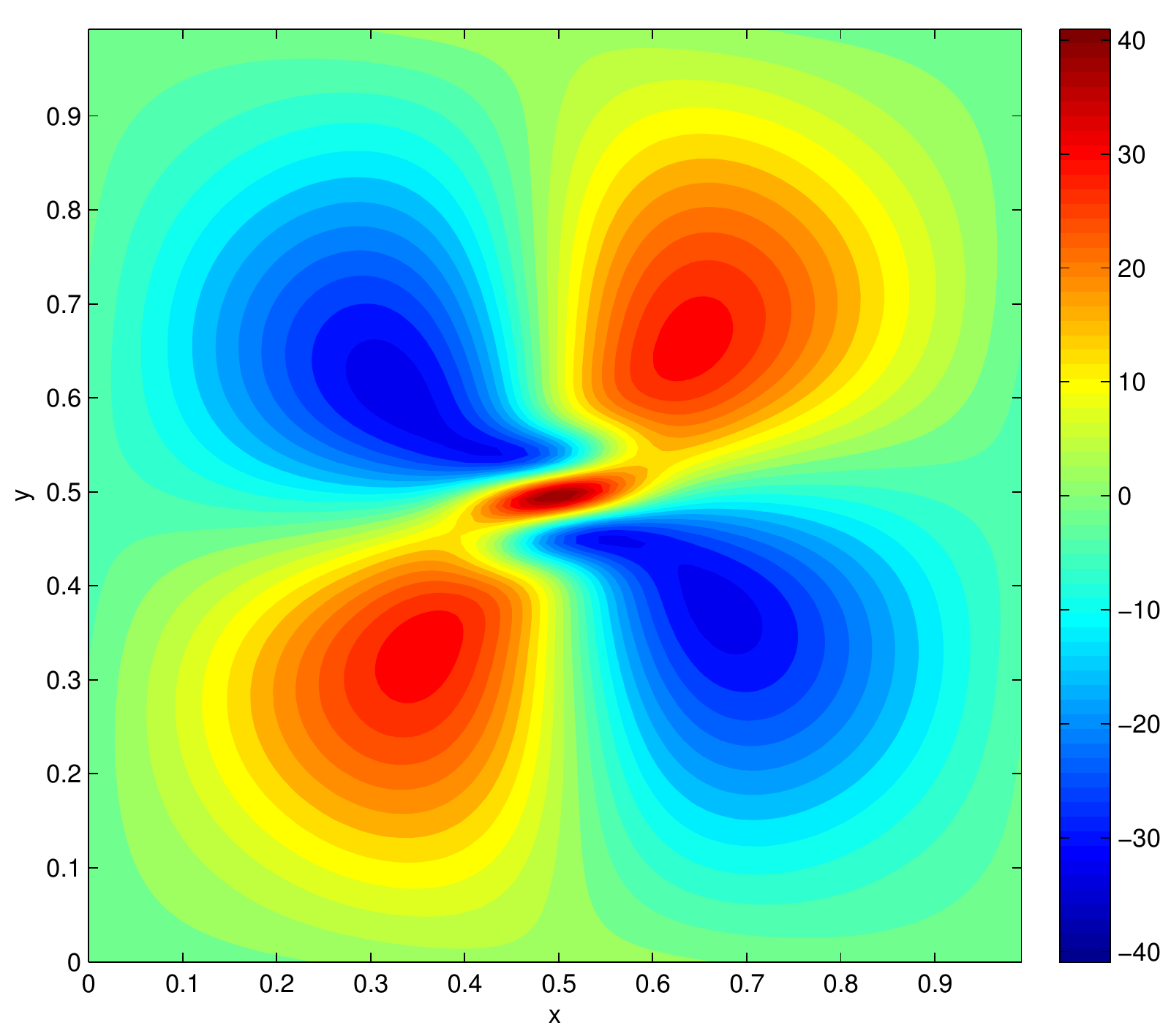}}
\subfigure[]{\includegraphics[width=0.3\textwidth]{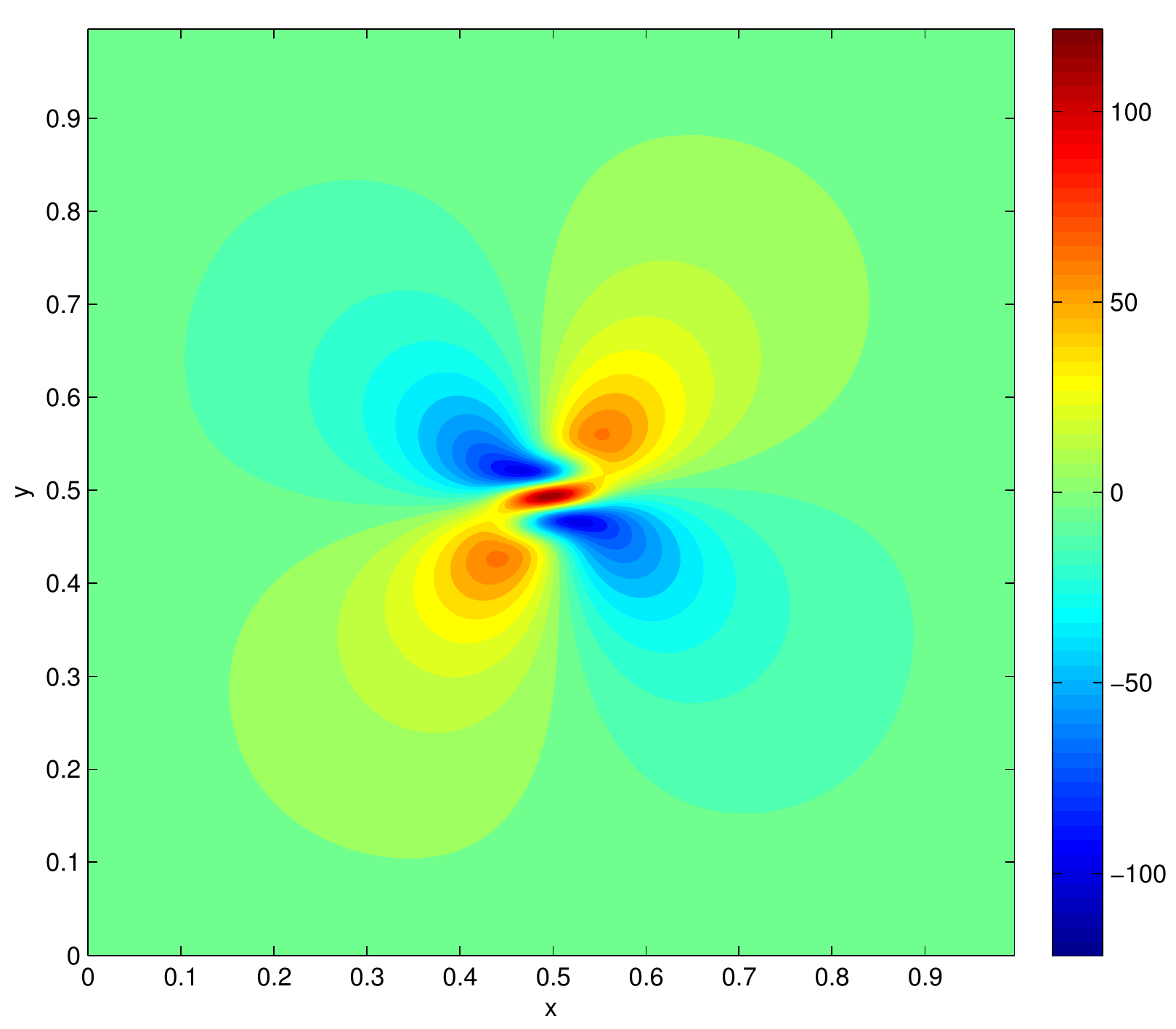}}
\subfigure[]{\includegraphics[width=0.3\textwidth]{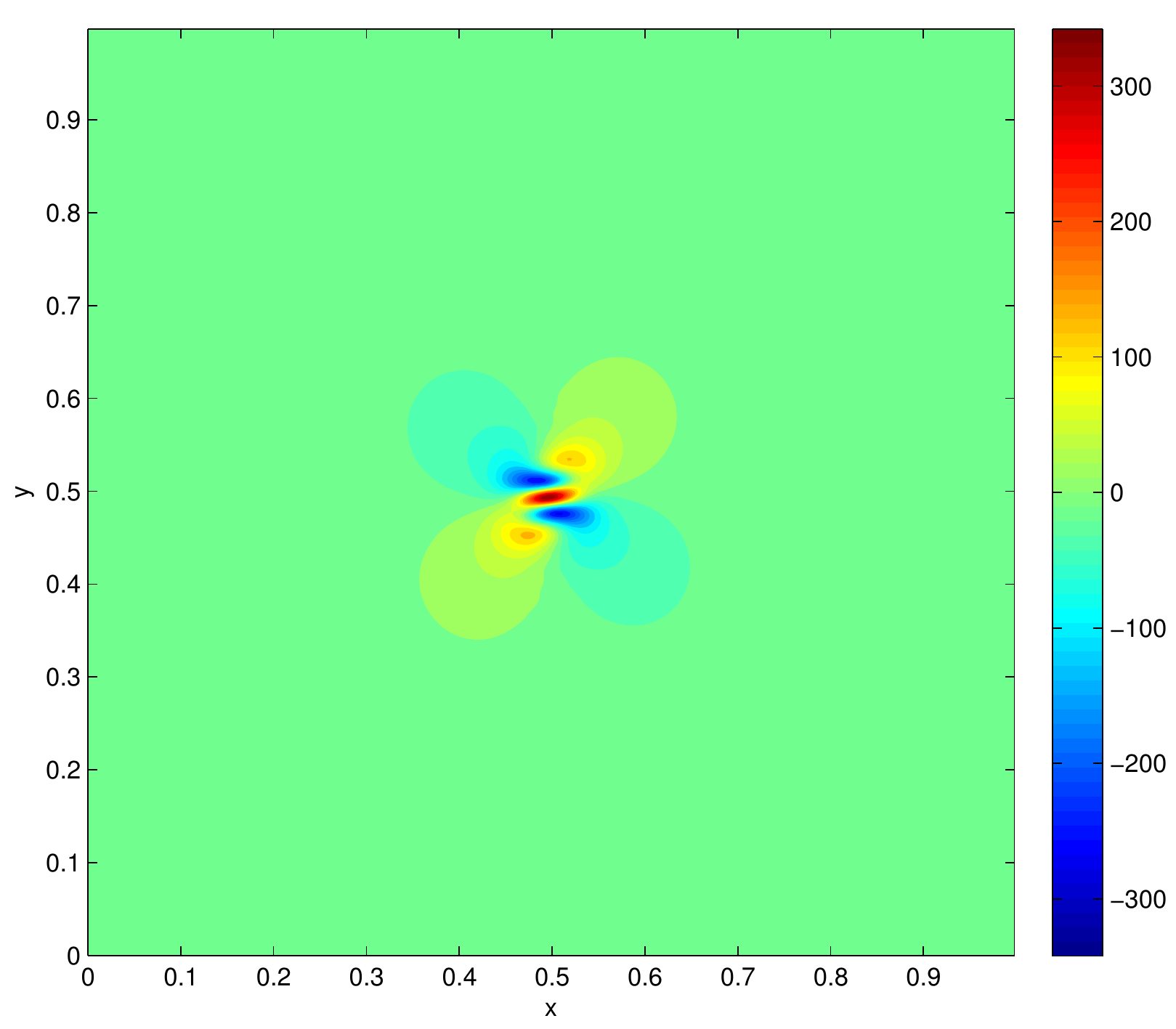}}
\caption{Dependence of the maximum palinstrophy rate of growth
  $\R_{\P_0}(\tpEP)$ on $\P_0$ for (a) $\E_0 = 100$ and (b) $\E_0 =
  10^2, 10^3$ and $10^4$. Figure (a) shows both solution branches,
  whereas figure (b) only the ones with larger values of $\R_{\P_0}$.
  Optimal vortex states corresponding to the two branches are shown in
  figures (c--e) and (f--h) for the following palinstrophy values:
  (c,f) $\P_0 \approx 10\P_c$, (d,g) $\P_0 \approx 10^2\P_c$ and (e,h)
  $\P_0 \approx 10^3\P_c$ (marked with short vertical dashes), where
  $\P_c = (2\pi)^2\E_0$ is the Poincar\'{e} limit indicated with
  vertical dash-dotted lines in figures (a) and (b). }
\label{fig:E0P0}
\end{center}
\end{figure}

\begin{figure}
\setcounter{subfigure}{0}
\begin{center}
\subfigure[]{\includegraphics[width=0.45\textwidth]{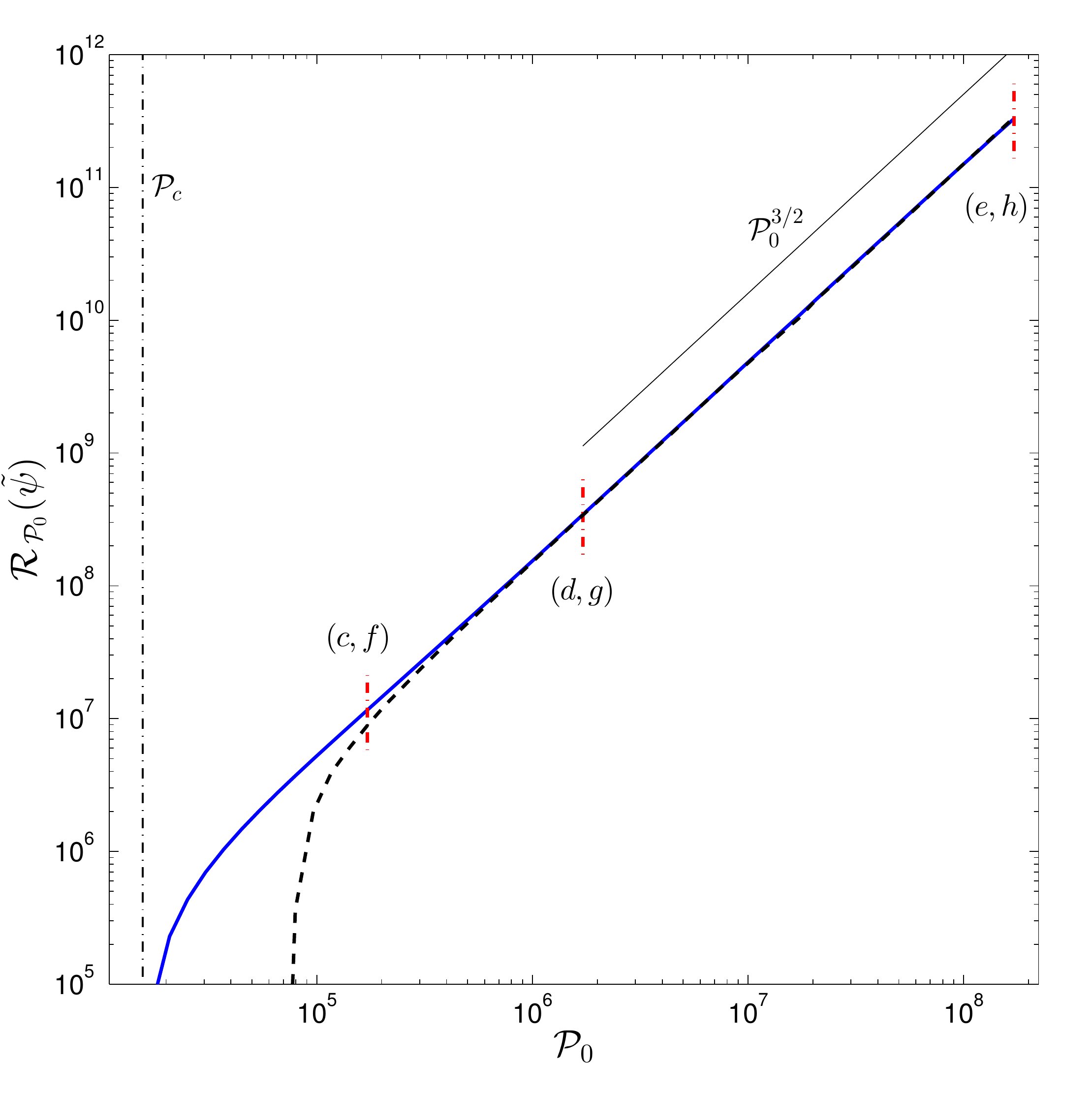}}
\subfigure[]{\includegraphics[width=0.45\textwidth]{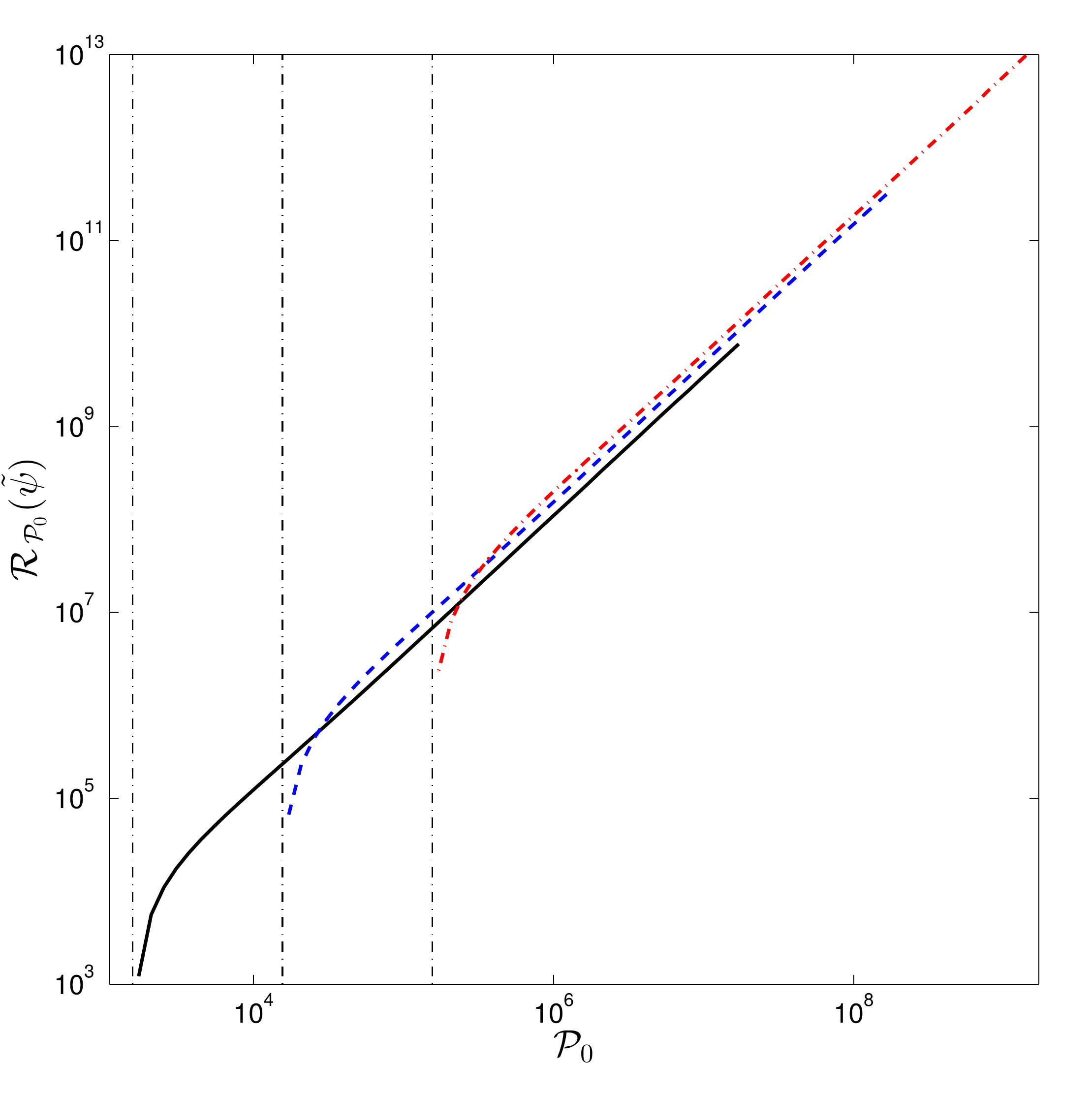}}\\
\subfigure[]{\includegraphics[width=0.3\textwidth]{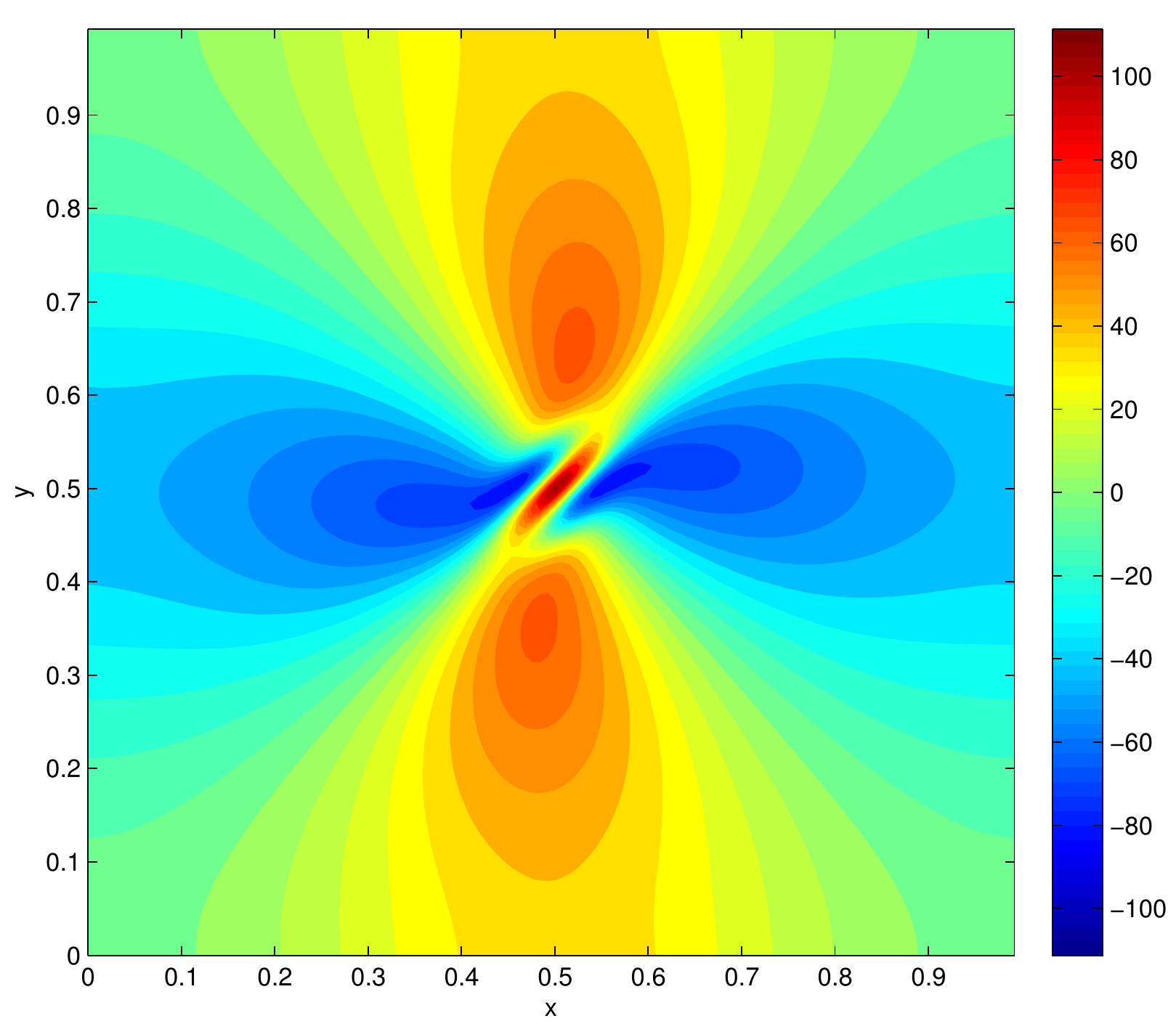}}
\subfigure[]{\includegraphics[width=0.3\textwidth]{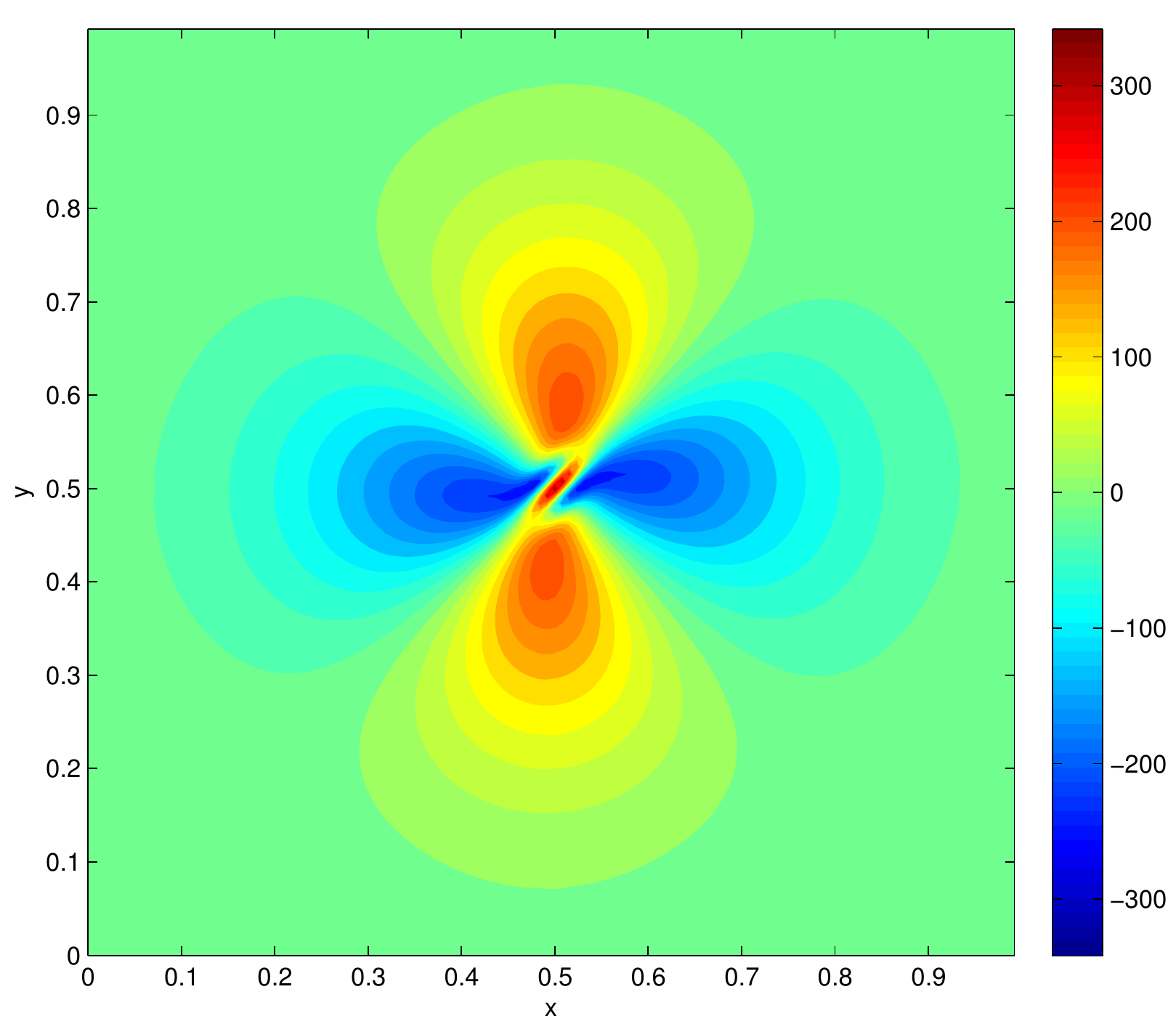}}
\subfigure[]{\includegraphics[width=0.3\textwidth]{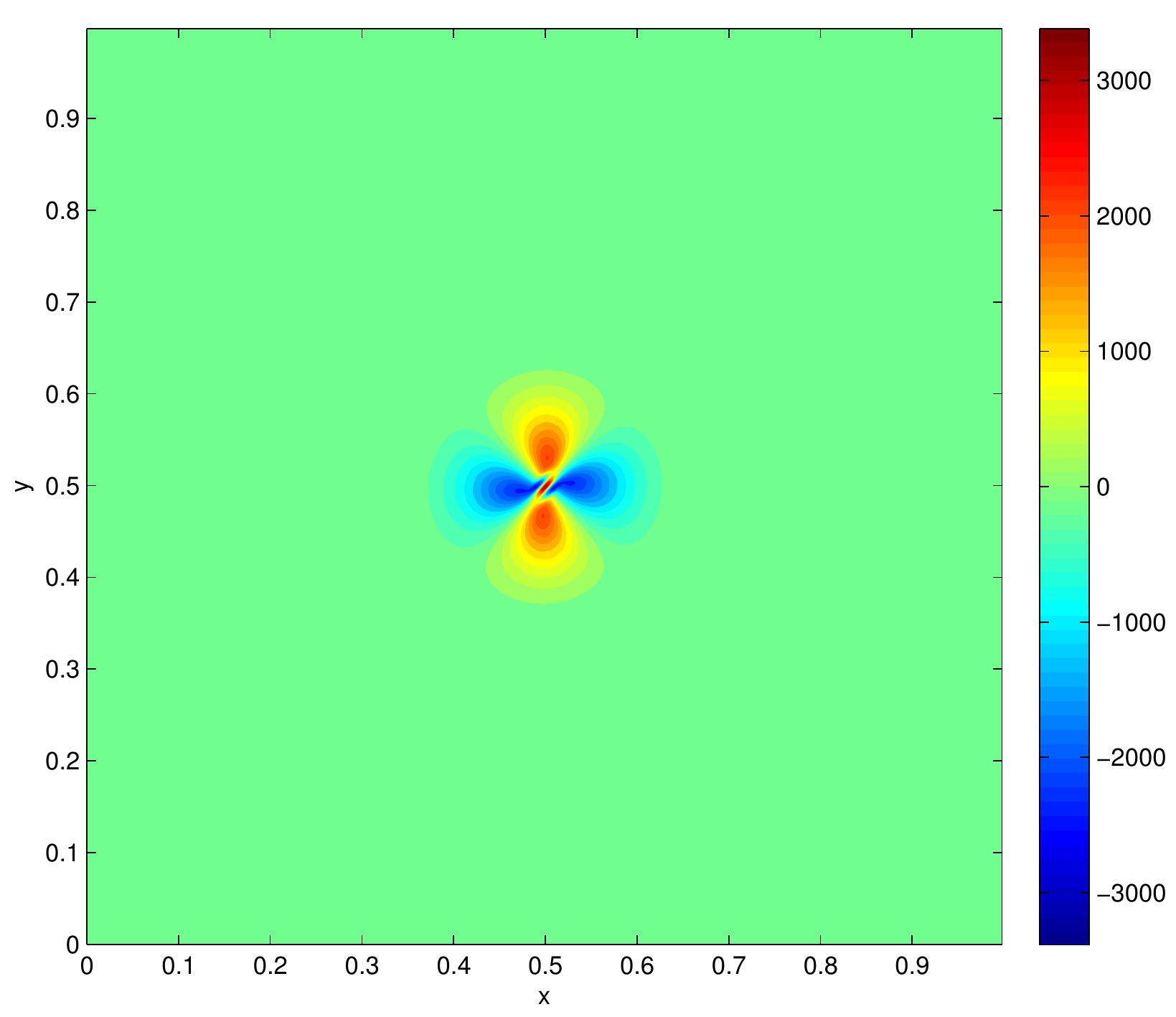}}\\
\subfigure[]{\includegraphics[width=0.3\textwidth]{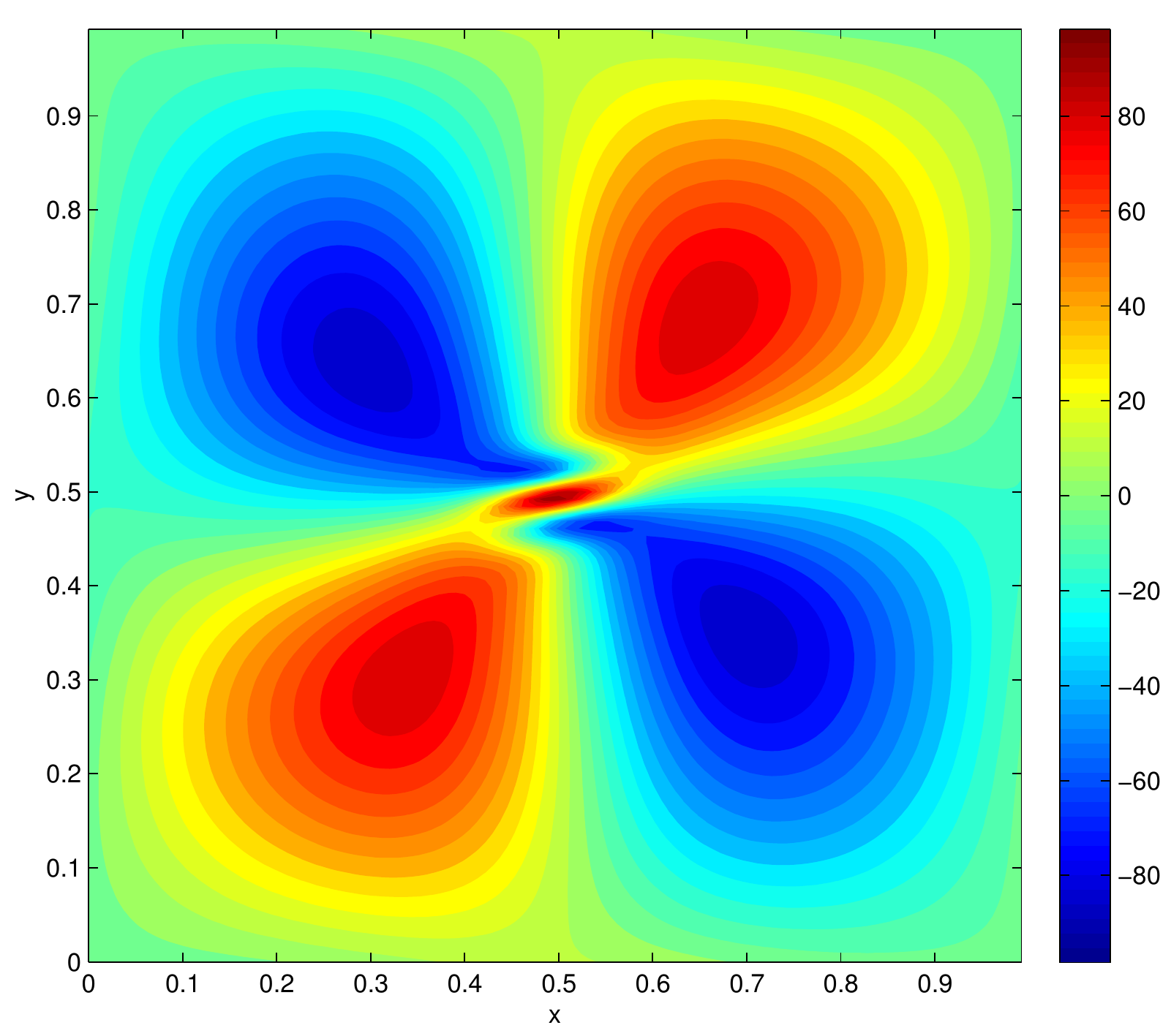}}
\subfigure[]{\includegraphics[width=0.3\textwidth]{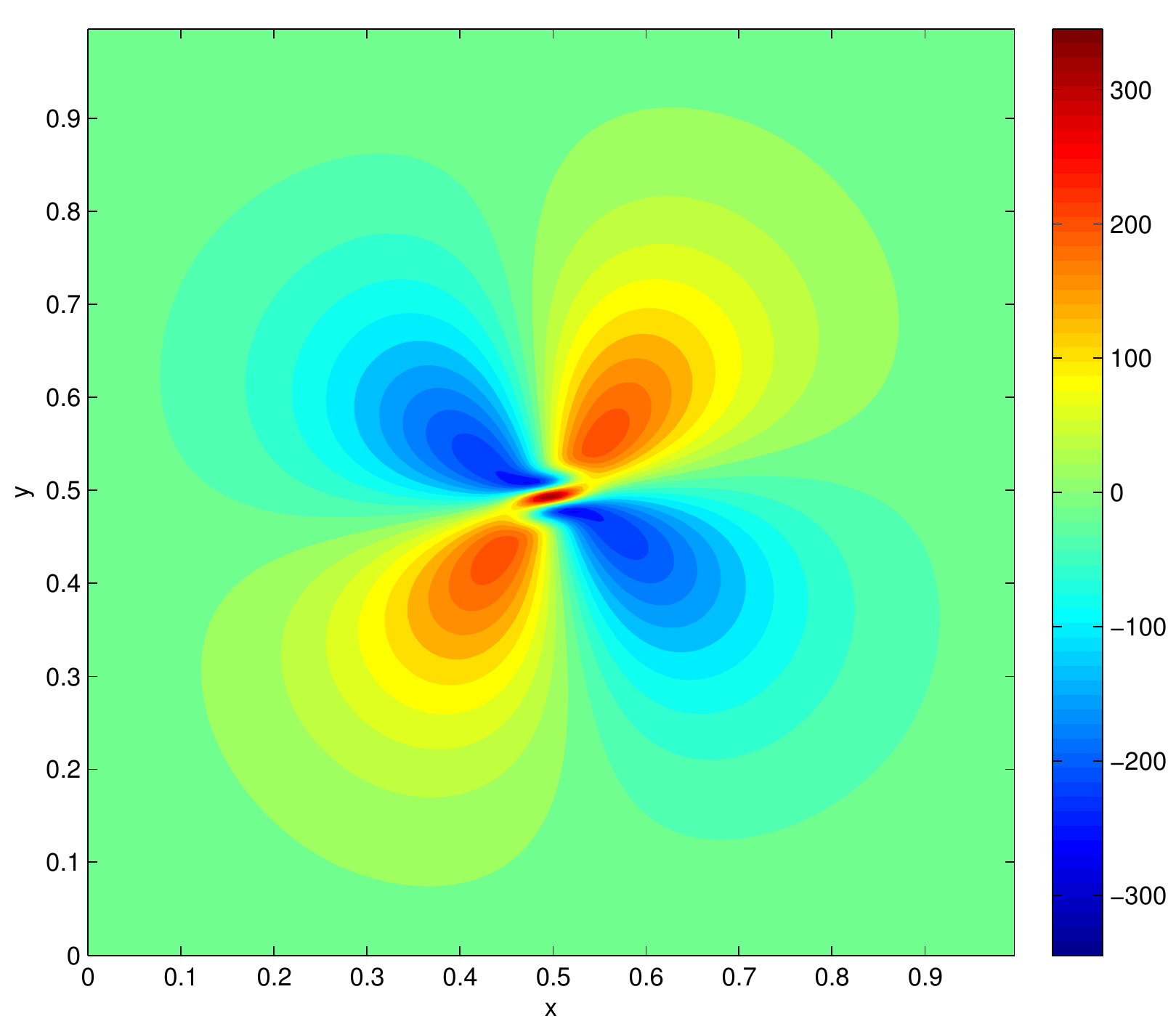}}
\subfigure[]{\includegraphics[width=0.3\textwidth]{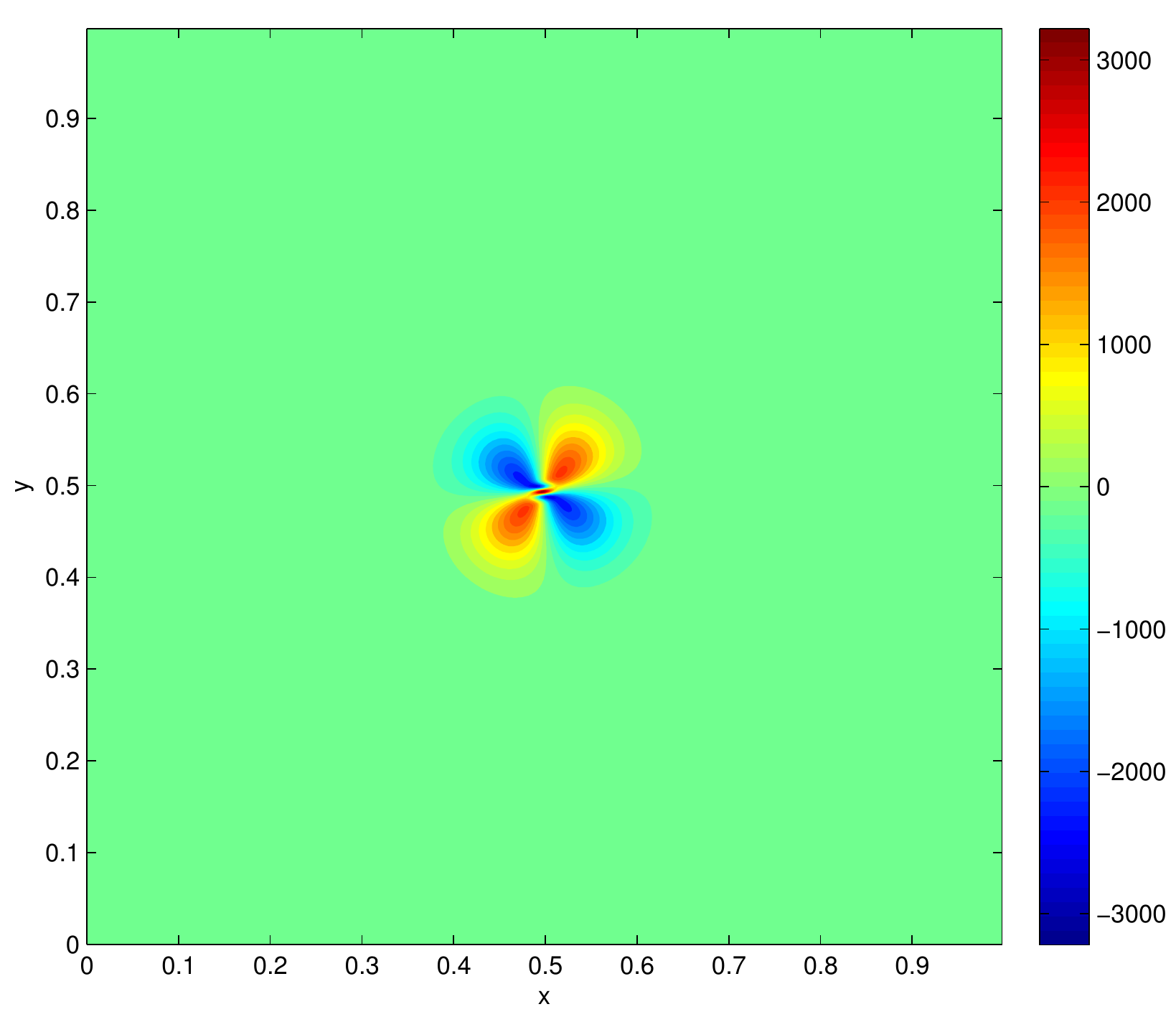}}
\caption{Dependence of the maximum palinstrophy rate of growth
  $\R_{\P_0}(\tpKP)$ on $\P_0$ for (a) $\K_0 = 10$ and (b) $\K_0 =
  10^0, 10^1$ and $10^2$. Figure (a) shows both solution branches,
  whereas figure (b) only the ones with larger values of $\R_{\P_0}$.
  Optimal vortex states corresponding to the two branches are shown in
  figures (c--e) and (f--h) for the following palinstrophy values:
  (c,f) $\P_0 = 10\P_c$, (d,g) $\P_0 = 10^2\P_c$ and (e,h)
  $\P_0 = 10^4\P_c$ (marked with short vertical dashes), where
  $\P_c = (2\pi)^4\K_0$ is the Poincar\'{e} limit indicated with
  vertical dash-dotted lines in figures (a) and (b).}
\label{fig:K0P0}
\end{center}
\end{figure}

We now discuss solutions of two-constraint optimization problems
\eqref{eq:optR_EP} and \eqref{eq:optR_KP}. In order to use these
solutions to assess the sharpness of estimates \eqref{eq:optR_EP} and
\eqref{eq:optR_KP}, in both cases we present the results by fixing one
of the constrained quantities, $\E_0$ or $\K_0$, and then studying the
maximum rate of growth of palinstrophy $\R_{\P_0}(\tpEP)$ and
$\R_{\P_0}(\tpKP)$ as a function of the palinstrophy $\P_0$ which is
allowed to vary over several orders of magnitude. The results are
shown in figure \ref{fig:E0P0}(a) for $\E_0 = 100$ and in figure
\ref{fig:K0P0}(a) for $\K_0 = 10$, each of which features two distinct
solution branches representing the local maximizers. In both cases the
values of the palinstrophy $\P_0$ for which the maximizing solutions
are found are bounded from below by Poincar\'e's inequalities. In
figures \ref{fig:E0P0}(c)--(h) and \ref{fig:K0P0}(c)--(h) we show the
vorticity fields $-\Delta\tpEP$ and $-\Delta\tpKP$ corresponding to
each of the two branches and different values of palinstrophy. We also
remark that, as predicted in section \ref{sec:smallP}, for values of
$\P_0$ approaching the Poincar\'e limit, respectively, $\P_c = (2\pi)^2\E_0$ in the
$(\E_0,\P_0)$-constrained problem and $\P_c = (2\pi)^4\K_0$ in the
$(\K_0,\P_0)$-constrained problem, the maximizers $\tpEP$ and $\tpKP$
approach the Laplacian eigenfunctions given in \eqref{eq:tfa} and
\eqref{eq:tfs}, and the pairs of branches shown in Figures
\ref{fig:E0P0}(a) and \ref{fig:K0P0}(a) correspond to the fields with
the aligned and staggered arrangements of the ``vortex cells'',
cf.~figure \ref{fig:tf0}. The localized vortex structures present in
the fields shown in figures \ref{fig:E0P0}(e) and \ref{fig:K0P0}(e)
are magnified in figures \ref{fig:vortex}(b,c).  In figures
\ref{fig:E0P0}(b) and \ref{fig:K0P0}(b) we present the maximum rate of
growth of palinstrophy for a few different values of the first
constraint, namely $\E_0 = 10^2, 10^3, 10^4$ and $\K_0 = 10^0, 10^1,
10^2$. For clarity, in these figures we only show the branches with
higher values of $\R_{\P_0}$ which in both cases correspond to the
maximizing fields with staggered vortex cells.

As is evident from figures \ref{fig:E0P0}(a,b) and
\ref{fig:K0P0}(a,b), the two two-constraint problems lead to very
different behavior of the maximum rate of growth of palinstrophy when
$\P_0 \rightarrow \infty$. In the case with the
$(\K_0,\P_0)$-constraint it exhibits a clear power-law characterized
by
\begin{equation}
\frac{d\P}{dt} \sim \P_0^{1.49 \pm 0.02}
\label{eq:dPdt_KP_num}
\end{equation}
which is consistent with estimate \eqref{eq:dPdt_KP}, thereby
confirming its sharpness. On the other hand, the case with the
$(\E_0,\P_0)$-constraint reveals a sharp decrease of $d\P/dt$ observed
for large values of $\P_0$ regardless of the value of $\E_0$. It
should be clarified, however, that this does not mean that the
branches cannot be continued, but only that for sufficiently large
values of $\P_0$ the corresponding values of $d\P/dt$ are negative and
thus are not shown in a log-log plot.  We emphasize that this behavior
is in fact consistent with estimate \eqref{eq:dPdt_EP} in which the
increase of $d\P/dt$ is at large values of $\P_0$ limited by the
negative-definite quadratic term. As shown in figure
\ref{fig:E0P0}(a), this behavior is qualitatively captured by the
dependence of $\R_{\P_0}(\tpEP)$ on $\P_0$.

\subsection{Palinstrophy Growth in Finite Time}
\label{sec:results_finite_time}

We consider solutions of Navier-Stokes system \eqref{eq:NS2D} with the
following initial data
\begin{itemize}
\item[(i)] $\,\,\omega_0 = -\triangle\tilde{\psi}_{\P_0}$,

\item[(ii)] $\,\omega_0 = -\triangle\tilde{\psi}_{\K_0,\P_0}$,

\item[(iii)] $\omega_0 = -\triangle\tilde{\psi}_{\E_0,\P_0}$. 

\end{itemize}

\begin{figure}
\setcounter{subfigure}{0}
\begin{center}
  \subfigure[]{\includegraphics[width=0.45\textwidth]{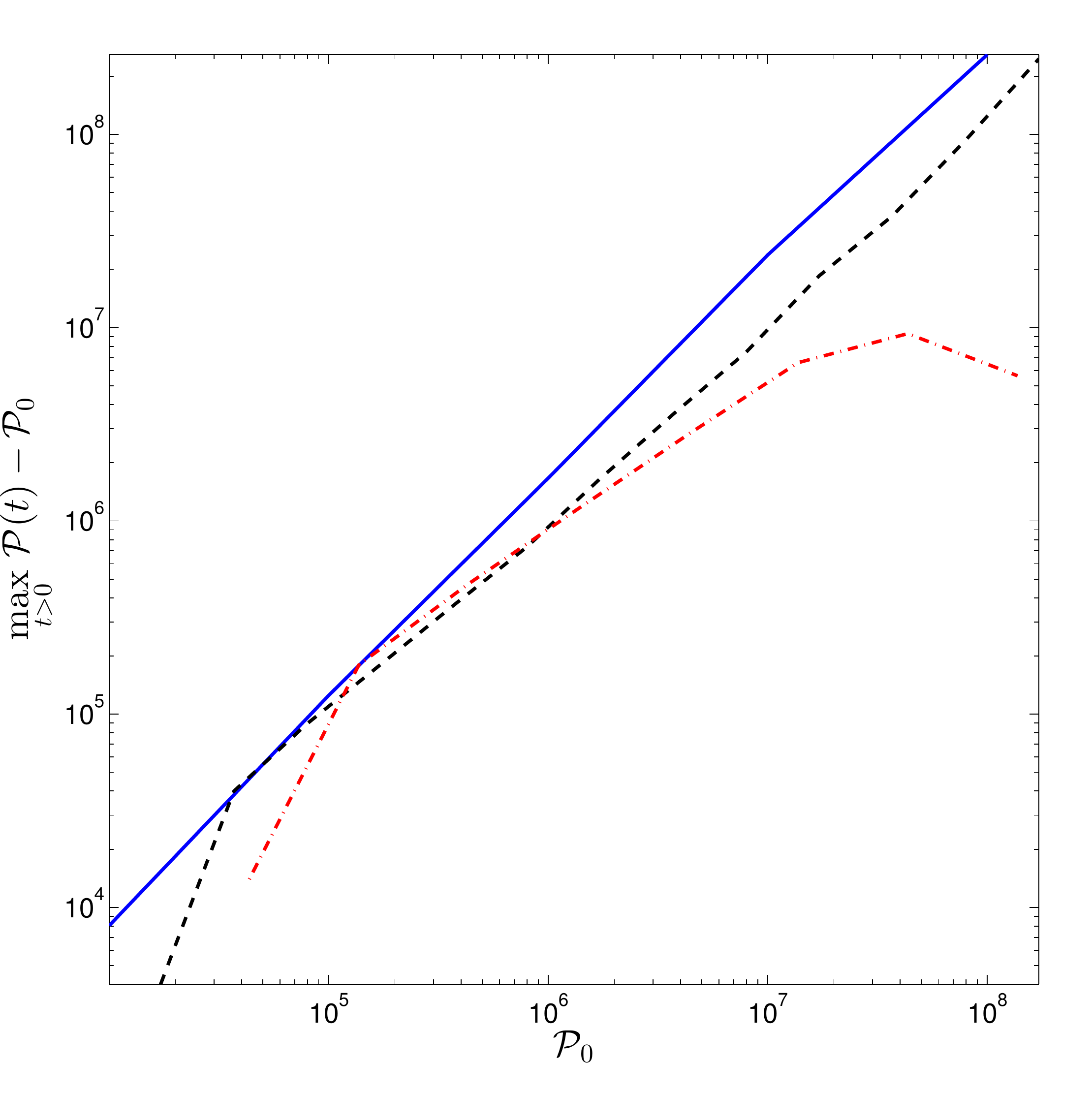}}
  \subfigure[]{\includegraphics[width=0.45\textwidth]{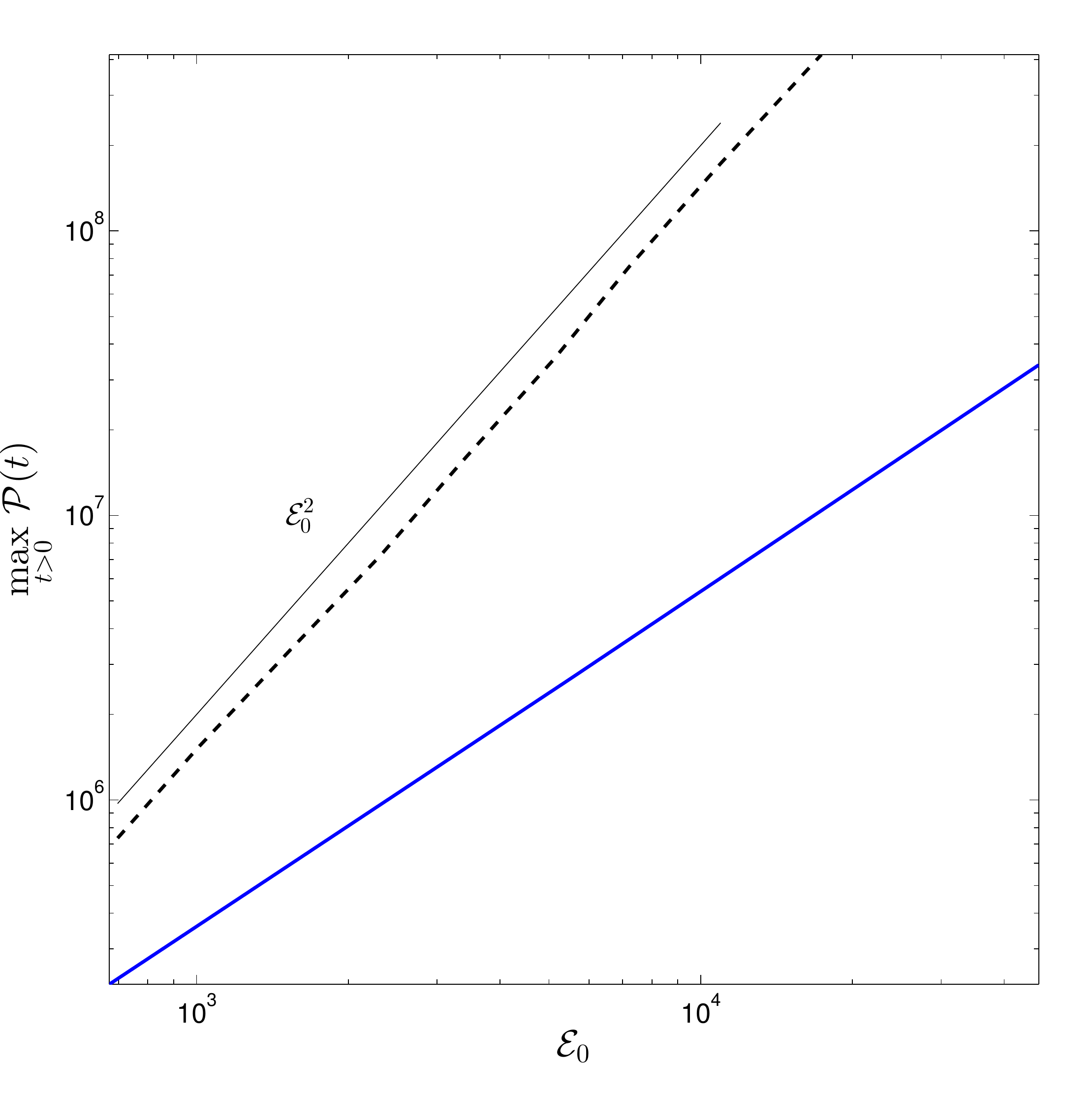}}
  \subfigure[]{\includegraphics[width=0.45\textwidth]{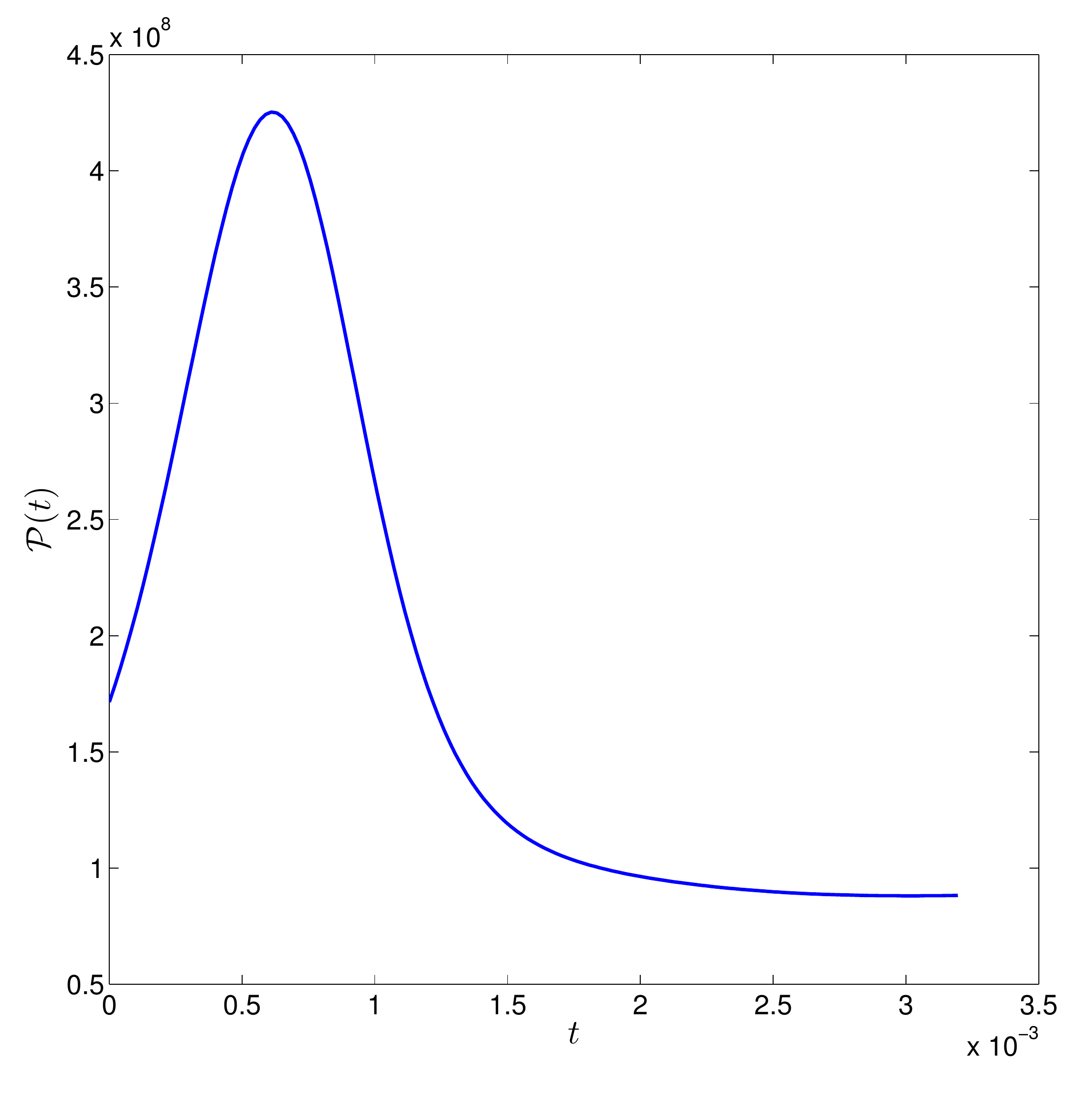}}
  \subfigure[]{\includegraphics[width=0.45\textwidth]{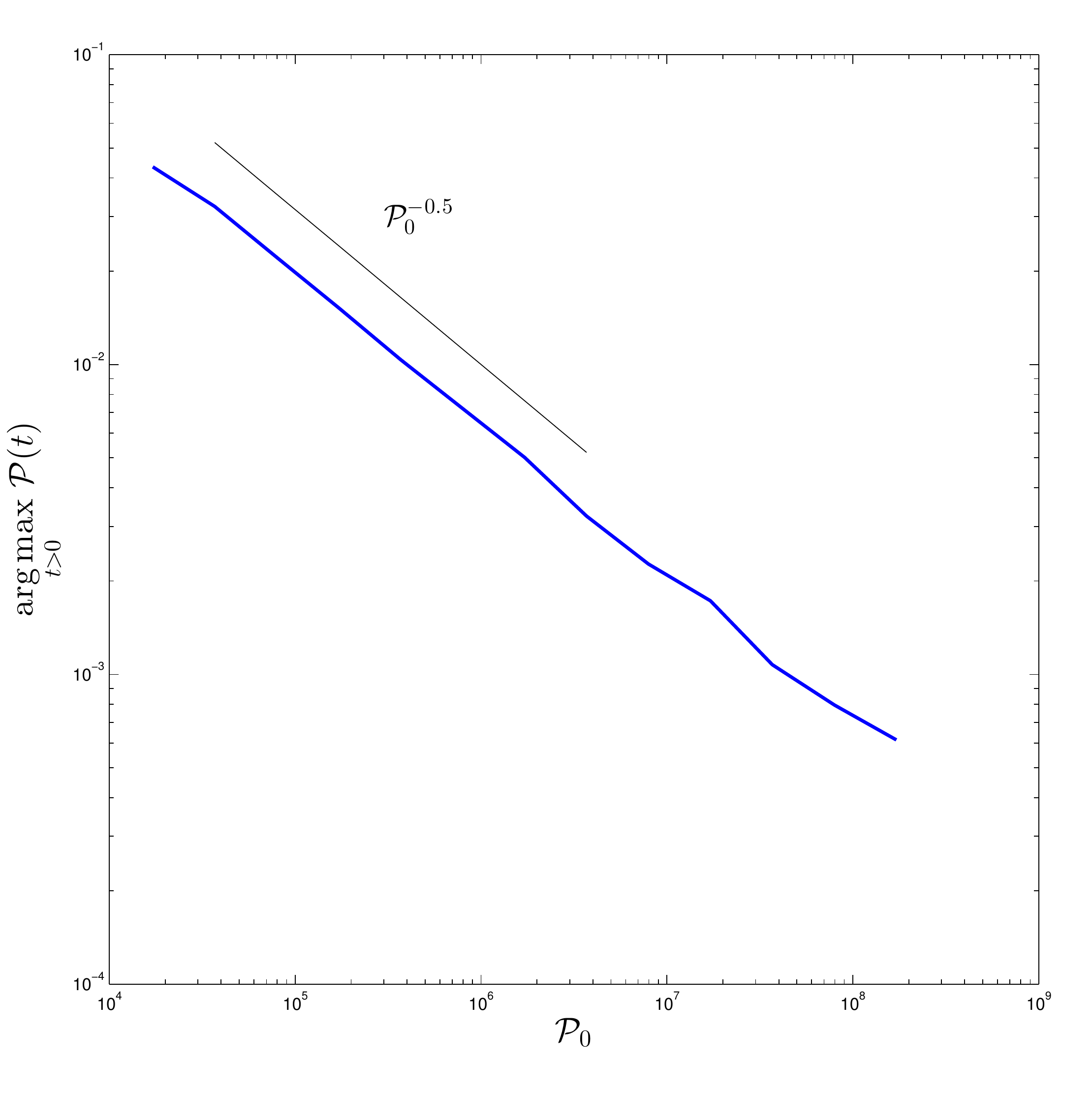}}
  \caption{(a) $\delta\P$ as a function of initial palinstrophy $\P_0$
    for (solid) case (i), (dashed) case (ii) and (dotted) case (iii);
    (b) $\P_{\max}$ as a function of initial enstrophy $\E_0$ for
    (solid) case (i) and (dashed) case (ii); (c) an individual
      time-history of $\P(t)$ in case (ii) corresponding to $\K_0 =
      10$ and $\P_0 = 10^4 \, \P_c$, where $\P_c = (2\pi)^4\K_0$ is
      the Poincar\'{e} limit; (d) time $T_{\max}$ when the maximum
      palinstrophy is attained as a function of the initial
      palinstrophy $\P_0$ in case (ii); see text in section
    \ref{sec:results_finite_time} for the definition of each case.}
\label{fig:maxPt_vsP0_num}
\end{center}
\end{figure}

\begin{figure}
\setcounter{subfigure}{0}
\begin{center}
\subfigure[]{\includegraphics[width=0.3\textwidth]{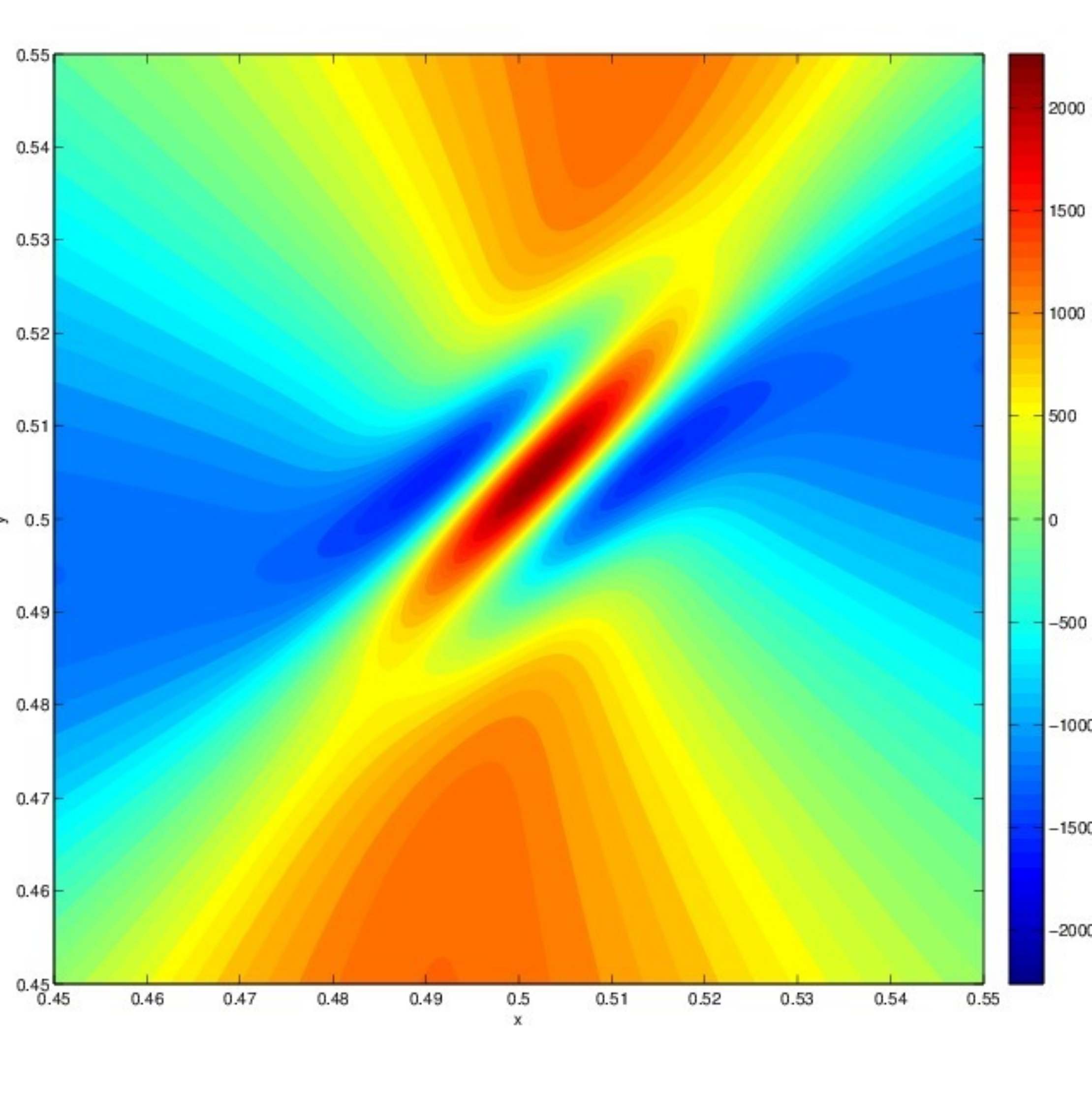}}\quad
\subfigure[]{\includegraphics[width=0.3\textwidth]{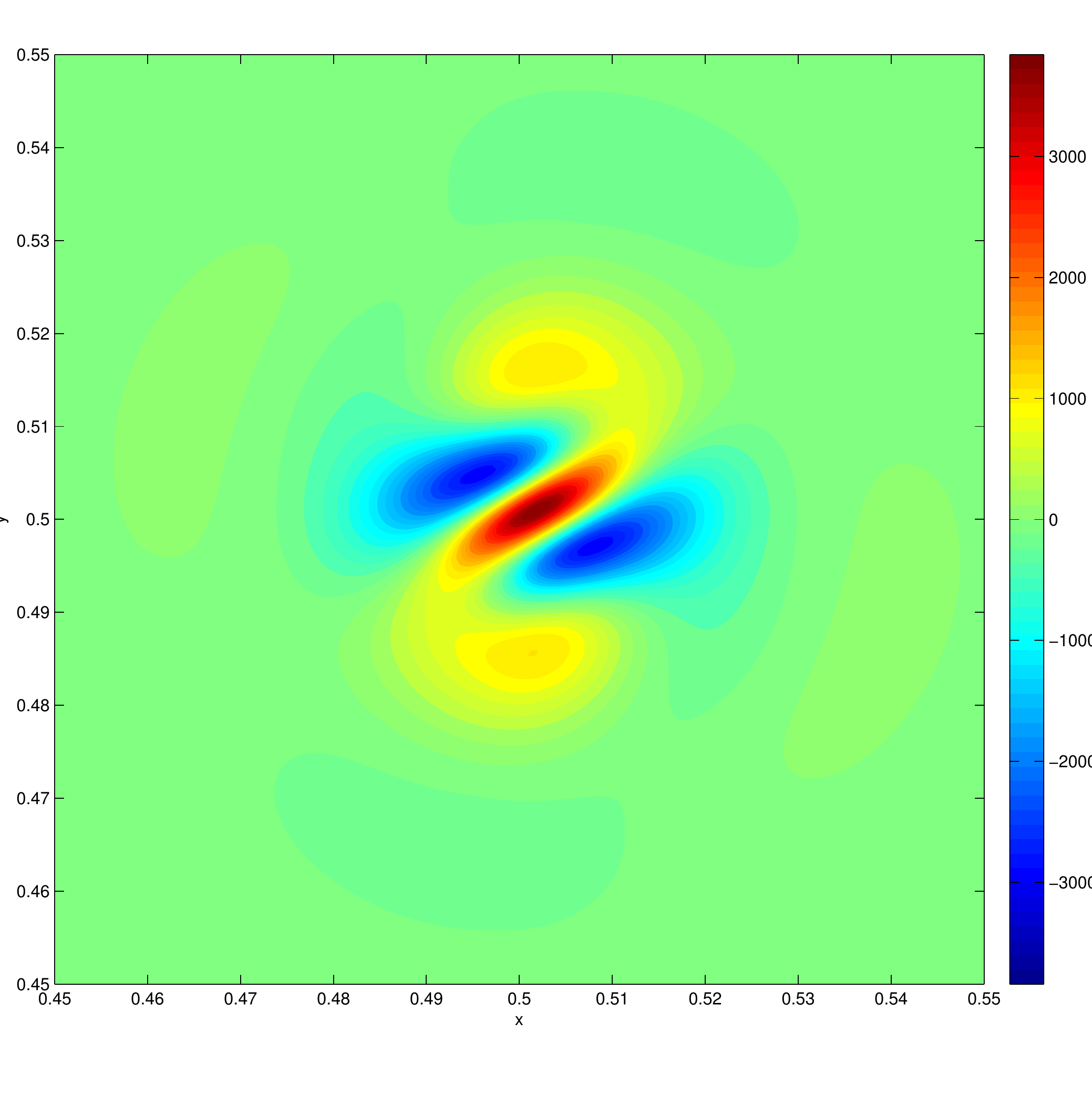}}\quad
\subfigure[]{\includegraphics[width=0.3\textwidth]{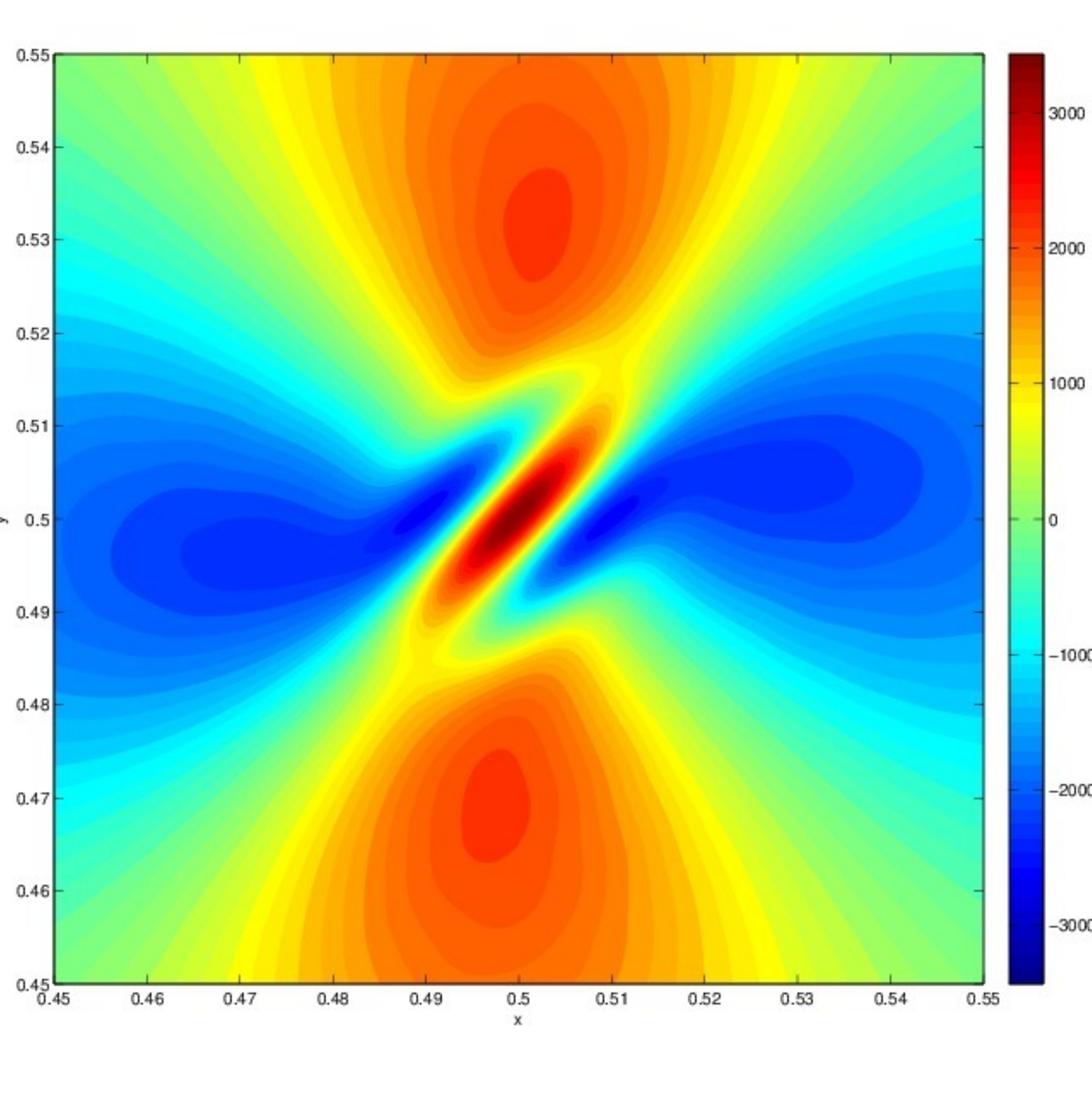}}
\caption{Magnifications of localized vortex structures present in the
  maximizing vorticity fields (a) $-\Delta\tpP$, (b) $-\Delta\tpEP$
  and (c) $-\Delta\tpKP$ with $\P_0 \approx 10^8$ in all three cases,
  $\E_0 = 10^3$ in (b) and $\K_0 = 10$ in (c).}
\label{fig:vortex}
\end{center}
\end{figure}

To obtain insights about the sharpness of finite-time estimates
\eqref{eq:maxPt_Doering} and \eqref{eq:maxPt_Ayala}, we are interested
in the maximum palinstrophy attained over time
$\P_{\max}:=\mathop{\max}_{t>0}\,\P(t)$ and its increment with respect
to the initial value $\delta\P := \P_{\max} - \P_0$ as functions of
the initial enstrophy $\E_0$ and palinstrophy $\P_0$ (the reason for
studying $\delta\P$ is that the term $\P_0$ may mask the behavior of
the other term on the RHS in \eqref{eq:maxPt_Doering} if it should
also scale with an exponent close to the unity). Figure
\ref{fig:maxPt_vsP0_num}(a) shows $\delta\P$ for each case (i)--(iii)
as a function of $\P_0$ with $\K_0 = 10$ in case (ii) and $\E_0 =
10^3$ in case (iii). These results exhibit a power-law behavior in
cases (i) and (ii), whereas in case (iii) a sharp decrease of
$\delta\P$ is observed as $\P_0\to\infty$.  The power laws for cases
(i) and (ii) are
\begin{equation}
\begin{aligned}
(i)\,\, & \delta\P \sim \P_0^{1.13 \pm 0.03}, \\  
(ii)\,\, & \delta\P \sim \P_0^{1.05 \pm 0.09}.   
\end{aligned}
\label{eq:Pmax_vs_P0}
\end{equation}
This behavior is parallel to the behavior reported in sections
\ref{sec:results1} and \ref{sec:results2} where power-law scaling
was observed in the single-constraint problem and in the
$(\K_0,\P_0)$-constrained problem, but not in the
$(\E_0,\P_0)$-constrained problem. The exponents characterizing 
power laws \eqref{eq:Pmax_vs_P0} are significantly smaller
than 2 predicted by estimate \eqref{eq:maxPt_Doering}. The
dependence of $\P_{\max}$ on $\E_0$ is cases (i) and (ii) is shown
in figure \ref{fig:maxPt_vsP0_num}(b) ($\E_0$ rather than $\P_0$ is
chosen as the abscissa, since this is the ``independent variable''
in the nonlinear term in estimate \eqref{eq:maxPt_Ayala}, and case
(iii) is not shown, because in this configuration $\E_0$ is fixed).
The following two distinct power laws are observed in the two cases
\begin{equation}
\begin{aligned}
(i)\,\, & \P_{\max} \sim \E_0^{1.17 \pm 0.02}, \\  
(ii)\,\, & \P_{\max} \sim \E_0^{1.98 \pm 0.07},   
\end{aligned}
\label{eq:Pmax_vs_E0}
\end{equation}
implying that the maximizing vortex states obtained under the
$(\K_0,\P_0)$-constraint lead to a finite-time palinstrophy evolution
which {\em also} saturates the finite-time estimate
\eqref{eq:maxPt_Ayala}.  The significance of this finding will be
discussed in more detail in the following section.  In figure
  \ref{fig:maxPt_vsP0_num}(c) we show the time evolution of the
  palinstrophy $\P(t)$ corresponding to case (ii) with $\K_0 = 10$ and
  $\P_0 = 10^4 \, \P_c$, where $\P_c = (2\pi)^4\K_0$ is the
  Poincar\'{e} limit, which is representative of the entire family. We
  note that the initially steep growth of the palinstrophy is followed
  by its eventual viscous decay. The time $T_{\max} := \argmax_{t\ge
    0} \P(t)$ when the maximum is attained depends on the initial
  palinstrophy $\P_0$ exhibiting a well-defined power-law, cf.~figure
  \ref{fig:maxPt_vsP0_num}(d),
\begin{equation}
T_{\max} \sim \P_0^{-0.47 \pm 0.06}.
\label{eq:Tmax_vs_P0}
\end{equation}
We remark that scaling with the same exponent was also observed in the
evolution leading to the maximum finite-time growth of enstrophy in
the 1D Burgers problem \citep{ap11a,p12}. For a detailed discussion of
the vortex dynamics mechanisms responsible for the evolution leading
to power-law \eqref{eq:Pmax_vs_E0}(ii), we refer the reader to the
companion paper by \citet{ap13b}.  The exponents characterizing all
the power laws discussed in this section are collected in table
\ref{tab:exp}.

\section{Discussion}
\label{sec:discuss}

In this section we comment on some of the theoretical results
introduced in section \ref{sec:NS} in the light of the findings
presented in section \ref{sec:results}. Our main interest here is
establishing the sharpness of estimates
\eqref{eq:dPdt_EP}--\eqref{eq:dPdt_P} for $d\P/dt$, and from
figures \ref{fig:K0P0}(a,b) and table \ref{tab:exp} we see that
solutions of optimization problem \eqref{eq:optR_KP} lead to the
growth of $\R_{\P_0}$ which saturates estimate \eqref{eq:dPdt_KP} for
sufficiently large $\P_0$, so that we conclude that this
estimate is sharp. As regards upper bound \eqref{eq:dPdt_EP}, due to
the presence of the negative quadratic term, this estimate does not
have the form of a power-law allowing for an arbitrary growth of
$\R_{\P_0}(\tpEP)$. This is indeed confirmed by the behavior of the
solutions of optimization problem \eqref{eq:optR_EP} shown in Figures
\ref{fig:E0P0}(a,b) where we see that on each branch the quantity
$\R_{\P_0}(\tpEP)$ eventually decreases with $\P_0$. Hence, in this
case a power-law cannot be in fact defined. Since the qualitative
features of estimate \eqref{eq:dPdt_EP} are reproduced by the actual
dependence of $\R_{\P_0}(\tpEP)$ on $\P_0$ observed in figures
\ref{fig:E0P0}(a,b), we can conclude that this estimate predicts the
correct behavior, although in the absence of a power-law, it is hard
to quantify this statement in terms of exponents. It is interesting
that optimization problems \eqref{eq:optR_EP} and \eqref{eq:optR_KP}
which have a rather similar structure lead to quite different global
behavior of the maximizing solutions. The reason for this is that, as
the palinstrophy $\P_0$ is increased, in the $(\E_0,\P_0)$-constrained
family of optimizers the energy $\K(\tpEP)$ can not increase
arbitrarily, as it is upper-bounded by $\E_0$ via Poincar\'{e}'s
inequality. On the other hand, in the $(\K_0,\P_0)$-constrained family
of optimizers the constraint on $\K_0$ does not limit the growth of
the enstrophy $\E(\tpKP)$ of the maximizing solutions. 

In regard to the finite-time estimates, the fact that upper bound
\eqref{eq:maxPt_Ayala} is saturated by the evolution corresponding to
the $(\K_0,\P_0)$-constrained maximizers is intriguing. We recall that
the maximizers found to saturate the instantaneous estimates in 1D
subject to {\em one} constraint only (on $\E$) did not lead to
evolution saturating the corresponding finite-time estimates
\citep{ld08,ap11a}. The role of the number of the constraints imposed
on the solutions in this type of optimization problems deserves
further study.
 
Moving on to estimate \eqref{eq:dPdt_P} and solutions of the
single-constraint optimization problem \eqref{eq:optR_P}, we observe
in figure \ref{fig:RvsP0_1constr}(b) that while $\R_{\P_0}(\tpP)$
exhibits a very clean power-law dependence on $\P_0$, the associated
exponent is in fact significantly less than 2 predicted by estimate
\eqref{eq:dPdt_P}, cf. table \ref{tab:exp}. This was in fact to be
expected, since upper bound \eqref{eq:dPdt_P} was obtained with the
use of Poincar\'e's inequality which is saturated only by the
eigenfunctions of the Laplacian operator and, as is evident from
figure \ref{fig:RvsP0_1constr}(e,h), the maximizing solutions $\tpP$
for large $\P_0$ are quite different from such eigenfunctions.  We add
that analogous instantaneous estimates in 1D and in 3D were in fact
found to be sharp by \citet{ld08}, cf. table \ref{tab:estimates}.

We now comment on the structure of the maximizing vorticity fields.
First, we observe that in all three optimization problems
there are two branches of locally maximizing solutions, cf.~ figures
\ref{fig:RvsP0_1constr}, \ref{fig:E0P0} and \ref{fig:K0P0}, and they
are obtained via continuation from the limiting, for small $\P_0$,
solutions which were characterized analytically in section
\ref{sec:smallP}. In that section we also observed that in the limit
of small $\P_0$ the cubic term in $\R_{\P_0}(\psi)$ vanishes,
cf.~\eqref{eq:R0}, so that the maximizing fields sustain no stretching
of the vorticity gradients.  Interestingly, since these limiting
maximizers satisfy equation \eqref{eq:EL0a} (which is a special case
of $\Delta\psi = F(\psi)$ with a particular $F \; : \; \RR \rightarrow
\RR$), they are also steady solutions of the 2D Euler equations
\citep{mb02}.  When used as the initial data in a 2D time-dependent
Navier-Stokes problem \eqref{eq:NS2D}, they give rise to the
Taylor-Green vortex flow characterized by a purely exponential decay
without any nonlinear interactions. We add that 3D generalizations of
this flow lead to nontrivial time evolution and have been investigated
in the context of the finite-time blow-up problem
\citep{bmonmu03,b91}. The relationship between the symmetry of
vortex configurations and the amplification of the vorticity via
stretching was studied by \citet{p01}.

\begin{figure}
\begin{center}
\includegraphics[width=0.45\textwidth]{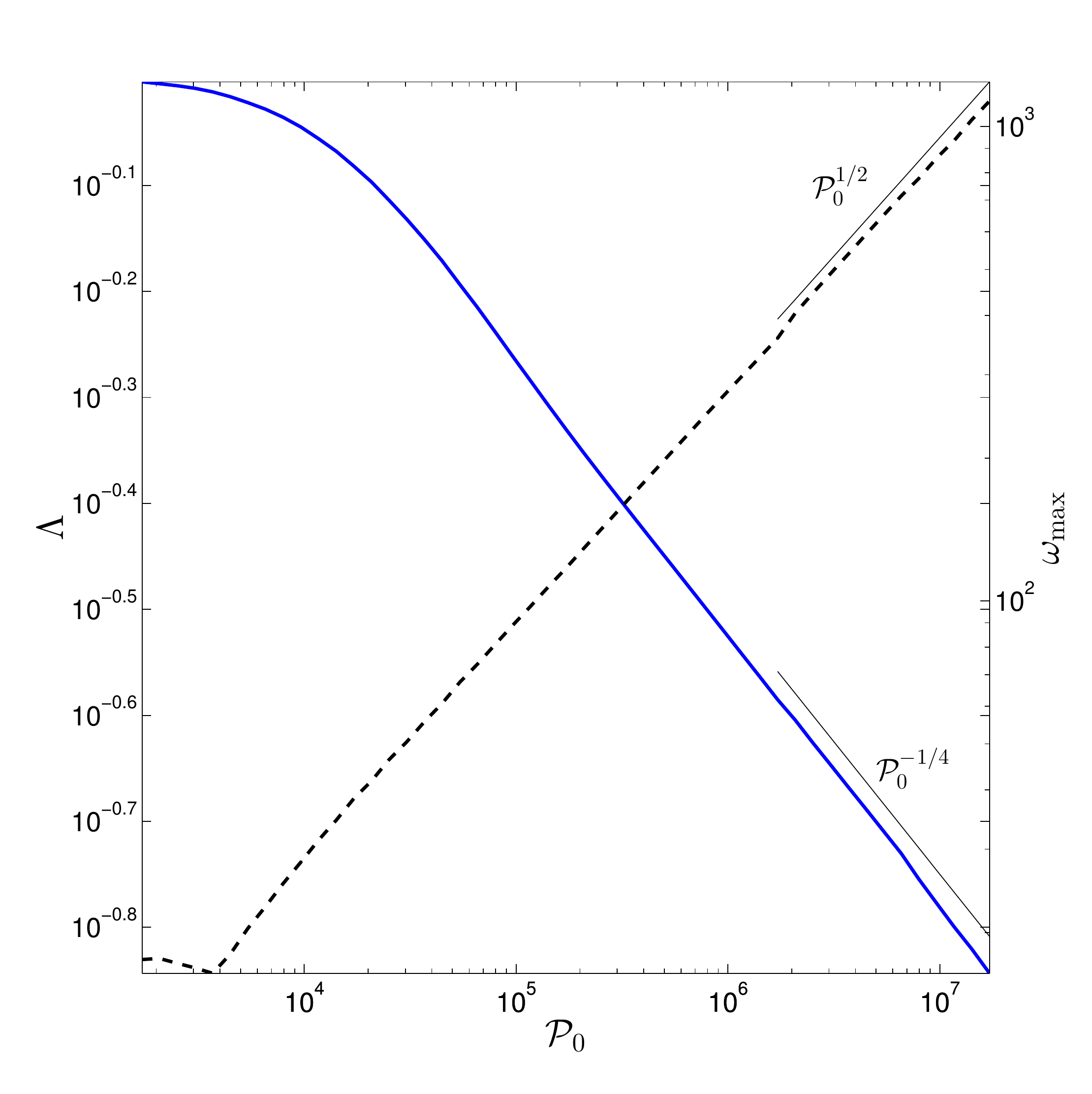}
\caption{Dependence of (solid) the characteristic length scale
    $\Lambda = 2\pi\sqrt{ \K(\tpKP) / \E(\tpKP) }$ and (dashed) the
    vorticity magnitude $\omega_{\max} = \| \Delta\tpKP
    \|_{L_{\infty}(\Omega)}$ on $\P_0$ with $\K=1$ for the family of
    maximizers obtained subject to the $(\K_0,\P_0)$-constraint
    (cf. figure \ref{fig:K0P0}).}
\label{fig:lengthscale_vsP0}
\end{center}
\end{figure}

As the palinstrophy $\P_0$ increases, the maximizers in all three
optimization problems become localized multipolar vortex structures
shown in figures \ref{fig:vortex}(a--c) and featuring a central
elongated filament stretched by four satellite vortices: two stronger
ones which are closer to the central filament and have the opposite
sign, and two weaker ones which are further away and have the same
sign as the central filament. In the absence of this central filament,
the four satellite vortices would resemble the axial vorticity
distribution in the meridional plane intersecting two parallel vortex
rings, which was in fact the optimal vortex state found by \citet[see
Figure 4.6a]{ld08} to saturate the 3D instantaneous estimate
\eqref{eq:dEdt3D}. This observation offers some analogy to the 3D
problem with the presence of the central vortex filament reflecting
the difference in the physical quantities maximized in the two
problems: vorticity (enstrophy) in 3D versus vorticity gradients
(palinstrophy) in 2D. In the limit of large $\P_0$, the vortex states
corresponding to the two branches appear very similar, except for the
rotation by a $45\deg$ angle.  The difference between the cases with
one and two constraints is that in the former case the satellite
vortices tend to be less localized (which is a consequence of the fact
that in that case $\K_0$ and $\E_0$ can change freely).  With
increasing palinstrophy $\P_0$, the optimal vortex structures in the
single-constraint and $(\K_0,\P_0)$-constrained cases shrink in a
shape-preserving manner with the characteristic dimension $\Lambda$ of
the vortex structure vanishing while its magnitude $\omega_{\max}$
grows, so that the vorticity fields can be empirically approximated by
the asymptotic formula $\omega_{\P_0}(\x) \sim
\omega_{\max}\,\Pi(\x/\Lambda)$ for some distribution $\Pi$
independent of $\P_0$. Figure \ref{fig:lengthscale_vsP0} shows the
dependence of the quantities $\Lambda = 2\pi\sqrt{ \K(\tpKP) /
  \E(\tpKP)}$, cf. \citep{dg95}, and $\omega_{\max} = \| \Delta\tpKP
\|_{L_{\infty}(\Omega)}$ computed for the upper branch of
$(\K_0,\P_0)$-constrained maximizers on $\P_0$ with $\K_0 = 1$, cf.
figure \ref{fig:K0P0}. This data reveals clear power laws $\Lambda
\sim \P_0^{-1/4}$ and $\omega_{\max} \sim \P_0^{1/2}$ holding for
sufficiently large $\P_0$ which confirms the scale-independent
structure of the maximizing vortex states.  Obtaining an analytical
characterization of the maximizers $\tpP$ and $\tpKP$ in the limit
$\P_0 \rightarrow \infty$ is an interesting open research problem in
mathematical analysis.

\begin{figure}
\begin{center}
\includegraphics[width=0.7\textwidth]{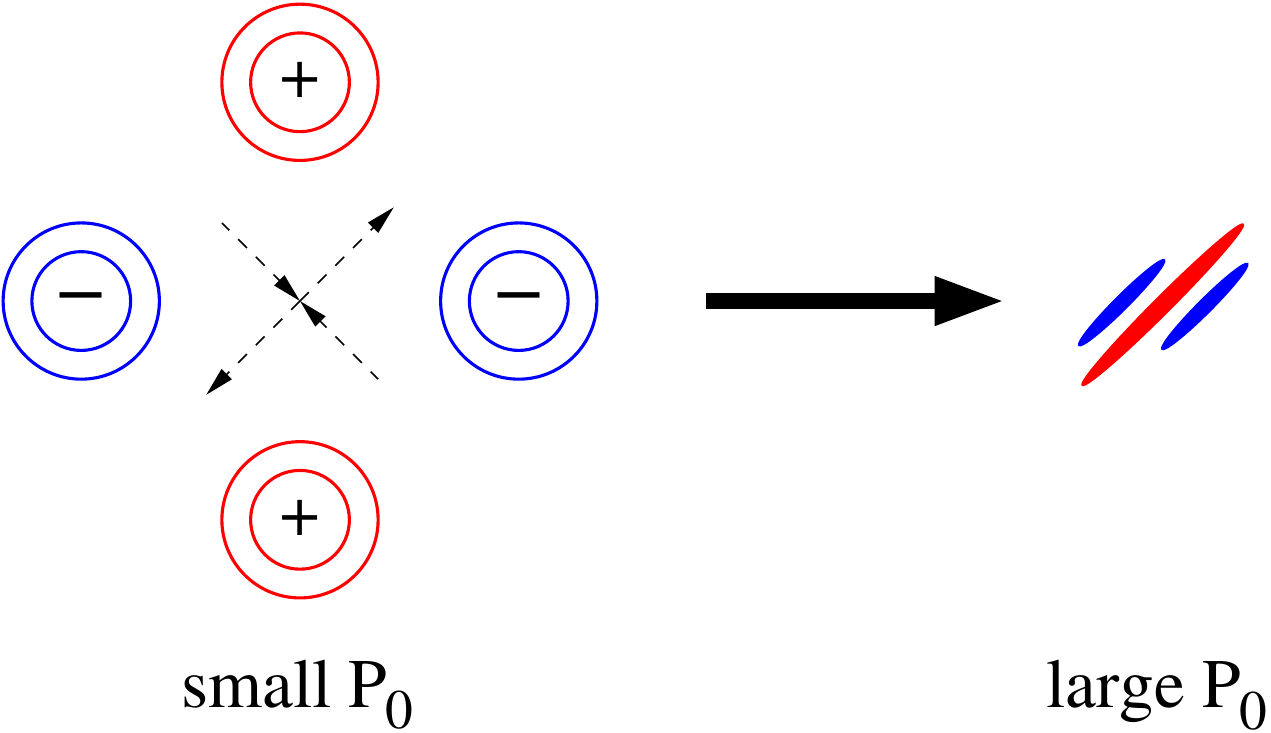}
\caption{Schematic illustration of the change of the structure of the
  local maximizers in the staggered arrangement of the vortex cells as
  $\P_0$ increases (dashed lines represent the principal directions of
  stretching and compression).}
\label{fig:pattern}
\end{center}
\end{figure}

Finally, we observe that both families of the maximizing solutions
shown in figures \ref{fig:RvsP0_1constr}--\ref{fig:K0P0} exhibit an
interesting pattern. While for decreasing $\P_0$ the maximizing
solutions approach the Laplacian eigenfunctions with either aligned or
staggered arrangement of the vortex cells, cf.
\eqref{eq:tfa}--\eqref{eq:tfs} and figure \ref{fig:tf0}, for
increasing $\P_0$ the dominating vortex structure is shifted to the
stagnation point of the maximizing fields corresponding to small
$\P_0$. Furthermore, the dominating vortex structure in the limit of
large $\P_0$ is aligned with the direction of the maximum stretching
characterizing the maximizer in the low $\P_0$ limit, i.e.,
vertically/horizontally for the aligned arrangement and inclined at
the angle of $45\deg$ for the staggered arrangement. This pattern is
schematically illustrated in figure \ref{fig:pattern}.

\begin{table}
\renewcommand{\arraystretch}{1.5}
\begin{center}
\hspace*{-1.1cm}
\begin{tabular}{c|c|c}       
  Estimate   &  Constraints & Computed Power-Law  \\ 
  \hline
  $\frac{d\P}{dt}  \le \nu\frac{\P^2}{\E} + \frac{C}{\nu} \E\, \P$  \ [cf.~\eqref{eq:dPdt_EP}] & $\E_0,\P_0$  & Not applicable \\
  \hline
  $\frac{d\P}{dt}  \le \frac{C}{\nu} \,\K^{\frac{1}{2}}\, \P^{\frac{3}{2}}$ \ [cf.~\eqref{eq:dPdt_KP}] & $\K_0,\P_0$  & $\frac{d\P}{dt} \sim \P_0^{1.49\pm0.02}$ \\
  \hline
  $\frac{d\P}{dt}  \le  \frac{C}{\nu} \, \P^2$ \ [cf.~\eqref{eq:dPdt_P}] & $\P_0$  & $\frac{d\P}{dt} \sim \P_0^{1.57 \pm 0.05}$ \\
  \hline
  $\max_{t > 0} \P(t) \le \P_0 + \frac{C}{\nu^2}\E_0^2$ \ [cf. \eqref{eq:maxPt_Doering}] & $\P_0$  & $\max_{t > 0} \P(t) \sim \E_0^{ 1.17 \pm 0.02}$ \\
  \hline
  $\max_{t>0} \P(t) \le \left(\P_0^{1/2} + \frac{C_2}{4\nu^2}\K_0^{1/2}\E_0\right)^2$ \ [cf. \eqref{eq:maxPt_Ayala}] & $\K_0,\P_0$  & $\max_{t > 0} \P(t) \sim \E_0^{ 1.98 \pm 0.07}$ \\
  \hline
  N/A & $\K_0,\P_0$  & $\argmax_{t\ge 0} \P(t) \sim \P_0^{-0.47 \pm 0.06}$
\end{tabular}
\end{center}
\caption{Comparison of the analytical estimates from section \ref{sec:NS} and 
  the power-laws discovered by solving the maximization problem in sections 
  \ref{sec:results1} and \ref{sec:results2}, and the initial-value 
    problems in section \ref{sec:results_finite_time}.}
\label{tab:exp}
\end{table}

\section{Conclusions and Outlook}
\label{sec:final}

In this investigation we addressed a problem which is a part of a
broader research program concerning characterization of the maximum
growth of certain quadratic quantities in the hydrodynamic systems in
different spatial dimensions (cf.~table \ref{tab:estimates}). Here we
focused on the upper bounds for the instantaneous rate of growth of
palinstrophy $d\P/dt$ and demonstrated that certain available
estimates are in fact sharp with respect to variations of the
palinstrophy and are saturated by families of vorticity fields with
nontrivial structure. Estimate \eqref{eq:dPdt_KP} was found to be
realizable, even though there exist other estimates,
cf.~\eqref{eq:dPdt_TD}, with the RHS involving a lower power of $\P$,
which is a consequence of the choice of the quantities constraining
the upper bound expression ($\K_0^{1/2}$ versus
$\|\omega_0\|_{L_{\infty}(\Omega)}$).  The optimal vortex states with
prescribed energy $\K_0$ and palinstrophy $\P_0$ were also found to
lead to a time-evolution saturating the finite-time estimate, which is
an interesting result highlighting the role the number of constraints
may play in this type of problems. Sharpness of finite-time estimates
can also be assessed by solving optimization problems defined over
{\em finite} windows of time, as done by \citet{ap11a} for the 1D
Burgers equation. It is interesting to see whether such an approach
could lead to an improved prefactor in power law
\eqref{eq:Pmax_vs_E0}(ii). We intend to address this question in the
future. As regards the present study, another interesting question is
how our findings would change if the optimization problems were
formulated in an unbounded, rather than periodic, domain. Sharpness of
estimates \eqref{eq:dPdt_TD}--\eqref{eq:Pmax_TD}, which have a rather
different structure than bounds \eqref{eq:dPdt_KP} and
\eqref{eq:maxPt_Ayala}, with respect to variations of the palinstrophy
$\P_0$ is also an interesting open question, however, due to the
presence of the nondifferentiable factor $\|\omega\|_{L_{\infty}}$ on
the RHS, the corresponding variational optimization problems will be
nonsmooth and require highly specialized methods for their numerical
solution. These estimates appear sharper than \eqref{eq:dPdt_KP} and
\eqref{eq:maxPt_Ayala} with respect to variations of the viscosity
$\nu$, and it would also be interesting to examine their realizability
in such terms. A separate set of questions concerns flows on domains
with rigid boundaries and some results relevant to the blow-up problem
in such setting were recently reported by \citet{gt13}.

Concerning the research program presented in Introduction, quantifying
the maximum finite-time growth of enstrophy in the 3D Navier-Stokes
system remains of course the ultimate goal, one which we hope is
within reach in the foreseeable future given the currently available
computational tools and resources. An interesting intermediate step is
to consider similar questions for the 2D surface quasi-geostrophic
(SQG) equation
\begin{equation}
\Dpartial{\theta}{t} + \left( \v \cdot \bnabla \right) \theta = - \nu \, ( - \Delta
)^{\alpha} \theta,
\label{eq:SQG}
\end{equation}
where $\v = \bnabla^{\perp} \, ( - \Delta )^{-1/2} \, \theta$ and $0
\le \alpha \le 1$. As is suspected \citep{k10,s11}, solutions of
\eqref{eq:SQG} may exhibit finite-time blow-up in the supercritical
case $\alpha < 1/2$. Therefore, aside from its own intrinsic interest,
this problem represents a useful testbed for development and
validation of methods to track singular solutions in the 2D setting
which is more computationally manageable than the full 3D
Navier-Stokes problem.

We wish to emphasize that the research methodology developed here,
relying on a systematic characterization of the extremal behavior,
appears applicable to other related open problems in the field of
theoretical fluid dynamics. An example of such a problem is obtaining
sharp bounds on the Nusselt number in the Raleigh-B\'enard convection
\citep{hcd13}.  There are also similar problems related to mixing.

\section*{Acknowledgements}

The authors are indebted to Charles Doering and Evelyn Lunasin for
many enlightening discussions concerning the research problems studied
in this work and, in particular, for providing estimates
\eqref{eq:dPdt_EP} and \eqref{eq:maxPt_Doering}. They are also
thankful to an anonymous referee for many insightful comments
concerning, in particular, estimate \eqref{eq:dPdt_log} and its
relationship to other bounds considered in this study. The authors are
grateful to Nicholas Kevlahan for making his parallel Navier-Stokes
solver available, which was used to obtain the results reported in
section \ref{sec:results_finite_time}.  This research was funded
through an Early Researcher Award (ERA) and an NSERC Discovery Grant.
The computational time was made available by SHARCNET.

\appendix 
\section{Derivation of Estimates \eqref{eq:dPdt_KP} and \eqref{eq:maxPt_Ayala}}
\label{sec:dPdt}

A key element necessary to derive the upper bound in \eqref{eq:dPdt_KP} is
the following estimate for the $L_{\infty}$ norm of doubly-periodic
functions $\u \; : \; \Omega\to\RR^2$
\begin{equation}\label{eq:Linf_2D}
||\uvec||_{\infty} \leq  C || \uvec ||^{1/2}_2 || \Delta \uvec ||^{1/2}_2.
\end{equation}
where $C>0$, which follows from Sobolev Interpolation Theorem
\citep{af05}. A calculation showing that for $\Omega=[0,1]^2$, $C =
1/\sqrt{\pi}$ can be found in \citet{a14}. 

We notice that the estimate depends only on the $L_2$ norms of the
function and some of its second derivatives. To obtain estimate
\eqref{eq:dPdt_KP} from section \ref{sec:NS}, we write the rate of
growth of palinstrophy as, cf. \eqref{eq:R},
\begin{equation}
\frac{d\P}{dt} = -\nu\int_\Omega |\Delta\omega|^2 \,d\Omega + \int_\Omega \mathbf{u}\cdot\nabla\omega\Delta\omega \,d\Omega.
\label{eq:R2}
\end{equation}
The second term on the RHS in \eqref{eq:R2} can be upper-bounded as
\begin{align*}
\left| \int_\Omega \mathbf{u}\cdot\nabla\omega\Delta\omega \,d\Omega\right | & \leq  ||\mathbf{u}\cdot\nabla\omega||_2||\Delta\omega||_2\\
 & \leq  ||\mathbf{u}||_{\infty}||\nabla\omega||_2||\Delta\omega||_2 \\
 & \leq  C ||\mathbf{u}||^{1/2}_2||\Delta\mathbf{u}||^{3/2}_2||\Delta\omega||_2,
\end{align*}
where inequality \eqref{eq:Linf_2D} has been used together with the
Cauchy-Schwarz inequality and the fact that $||\nabla\omega||_2 =
||\Delta\mathbf{u}||_2$. The application of Young's inequality
\begin{displaymath}
ab \leq \frac{\beta^pa^p}{p} + \frac{b^q}{q\beta^q}
\end{displaymath}
to \eqref{eq:R2} with $p = q = 2$ and $\beta^2 = (2\nu)^{-1}$ yields
\begin{equation}
\begin{aligned}
\frac{d\P}{dt} & \leq  -\nu||\Delta\omega||^2_2 + \frac{C^2}{4\nu}||\mathbf{u}||_2||\Delta\mathbf{u}||^3_2 + \nu||\Delta\omega||^2_2  \\
 & \leq  \frac{C^2}{4\nu}||\mathbf{u}||_2||\Delta\mathbf{u}||^3_2.
\end{aligned}
\label{eq:bound_dPdt}
\end{equation}
Finally, inequality \eqref{eq:bound_dPdt} can be rewritten in
terms of energy and palinstrophy as, cf.~\eqref{eq:dPdt_KP},
\begin{equation}\label{eq:dPdt_estimate}
\frac{d\P}{dt} \leq \frac{C^2}{\nu}\K^{1/2}\P^{3/2}.
\end{equation}
It follows from Navier-Stokes system \eqref{eq:NS2D} that $d\K/dt =
-2\nu\E$ and $d\E/dt = -2\nu\P$.  Therefore, $\K(t) \le \K(0) = \K_0$
and $\E(t) \le \E(0) = \E_0$ for all $t>0$.  Estimate
\eqref{eq:dPdt_estimate} can be transformed as
\begin{alignat*}{2}
\frac{d\P}{dt} & \leq \frac{C^2}{\nu}\K_0^{1/2}\P^{3/2} & \quad &\Longrightarrow \\
\P^{-1/2}\frac{d\P}{dt} & \, \leq \, \frac{C^2}{\nu}\K_0^{1/2}\P & \quad\quad&\Longrightarrow \\
\frac{d}{dt}(2\P^{1/2}) & \, \leq \, \frac{C^2}{\nu}\K_0^{1/2}\P. &&
\end{alignat*}
Integrating the last inequality over time and using the fact that
\begin{displaymath}
\int_0^t \P(s)\,ds = \frac{\E(0) - \E(t)}{2\nu} \leq \frac{\E_0}{2\nu},
\end{displaymath}
it is possible to obtain
\begin{equation}
\P(t) \leq \left[ \P^{1/2}_0 + \left(\frac{C}{2\nu}\right)^2\K_0^{1/2}\E_0 \right]^2.  
\label{eq:Pt}
\end{equation}
This upper bound is valid for all $t>0$ and the right-hand side depends
only on the initial values of energy, enstrophy and palinstrophy.
Estimate \eqref{eq:maxPt_Ayala} is then obtained by taking the
maximum over time on the LHS in \eqref{eq:Pt}.

\bibliography{ap_13a_jfm}

\begin{thebibliography}{46}
\expandafter\ifx\csname natexlab\endcsname\relax\def\natexlab#1{#1}\fi

\bibitem[Adams \& Fournier(2005)]{af05}
{\sc Adams, R.~A. \& Fournier, J.~F.} 2005 {\em {Sobolev} Spaces\/}. Elsevier.

\bibitem[Ayala(2014)]{a14}
{\sc Ayala, D.} 2014 Extreme vortex states and singularity formation. PhD
  thesis, McMaster University, in preparation.

\bibitem[Ayala \& Protas(2011)]{ap11a}
{\sc Ayala, D. \& Protas, B.} 2011 On maximum enstrophy growth in a
  hydrodynamic system. {\em Physica D\/} {\bf 240}, 1553--1563.

\bibitem[Ayala \& Protas(2013)]{ap13b}
{\sc Ayala, D. \& Protas, B.} 2013 Vortices, maximum growth and the problem of
  finite--time singularity formation. {\em Fluid Dynamics Research\/} (accepted
  for publication).

\bibitem[Brachet(1991)]{b91}
{\sc Brachet, M.~E.} 1991 Direct simulation of three-dimensional turbulence in
  the {Taylor--Green} vortex. {\em Fluid Dynamics Research\/} {\bf 8}, 1--8.

\bibitem[Brachet {\em et~al.\/}(1983)Brachet, Meiron, Orszag, Nickel, Morf \&
  Frisch]{bmonmu03}
{\sc Brachet, M.~E., Meiron, D.~I., Orszag, S.~A., Nickel, B.~G., Morf, R.~H.
  \& Frisch, U.} 1983 Small-scale structure of the {Taylor--Green} vortex. {\em
  Journal of Fluid Mechanics\/} {\bf 130}, 411--452.

\bibitem[Bustamante \& Brachet(2012)]{bb12}
{\sc Bustamante, M.~D. \& Brachet, M.} 2012 Interplay between the
  {Beale-Kato-Majda} theorem and the analyticity-strip method to investigate
  numerically the incompressible {Euler} singularity problem. {\em Phys. Rev.
  E\/} {\bf 86}, 066302.

\bibitem[Dascaliuc {\em et~al.\/}(2010)Dascaliuc, Foias \& Jolly]{dfj10}
{\sc Dascaliuc, R., Foias, C. \& Jolly, M.~S.} 2010 Estimates on enstrophy,
  palinstrophy, and invariant measures for {2D} turbulence. {\em Journal of
  Differential Equations\/} {\bf 248}, 792--819.

\bibitem[Doering(2009)]{d09}
{\sc Doering, C.~R.} 2009 The {3D Navier-Stokes} problem. {\em Annual Review of
  Fluid Mechanics\/} pp. 109--128.

\bibitem[Doering \& Gibbon(1995)]{dg95}
{\sc Doering, C.~R. \& Gibbon, J.~D.} 1995 {\em Applied Analysis of the
  {Navier-Stokes} Equations\/}. Cambridge University Press.

\bibitem[Doering \& Lunasin(2011)]{dl11}
{\sc Doering, C.~R. \& Lunasin, E.} 2011 Limits on palinstrophy growth for
  solutions of the two--dimensional {Navier--Stokes} equations. Personal
  communication.

\bibitem[Edwards {\em et~al.\/}(1994)Edwards, Tuckerman, Friesner \&
  Sorensen]{edwards:KrylovMethod}
{\sc Edwards, W.~S., Tuckerman, L.~S., Friesner, R.~A. \& Sorensen, D.~C.} 1994
  Krylov method for the incompressible {Navier--Stokes} equation. {\em Journal
  of Computational Physics\/} {\bf 110}, 82--102.

\bibitem[Farazmand {\em et~al.\/}(2011)Farazmand, Kevlahan \& Protas]{fkp11}
{\sc Farazmand, M., Kevlahan, N. K.~R. \& Protas, B.} 2011 Controlling the dual
  cascade of two-dimensional turbulence. {\em Journal of Fluid Mechanics\/}
  {\bf 668}, 202--222.

\bibitem[Fefferman(2000)]{f00}
{\sc Fefferman, C.~L.} 2000 Existence and smoothness of the {Navier-Stokes}
  equation. available at
  \url{http://www.claymath.org/millennium/Navier-Stokes_Equations/navierstokes.pdf},
  {Clay Millennium Prize Problem Description}.

\bibitem[Foias \& Temam(1989)]{ft89}
{\sc Foias, C. \& Temam, R.} 1989 Gevrey class regularity for the solutions of
  the {Navier--Stokes} equations. {\em Journal of Functional Analysis\/} {\bf
  87}, 359--369.

\bibitem[Gibbon {\em et~al.\/}(2008)Gibbon, Bustamante \& Kerr]{gbk08}
{\sc Gibbon, J.~D., Bustamante, M. \& Kerr, R.~M.} 2008 The three--dimensional
  {Euler} equations: singular or non--singular? {\em Nonlinearity\/} {\bf 21},
  123--129.

\bibitem[Gibbon \& Titi(2013)]{gt13}
{\sc Gibbon, J.~D. \& Titi, E.~S.} 2013 The {3D} incompressible {Euler}
  equations with a passive scalar: a road to blow-up? ArXiv:1211.3811.

\bibitem[Grafke {\em et~al.\/}(2008)Grafke, Homann, Dreher \& Grauer]{ghdg08}
{\sc Grafke, T., Homann, H., Dreher, J. \& Grauer, R.} 2008 Numerical
  simulations of possible finite-time singularities in the incompressible
  {Euler} equations: comparison of numerical methods. {\em Physica D\/} {\bf
  237}, 1932--1936.

\bibitem[Gunzburger(2003)]{g03}
{\sc Gunzburger, M.~D.} 2003 {\em Perspectives in Flow Control and
  Optimization\/}. SIAM.

\bibitem[Hassanzadeh {\em et~al.\/}(2013)Hassanzadeh, Chini \& Doering]{hcd13}
{\sc Hassanzadeh, P., Chini, G.~P. \& Doering, C.~R.} 2013 Wall to wall optimal
  transport. ArXiv:1309.5542.

\bibitem[Hou(2009)]{h09}
{\sc Hou, T.~Y.} 2009 Blow--up or no blow--up? a unified computational and
  analytic approach to {3D} incompressible {Euler} and {Navier--Stokes}
  equations. {\em Acta Numerica\/} pp. 277--346.

\bibitem[Kerr(1993)]{k93}
{\sc Kerr, R.~M.} 1993 Evidence for a singularity of the three--dimensional,
  incompressible {Euler} equations. {\em Phys. Fluids A\/} {\bf 5}, 1725--1746.

\bibitem[Kiselev(2010)]{k10}
{\sc Kiselev, A.} 2010 Regularity and blow up for active scalars. {\em Math.
  Model. Nat. Phenom.\/} {\bf 5}, 225--255.

\bibitem[Kreiss \& Lorenz(2004)]{kl04}
{\sc Kreiss, H. \& Lorenz, J.} 2004 {\em Initial-Boundary Value Problems and
  the {Navier-Stokes} Equations\/}, {\em Classics in Applied Mathematics\/},
  vol.~47. SIAM.

\bibitem[Ladyzhenskaya(1969)]{l69b}
{\sc Ladyzhenskaya, O.~A.} 1969 {\em The Mathematical Theory of Viscous
  Incompressible Flow\/}. Gordon and Breach.

\bibitem[Lu(2006)]{l06}
{\sc Lu, L.} 2006 Bounds on the enstrophy growth rate for solutions of the {3D
  Navier-Stokes} equations. PhD thesis, University of Michigan.

\bibitem[Lu \& Doering(2008)]{ld08}
{\sc Lu, L. \& Doering, C.~R.} 2008 Limits on enstrophy growth for solutions of
  the three-dimensional {Navier--Stokes} equations. {\em Indiana University
  Mathematics Journal\/} {\bf 57}, 2693--2727.

\bibitem[Luenberger(1969)]{l69}
{\sc Luenberger, D.} 1969 {\em Optimization by Vector Space Methods\/}. John
  Wiley and Sons.

\bibitem[Majda \& Bertozzi(2002)]{mb02}
{\sc Majda, A.~J. \& Bertozzi, A.~L.} 2002 {\em Vorticity and Incompressible
  Flow\/}. Cambridge University Press.

\bibitem[Matsumoto {\em et~al.\/}(2008)Matsumoto, Bec \& Frisch]{mbf08}
{\sc Matsumoto, T., Bec, J. \& Frisch, U.} 2008 Complex--space singularities of
  {2D Euler} flow in lagrangian coordinates. {\em Physica D\/} {\bf 237},
  1951--1955.

\bibitem[Ohkitani(2008)]{o08}
{\sc Ohkitani, K.} 2008 A miscellany of basic issues on incompressible fluid
  equations. {\em Nonlinearity\/} {\bf 21}, 255--271.

\bibitem[Ohkitani \& Constantin(2008)]{oc08}
{\sc Ohkitani, K. \& Constantin, P.} 2008 Numerical study of the
  eulerian--lagrangian analysis of the {Navier-Stokes} turbulence. {\em Phys.
  Fluids\/} {\bf 20}, 1--11.

\bibitem[Orlandi {\em et~al.\/}(2012)Orlandi, Pirozzoli \& Carnevale]{opc12}
{\sc Orlandi, P., Pirozzoli, S. \& Carnevale, G.~F.} 2012 Vortex events in
  {Euler} and {Navier-Stokes} simulations with smooth initial conditions. {\em
  Journal of Fluid Mechanics\/} {\bf 690}, 288--320.

\bibitem[Pelinovsky(2012{\natexlab{{\em a\/}}})]{p12b}
{\sc Pelinovsky, D.} 2012{\natexlab{{\em a\/}}} Enstrophy growth in the viscous
  {Burgers} equation. {\em Dynamics of Partial Differential Equations\/} {\bf
  9}, 305--340.

\bibitem[Pelinovsky(2012{\natexlab{{\em b\/}}})]{p12}
{\sc Pelinovsky, D.} 2012{\natexlab{{\em b\/}}} Sharp bounds on enstrophy
  growth in the viscous {Burgers} equation. {\em Proceedings of Royal Society
  A\/} {\bf 468}, 3636--3648.

\bibitem[Pelz(2001)]{p01}
{\sc Pelz, R.~B.} 2001 Symmetry and the hydrodynamic blow--up problem. {\em
  Journal of Fluid Mechanics\/} {\bf 444}, 299--320.

\bibitem[Pouquet {\em et~al.\/}(1975)Pouquet, Lesieur, Andr\'e \&
  Basdevant]{plab75}
{\sc Pouquet, A., Lesieur, M., Andr\'e, J.~C. \& Basdevant, C.} 1975 Evolution
  of high {Reynolds} number two-dimensional turbulence. {\em Journal of Fluid
  Mechanics\/} {\bf 72}, 305--319.

\bibitem[Protas {\em et~al.\/}(1999)Protas, Babiano \& Kevlahan]{pkb99}
{\sc Protas, B., Babiano, A. \& Kevlahan, N. K.~R.} 1999 On geometrical
  alignment properties of two--dimensional forced turbulence. {\em Physica D\/}
  {\bf 128}, 169--179.

\bibitem[Protas {\em et~al.\/}(2004)Protas, Bewley \& Hagen]{pbh04}
{\sc Protas, B., Bewley, T. \& Hagen, G.} 2004 A comprehensive framework for
  the regularization of adjoint analysis in multiscale {PDE} systems. {\em
  Journal of Computational Physics\/} {\bf 195}, 49--89.

\bibitem[Rabin {\em et~al.\/}(2012)Rabin, Caulfield \& Kerswell]{rck12}
{\sc Rabin, S. M.~E., Caulfield, C.~P. \& Kerswell, R.~R.} 2012 Variational
  identification of minimal seeds to trigger transition in plane {Couette}
  flow. {\em Journal of Fluid Mechanics\/} {\bf 712}, 244--272.

\bibitem[Ruszczy\'nski(2006)]{r06}
{\sc Ruszczy\'nski, A.} 2006 {\em Nonlinear Optimization\/}. Princeton
  University Press.

\bibitem[Scott(2011)]{s11}
{\sc Scott, R.~K.} 2011 A scenario for finite-time singularity in the
  quasigeostrophic model. {\em Journal of Fluid Mechanics\/} {\bf 687},
  492--502.

\bibitem[Serrin(1963)]{s63}
{\sc Serrin, J.} 1963 The initial value problem for the {Navier-Stokes}
  equations. In {\em Nonlinear Problems\/} (ed. Rudolph~E. Langer), pp. 69--98.
  Mathematics Research Center, United States Army, The University of Wisconsin
  Press.

\bibitem[Siegel \& Caflisch(2009)]{sc09}
{\sc Siegel, M. \& Caflisch, R.~E.} 2009 Calculation of complex singular
  solutions to the {3D} incompressible {Euler} equations. {\em Physica D\/}
  {\bf 238}, 2368--2379.

\bibitem[Sulem {\em et~al.\/}(1983)Sulem, Sulem \& Frisch]{ssf83}
{\sc Sulem, C., Sulem, P.~L. \& Frisch, H.} 1983 Tracing complex singularities
  with spectral methods. {\em Journal of Computational Physics\/} {\bf 50},
  138--161.

\bibitem[Tran \& Dritschel(2006)]{td06}
{\sc Tran, Ch.~V. \& Dritschel, D.~G.} 2006 Vanishing enstrophy dissipation in
  two-dimensional {Navier--Stokes} turbulence in the inviscid limit. {\em
  Journal of Fluid Mechanics\/} {\bf 559}, 107--116.

\end{thebibliography}
\bibliographystyle{jfm}

\end{document}